\documentclass[twocolumn]{article}

\usepackage{geometry}
\usepackage{authblk}
\usepackage{cite}
\usepackage{amsmath}
\usepackage{braket}
\usepackage{qcircuit}
\usepackage{graphicx}
\usepackage{subfig}
\usepackage{multirow}

\title{Parametrized Constant-Depth Quantum Neuron}

\author[1]{Jonathan H. A. de Carvalho\thanks{Corresponding author}}
\author[1]{Fernando M. de Paula Neto}

\affil[1]{Centro de Informática, Universidade Federal de Pernambuco, Recife, Brazil}
\affil[ ]{\textit{\{jhac,fernando\}@cin.ufpe.br}}

\date{}

\begin{document}

\maketitle

\begin{abstract}

Quantum computing has been revolutionizing the development of algorithms. However, only noisy intermediate-scale quantum devices are available currently, which imposes several restrictions on the circuit implementation of quantum algorithms. In this paper, we propose a framework that builds quantum neurons based on kernel machines, where the quantum neurons differ from each other by their feature space mappings. Besides contemplating previous quantum neurons, our generalized framework has the capacity to instantiate other feature mappings that allow us to solve real problems better. Under that framework, we present a neuron that applies a tensor-product feature mapping to an exponentially larger space. The proposed neuron is implemented by a circuit of constant depth with a linear number of elementary single-qubit gates. The previous quantum neuron applies a phase-based feature mapping with an exponentially expensive circuit implementation, even using multi-qubit gates. Additionally, the proposed neuron has parameters that can change its activation function shape. Here, we show the activation function shape of each quantum neuron. It turns out that parametrization allows the proposed neuron to optimally fit underlying patterns that the existing neuron cannot fit, as demonstrated in the nonlinear toy classification problems addressed here. The feasibility of those quantum neuron solutions is also contemplated in the demonstration through executions on a quantum simulator. Finally, we compare those kernel-based quantum neurons in the problem of handwritten digit recognition, where the performances of quantum neurons that implement classical activation functions are also contrasted here. The repeated evidence of the parametrization potential achieved in real-life problems allows concluding that this work provides a quantum neuron with improved discriminative abilities. As a consequence, the generalized framework of quantum neurons can contribute toward practical quantum advantage.

\end{abstract}
\section{Introduction}

Quantum computing~\cite{nielsen_QC_QI} is expected to achieve the so-called supremacy over classical computing, which will allow us to solve previously intractable problems in a reasonable time~\cite{preskill_quantum_supremacy}. Specifically, that quantum advantage also allows the development of more efficient neural networks~\cite{nguyen_benchmarking_NNs}. Quantum neurons can implement arbitrary non-linear functions while taking advantage of quantum properties like superposition and entanglement~\cite{paula-neto_any_nonlinear-QN}. However, fault-tolerant quantum computations require encoding the information in many redundant qubits so that error correction codes can be employed~\cite{devitt_QEC_review}.

Currently, only noisy intermediate-scale quantum devices are available, so there are not enough resources to protect the system by quantum error correction~\cite{preskill_NISQ_era}. Depth, width, and number of operations become critical factors to successfully execute quantum circuits on the current hardware. Thus, aspects like state preparation, oracle expansion, connectivity, circuit rewriting, decoherence, gate infidelity, and measurement errors can compromise the computation~\cite{leymann_bitter_NISQ-era}.

Despite significant efforts at the quantum compiler level~\cite{zulehner_mapping_to_IBMQX, larose_qplatforms_overview}, recent quantum algorithms are developed considering those present restrictions~\cite{acasiete_QW_on_qdev, acampora_GA_on_qdev}, including the development of quantum neural networks~\cite{tacchino_qneuron_on_qdev, mangini_CVQN, tacchino_QNN_on_qdev, grant_hierarchical_qclassifiers, cong_convolutional_QNN, dallaire-demers_gen-adversarial_QNN, zoufal_QGAN_for_state-preparation, konar_QFS-Net}. Particularly, this work focuses on quantum perceptrons implemented efficiently on actual quantum devices. Tacchino et al.~\cite{tacchino_qneuron_on_qdev} took the first step by proposing a scheme that computes the inner product between binary-valued vectors. After, Mangini et al.~\cite{mangini_CVQN} reformulated that scheme to accept continuous-valued vectors instead of only binary-valued ones. Both schemes have an exponential advantage in terms of circuit width, although the circuit depth and the number of operations grow exponentially.

In this work, we propose a framework of quantum neurons where those previous schemes~\cite{tacchino_qneuron_on_qdev, mangini_CVQN} are particular cases. Based on that framework, each quantum neuron implements a kernel machine with a non-deterministic activation function. That activation function depends only on the kernel trick each quantum neuron applies. Thus, the quantum neurons differ from each other by their feature space mappings. It makes room to instantiate other quantum neurons under that generalized framework, including for actual quantum devices.

Generally, quantum kernel methods are used to estimate inner products for classical models in a hybrid setup. The advantage emerges from the fact that the inner products are estimated from classically intractable feature mappings. Those inner products can be estimated by the standard swap-test~\cite{buhrman_swap-test} or, more recently, by quantum kernel estimators~\cite{havlicek_QKE, schuld_QKE}. Speedups in classical models can be also obtained by classical sampling techniques inspired by quantum models that generate the kernel matrix of inner products~\cite{ding_q-inspired_SVM}.

The framework of quantum neurons proposed here computes the inner product between an input vector and a weight vector, and then explicitly extracts such information to an ancillary qubit. Measuring the ancilla gives a standalone fully-quantum classifier based on kernel machines. If the ancilla is not measured, the inner product can be propagated forward to other quantum neurons, which gives a quantum neural network~\cite{tacchino_QNN_on_qdev}. Finally, the proposed framework of quantum neurons is expected to achieve practical quantum advantage, as is expected with quantum kernel estimators~\cite{liu_qspeedup_in_ML}.

Based on that framework, we also propose a quantum neuron of constant depth, i.e., its circuit depth is independent of the input size. Constant-depth quantum circuits can demonstrate quantum advantage~\cite{bravyi_qadvantage_shallow-circ, bravyi_qadvantage_noisy-shallow-circ} and benefit from error mitigation techniques~\cite{li_variational_error_minimization, temme_error_mitigation, kandala_error_mitigation}. The proposed quantum neuron implements local feature mappings~\cite{stoudenmire_local_feature-map} by taking advantage of qubit encoding~\cite{grant_hierarchical_qclassifiers, konar_QFS-Net}. We demonstrate that encoding strategy actually implements a tensor-product feature mapping to an exponentially larger space, which improves the separating capacity~\cite{cover_theorem}.

We further improve that neuron capacity by including two parameters in its activation function. That parametrization can change the activation function shape of the proposed quantum neuron in order to fit different underlying patterns with no additional cost in the circuit implementation. Therefore, we propose a flexible quantum neuron of constant depth implemented with a linear number of elementary single-qubit gates. The existing quantum neuron~\cite{mangini_CVQN} is inflexible and exponentially expensive, even with multi-qubit gates in its circuit implementation.

Then, we proceed to a visual study that relates some interactions between the input and weight vectors in the original space with the respective neuron outputs in the feature space. Those activation function shapes reveal the problem structures that each quantum neuron can solve. By comparing the best solutions of each quantum neuron in toy classification problems, we demonstrate that parametrization can change the activation function shape in order to optimally fit all cases, even those that the existing quantum neuron cannot fit. We also demonstrate the feasibility of those quantum neuron solutions through a proof-of-concept experiment on a quantum simulator. A conclusive experiment about the neuron capabilities is finally conducted here as we address the recognition of handwritten digits and compare the results against quantum neurons based on another architecture. Those results in real-life problems attest that the proposed quantum neuron, especially due to the activation function parametrization, has better discriminative power than the previous proposals of quantum neurons.

This paper is organized as follows. Firstly, Section~\ref{sec:qn_framework} formalizes the quantum neuron framework based on kernel machines. The reader that is not familiar with the basic concepts of quantum computing should refer to~\cite{nielsen_QC_QI, yanofsky_QC_for_comp-sci}. Under the proposed framework, Section~\ref{sec:pcdqn} presents the parametrized quantum neuron of constant circuit depth. The activation function shapes of the quantum neurons are presented in Section~\ref{sec:act_functions}. Then, Section~\ref{sec:qn_demo} compares the quantum neurons in solving some toy classification problems. Section~\ref{sec:mnist_exp} addresses the task of handwritten digit recognition, including comparisons with another model of quantum neurons. Final remarks and future directions are discussed in Section~\ref{sec:conclusions}.
\section{Quantum Neuron Framework}
\label{sec:qn_framework}

The classical neuron model basically consists of two steps. First, the inner product between the input vector $\vec{i}$ and the weight vector $\vec{w}$ is computed, and then that inner product $\sum_j w_j i_j$ is passed to an activation function $\varphi(\cdot)$ in order to define the neuron output $y$~\cite{haykin_neural-nets}. Based on that model, one can construct different classical neurons by only changing the activation function that processes the obtained inner product. Examples of common activation functions are the Heaviside function and the logistic sigmoid function, which are presented in~(\ref{eq:heaviside_function}) and~(\ref{eq:logsig_function}) respectively.

\begin{equation}
    \varphi(\vec{w} \cdot \vec{i}) =
        \begin{cases}
            1, & \text{if}\ \vec{w} \cdot \vec{i} \geq 0 \\ 
            0, & \text{if}\ \vec{w} \cdot \vec{i} < 0
        \end{cases}
\label{eq:heaviside_function}
\end{equation}

\begin{equation}
    \varphi(\vec{w} \cdot \vec{i}) = \frac{1}{1 + e^{-(\vec{w} \cdot \vec{i})}}
\label{eq:logsig_function}
\end{equation}

Inspired by that classical neuron model, Tacchino et al.~\cite{tacchino_qneuron_on_qdev} proposed a scheme for actual quantum devices that encodes input and weight vectors, computes the inner product between them, and finally extracts the activation function output. Here, we generalize that scheme to a framework of quantum neurons based on kernel machines. Given two vectors $\boldsymbol{\phi}$ and $\boldsymbol{\theta}$ in an input space, a kernel machine $k(\cdot, \cdot)$ maps those vectors to other two vectors $\vec{w}$ and $\vec{i}$ in a feature space by a nonlinear transformation $\Phi(\cdot)$, and then computes the inner product between the transformed vectors, i.e., $k(\boldsymbol{\phi}, \boldsymbol{\theta}) = \Phi(\boldsymbol{\phi})^T \Phi(\boldsymbol{\theta}) = \vec{w} \cdot \vec{i}$~\cite{bishop_PRML}. The quantum neurons constructed by the framework are deeply related to that kernel trick, which paves the way to the quantum neuron that we propose in this work.

The quantum neurons implement kernel methods because the given $m$-dimensional classical vectors $\boldsymbol{\theta}$ and $\boldsymbol{\phi}$ are first mapped to $N$-dimensional quantum vectors $\vec{i} = \Phi(\boldsymbol{\theta})$ and $\vec{w} = \Phi(\boldsymbol{\phi})$. Those quantum vectors can be directly encoded in legitimate quantum states in the following way:

\begin{equation*}
    \ket{\psi_i} = \sum_{j=0}^{N-1} i_j \ket{j}
    \qquad \text{and} \qquad
    \ket{\psi_w} = \sum_{j=0}^{N-1} w_j \ket{j}.
\end{equation*}

Then, the inner product $\vec{w}^* \cdot \vec{i} = \langle \psi_w | \psi_i \rangle$ is computed by the quantum neurons. The final neuron output is given by the non-deterministic activation function presented in~(\ref{eq:framework_act_func}).

\begin{equation}
    \varphi(\vec{w}^* \cdot \vec{i}) = 
        \begin{cases}
            1, & \text{with probability}\ |\vec{w}^* \cdot \vec{i}|^2 \\
            0, & \text{with probability}\ 1 - |\vec{w}^* \cdot \vec{i}|^2
        \end{cases}
\label{eq:framework_act_func}
\end{equation}

In this way, the quantum activation function depends on the inner product $\vec{w}^* \cdot \vec{i}$ that in turn depends on the feature space mapping $\Phi(\cdot)$. Based on that framework, one can realize different quantum neurons, including on actual quantum computers, by only changing the mapping $\Phi(\cdot)$ that each quantum neuron implements.

\begin{figure}[!t]
    \centering
    \leavevmode
    \scriptsize
    \Qcircuit @C=1.0em @R=0.6em @!R {
        & \makebox[5.1\width][r]{$U_i$} & & & & \makebox[2.3\width][l]{$U_w$} & & & & \\
        \lstick{\ket{0}} & \gate{H} & \multigate{4}{E(\boldsymbol{\theta})} & \qw & \multigate{4}{D(\boldsymbol{\phi})} & \gate{H} & \gate{X} & \ctrl{1} & \qw & \qw \\
        \lstick{\ket{0}} & \gate{H} & \ghost{E(\boldsymbol{\theta})} & \qw & \ghost{D(\boldsymbol{\phi})} & \gate{H} & \gate{X} & \control \qw & \qw & \qw \\
        \lstick{\vdots \phantom{0}} & \vdots & \nghost{E(\boldsymbol{\theta})} & & \nghost{D(\boldsymbol{\phi})} & \vdots & \vdots & \vdots & & \\
        \lstick{\ket{0}} & \gate{H} & \ghost{E(\boldsymbol{\theta})} & \qw & \ghost{D(\boldsymbol{\phi})} & \gate{H} & \gate{X} & \ctrl{1} & \qw & \qw \\
        \lstick{\ket{0}} & \gate{H} & \ghost{E(\boldsymbol{\theta})} & \qw & \ghost{D(\boldsymbol{\phi})} & \gate{H} & \gate{X} & \ctrl{1} & \qw & \qw \\
        \lstick{\ket{0}} & \dstick{\text{{\scriptsize Ancilla \phantom{0}}}} \qw & \qw & \qw & \qw & \qw & \qw & \targ & \meter & \cw \gategroup{2}{2}{6}{3}{1em}{--} \gategroup{2}{5}{6}{7}{1em}{--} \\
    }
    \vspace{0.5em}
    \caption{Framework circuit to build quantum neurons based on kernel machines. First, $U_i$ encodes $\vec{i} = \Phi(\boldsymbol{\theta})$. Then, $U_w$ computes the inner product between $\vec{i}$ and $\vec{w} = \Phi(\boldsymbol{\phi})$. Such information is extracted to an ancillary qubit that gives a non-deterministic activation function when finally measured.}
    \label{fig:qn_framework}
\end{figure}
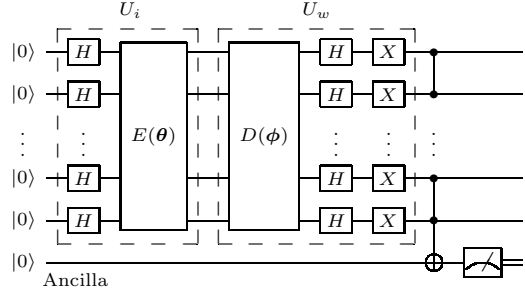

Figure~\ref{fig:qn_framework} shows the circuit implementation of the framework of quantum neurons based on kernel machines. Specifically, the quantum neurons differ from each other by the quantum operators $E$ and $D$, which depend on the classical vectors $\boldsymbol{\theta}$ and $\boldsymbol{\phi}$ respectively. The operator $E$ maps $\boldsymbol{\theta}$ to $\vec{i}$ and then encodes it in $\ket{\psi_i}$ from $\ket{+}^{\otimes n}$, where $\ket{+} = \frac{1}{\sqrt{2}} (\ket{0} + \ket{1})$ and $n$ is the number of encoding qubits. Thus,

\begin{equation*}
    E(\boldsymbol{\theta}) \ket{+}^{\otimes n} = \ket{\psi_i}.
\end{equation*}

The operator $D$ maps $\boldsymbol{\phi}$ to $\vec{w}$ and then decodes it from $\ket{\psi_w}$ to $\ket{+}^{\otimes n}$. Thus,

\begin{equation*}
    D(\boldsymbol{\phi}) \ket{\psi_w} = \ket{+}^{\otimes n}.
\end{equation*}

At a higher level of abstraction, the operator $U_i$ encodes $\ket{\psi_i}$ from a blank register $\ket{0}^{\otimes n}$:

\begin{equation*}
    U_i \ket{0}^{\otimes n} = E(\boldsymbol{\theta}) H^{\otimes n} \ket{0}^{\otimes n} = E(\boldsymbol{\theta}) \ket{+}^{\otimes n} = \ket{\psi_i}.
\end{equation*}

On the other hand, $U_w$ decodes $\ket{\psi_w}$ to $\ket{1}^{\otimes n}$:

\begin{align*}
    U_w \ket{\psi_w} = X^{\otimes n} H^{\otimes n} D(\boldsymbol{\phi}) \ket{\psi_w} &= X^{\otimes n} H^{\otimes n} \ket{+}^{\otimes n} \\
    &= X^{\otimes n} \ket{0}^{\otimes n} \\
    &= \ket{1}^{\otimes n}.
\end{align*}

The operator $D(\boldsymbol{\phi})$ and then $U_w$ are defined according to the action on $\ket{\psi_w}$, but those operators are applied in $\ket{\psi_i}$ in the circuit flow. Applying $U_w$ to $\ket{\psi_i}$ actually computes the inner product between $\vec{i}$ and $\vec{w}$ by taking advantage of an interesting property of unitary operators. Unitary operators preserve the space geometry, which includes inner products~\cite{yanofsky_QC_for_comp-sci}. If $A$ is a unitary operator, the inner product between a quantum state $\ket{v}$ and another quantum state $\ket{v'}$ is equal to the inner product between $A \ket{v}$ and $A \ket{v'}$, i.e., $\langle v' | v \rangle = \langle v' A^\dagger | A v \rangle$. Let $A = U_w$, $\ket{v} = \ket{\psi_i}$, and $\ket{v'} = \ket{\psi_w}$. We can rewrite the expression as:

\begin{equation*}
    \langle \psi_w | \psi_i \rangle = \langle \psi_w U_w^\dagger | U_w \psi_i \rangle = \langle 1^{\otimes n} | U_w \psi_i \rangle.
\end{equation*}

Thus, the projection of $U_w \ket{\psi_i}$ on the basis state $\ket{1}^{\otimes n}$ stores the inner product $\langle \psi_w | \psi_i \rangle$. In other words, that inner product is stored in the component of $\ket{1}^{\otimes n}$ into the superposition of $U_w \ket{\psi_i} = \ket{\psi_{i,w}}$. In this way,

\begin{multline*}
    \ket{\psi_{i,w}} = c_0 \ket{0} + c_1 \ket{1} + \cdots \\
    + c_{N-2} \ket{N-2} + \langle \psi_w | \psi_i \rangle \ket{N-1}.
\end{multline*}

To extract such information, a multi-controlled NOT gate is applied targeting an ancillary qubit. Note that only the basis state $\ket{N-1}$ in the superposition flips the ancilla to the state $\ket{1}$. Consequently, the final quantum state $\ket{\psi_f}$ is:

\begin{multline*}
    \ket{\psi_f} = c_0 \ket{0}\ket{0} + c_1 \ket{1}\ket{0} + \cdots \\
    + c_{N-2} \ket{N-2}\ket{0} + \langle \psi_w | \psi_i \rangle \ket{N-1}\ket{1}.
\end{multline*}

Since $\langle \psi_w | \psi_i \rangle = \vec{w}^* \cdot \vec{i}$, measuring that ancillary qubit finally gives the activation function already presented in~(\ref{eq:framework_act_func}). Therefore, that quantum neuron framework applies the kernel trick and activates following a non-deterministic function of the computed inner product, which was to be shown. To instantiate a quantum neuron, one only needs to realize a feature mapping $\Phi(\cdot)$ by means of the operators $E$ and $D$.

\subsection{Binary-Valued Quantum Neuron}

That generalized framework can be first instantiated in the binary-valued quantum neuron (BVQN)~\cite{tacchino_qneuron_on_qdev}. Given $m$-dimensional classical vectors $\boldsymbol{\theta} = (\theta_0, \theta_1, \cdots, \theta_{m-2}, \theta_{m-1})$ and $\boldsymbol{\phi} = (\phi_0, \phi_1, \cdots, \phi_{m-2}, \phi_{m-1})$, where $\theta_j, \phi_j \in \{ -1,1 \}$, the BVQN feature mapping simply normalizes those classical vectors, which leads to the following $N$-dimensional quantum vectors, where $N = m$:

\begin{align*}
    \vec{i} &= \Phi(\boldsymbol{\theta}) = \frac{1}{\sqrt{N}} \Big( \theta_0, \theta_1, \cdots, \theta_{m-2}, \theta_{m-1} \Big)
    \text{ and} \\
    \vec{w} &= \Phi(\boldsymbol{\phi}) = \frac{1}{\sqrt{N}} \Big( \phi_0, \phi_1, \cdots, \phi_{m-2}, \phi_{m-1} \Big).
\end{align*}

Those quantum vectors live in the Hilbert space spanned by $n$ qubits, where $N = 2^n$. As $N$ equals $m$, the BVQN has an exponential advantage in storing the information.

Note that $\ket{+}^{\otimes n} = \frac{1}{\sqrt{N}} \sum_{j=0}^{N-1} \ket{j}$, so $E(\boldsymbol{\theta})$ only needs to flip the sign of the basis states $\ket{j}$ where $\theta_j = -1$ to encode $\vec{i}$ in $\ket{\psi_i}$. Similarly, $D(\boldsymbol{\phi})$ only needs to cancel the sign of the basis states $\ket{j}$ where $\phi_j = -1$ to decode $\vec{w}$ from $\ket{\psi_w}$. Those operations are accomplished by applying sign-flip blocks one by one or, more efficiently, by applying the hypergraph states generation subroutine~\cite{tacchino_qneuron_on_qdev, rossi_qhypergraph-states}. However, those two strategies are exponentially expensive in terms of circuit depth. Finally, the BVQN fires with the probability presented in~(\ref{eq:bvqn_act_func}).

\begin{equation}
    |\vec{w}^* \cdot \vec{i}|^2 = \Bigg| \frac{1}{N} \sum_{j=0}^{m-1} \phi_j \theta_j \Bigg| ^2
\label{eq:bvqn_act_func}
\end{equation}

\subsection{Continuous-Valued Quantum Neuron}

A first alternative to accepting real vectors instead of only binary ones is to instantiate the framework in the continuous-valued quantum neuron (CVQN)~\cite{mangini_CVQN}, which implements a phase-based feature mapping. That mapping encodes each classical component in the phase of a complex number written in the exponential form with modulus $1/\sqrt{N}$. Thus, the classical vectors are mapped to quantum vectors of equal size in the following manner:

\begin{equation*}
    \vec{i} = \Phi(\boldsymbol{\theta}) = \frac{1}{\sqrt{N}} \Big( e^{i \theta_0}, e^{i \theta_1}, \cdots, e^{i \theta_{m-2}}, e^{i \theta_{m-1}} \Big)
\end{equation*}
and
\begin{equation*}
    \vec{w} = \Phi(\boldsymbol{\phi}) = \frac{1}{\sqrt{N}} \Big( e^{i \phi_0}, e^{i \phi_1}, \cdots, e^{i \phi_{m-2}}, e^{i \phi_{m-1}} \Big).
\end{equation*}

Periodicity in that phase-based scheme and the square modulus in the activation function make the CVQN equally recognize values that are actually different. To really distinguish such values, the classical vectors are scaled to the interval $[0, \pi/2]$. Thus, the feature mapping is applied to classical vectors $\boldsymbol{\theta}$ and $\boldsymbol{\phi}$, where $\theta_j, \phi_j \in [0, \pi/2]$. The $N$-dimensional quantum vectors, where $N = m$, are described by $n$ qubits, where $N = 2^n$, which represents an exponential advantage again in information storage.

To implement $E(\boldsymbol{\theta})$ and $D(\boldsymbol{\phi})$, the blocks that flip the sign in the BVQN are used here to encode each $e^{i \theta_j}$ and to decode each $e^{i \phi_j}$, respectively, by applying multi-controlled phase gates instead of multi-controlled $Z$ gates. The phase gate $P(\lambda)$ is represented in the matrix form by 
$\begin{pmatrix}
    1 & 0 \\
    0 & e^{i \lambda}
\end{pmatrix}$.
As sign-flip blocks, phase-shift blocks are also exponentially expensive in terms of circuit depth. Finally, the CVQN fires with the probability presented in~(\ref{eq:cvqn_act_func}).

\begin{equation}
    |\vec{w}^* \cdot \vec{i}|^2 = \Bigg| \frac{1}{N} \sum_{j=0}^{m-1} e^{i(\theta_j - \phi_j)} \Bigg| ^2
\label{eq:cvqn_act_func}
\end{equation}
\section{Proposed Quantum Neuron}
\label{sec:pcdqn}

Encoding data in the amplitudes of quantum states provides a gain in storage, but the circuit depth grows exponentially. The implication is that amplitude encoding requires a few qubits that would be controlled for a long time and would suffer many manipulations. However, in the current era of quantum computing, the circuit depth and the number of operations really matter due to the poor quality of the qubits available in the quantum devices.

Here, we propose to use a qubit encoding scheme under that framework of quantum neurons, differently from the amplitude encoding used by the BVQN and the CVQN. In this way, an $m$-dimensional classical vector is encoded in $m$ qubits, which represents less efficient storage, but the state preparation is definitely efficient in terms of circuit depth as each qubit only requires a single rotation to encode the data~\cite{grant_hierarchical_qclassifiers, konar_QFS-Net}. That is, qubit encoding requires more qubits that would be controlled for less time and would suffer fewer manipulations. In light of the present circuit requirements, qubit encoding emerges as a suitable encoding strategy to explore. By taking advantage of the qubit encoding efficiency, we can propose the constant-depth quantum neuron (CDQN).

Instead of a qubit encoding from rotations about the y-axis~\cite{grant_hierarchical_qclassifiers, konar_QFS-Net}, we apply that phase shift gate $P(\lambda)$ already used in amplitude encoding previously~\cite{mangini_CVQN}. Thus, the classical input vector $\boldsymbol{\theta}$ is directly encoded by the operator $E$ as follows:

\begin{equation*}
    E(\boldsymbol{\theta}) = \otimes_{j=0}^{m-1} P(\theta_j).
\end{equation*}

Similarly, the classical weight vector $\boldsymbol{\phi}$ is directly decoded by the operator $D$ in the following way:

\begin{equation*}
    D(\boldsymbol{\phi}) = \otimes_{j=0}^{m-1} P(-\phi_j).
\end{equation*}

As a result, $U_i$ comes down to:

\begin{equation*}
    U_i = \otimes_{j=0}^{m-1} P(\theta_j) H^{\otimes m}.
\end{equation*}

In turn, $U_w$ comes down to:

\begin{equation*}
    U_w = X^{\otimes m} H^{\otimes m} \otimes_{j=0}^{m-1} P(-\phi_j).
\end{equation*}

Therefore, with a circuit depth independent of the input size, the proposed quantum neuron computes the inner product between two vectors that are transformed by a tensor product of local feature mappings~\cite{stoudenmire_local_feature-map}. Those classical vectors are also pre-processed to the interval $[0, \pi/2]$, as for the CVQN.

Applying that $E(\boldsymbol{\theta})$ in a basis state $\ket{s}$ produces the following state:

\begin{equation*}
    E(\boldsymbol{\theta}) \ket{s} = 
    \prod_{j=0}^{m-1} (e^{i \theta_j})^{b_{j,s,m}} \ket{s},
\end{equation*}
where $b_{j,s,m}$ is the $j$-th bit in the binary representation of the integer value $s$ with $m$ bits. In this way, $b_{j,s,m}$ selects the phase shifts to be applied. For example, with $m=4$, applying $E(\boldsymbol{\theta})$ in $\ket{5}$ produces $e^{i \theta_1} e^{i \theta_3} \ket{5}$, using the big-endian convention. Applying $E(\boldsymbol{\theta})$ in $\ket{+}^{\otimes m}$ stores a conditioned product of $e^{i \theta_j}$ on the amplitude of each possible state $\ket{s}$, which represents the following feature mapping, where $N = 2^m$:

\begin{multline*}
    \vec{i} = \Phi(\boldsymbol{\theta}) = \\ \frac{1}{\sqrt{N}} \Big(
    \prod_{j=0}^{m-1} (e^{i \theta_j})^{b_{j,0,m}},
    \prod_{j=0}^{m-1} (e^{i \theta_j})^{b_{j,1,m}}, \cdots, \\
    \prod_{j=0}^{m-1} (e^{i \theta_j})^{b_{j, N-2, m}},
    \prod_{j=0}^{m-1} (e^{i \theta_j})^{b_{j, N-1, m}} \Big)
\end{multline*}
and

\begin{multline*}
    \vec{w} = \Phi(\boldsymbol{\phi}) = \\ \frac{1}{\sqrt{N}} \Big(
    \prod_{j=0}^{m-1} (e^{i \phi_j})^{b_{j,0,m}},
    \prod_{j=0}^{m-1} (e^{i \phi_j})^{b_{j,1,m}}, \cdots, \\
    \prod_{j=0}^{m-1} (e^{i \phi_j})^{b_{j, N-2, m}},
    \prod_{j=0}^{m-1} (e^{i \phi_j})^{b_{j, N-1, m}} \Big).
\end{multline*}

Different from the previous quantum neurons, the CDQN maps $m$-dimensional classical vectors to $N$-dimensional quantum vectors with $N > m$. Specifically, the CDQN maps the vectors to an exponentially larger space. Thus, the CDQN is a kernel method that takes advantage of Cover's theorem~\cite{cover_theorem}, which states that mapping to feature spaces of high dimensionality improves the separating capacity. Non-linearly separable problems are more likely to become linearly separable in a high-dimensional space. Finally, the CDQN fires with the probability presented in~(\ref{eq:cdqn_act_func}).

\begin{equation}
    |\vec{w}^* \cdot \vec{i}|^2 = \Bigg| \frac{1}{N} \sum_{s=0}^{N-1} \prod_{j=0}^{m-1} (e^{i (\theta_j - \phi_j)})^{b_{j,s,m}} \Bigg| ^2
\label{eq:cdqn_act_func}
\end{equation}

We can simplify the relation presented in~(\ref{eq:cdqn_act_func}) by observing that it is actually a combination of $m$ local inner products. Each local inner product is computed with respect to a local feature mapping that maps a $\theta_j$ to $\vec{i_j} = \frac{1}{\sqrt{2}} (1, e^{i \theta_j})$ and a $\phi_j$ to $\vec{w_j} = \frac{1}{\sqrt{2}} (1, e^{i \phi_j})$. As a result, the relation presented in~(\ref{eq:cdqn_act_func}) is equivalent to a product of $m$ local inner products in the form $\vec{w_j}^* \cdot \vec{i_j} = \frac{1}{2} (1 + e^{i (\theta_j - \phi_j)})$. Therefore, the CDQN fires with a probability that can be simplified as presented in~(\ref{eq:cdqn_simplified_act_func}).

\begin{equation}
    |\vec{w}^* \cdot \vec{i}|^2 = \Bigg| \frac{1}{N} \prod_{j=0}^{m-1} (1 + e^{i (\theta_j - \phi_j)}) \Bigg| ^2
\label{eq:cdqn_simplified_act_func}
\end{equation}

Building on the CDQN, we also propose to parametrize its activation function by a multiplicative factor $\tau$ and an additive constant $\delta$. Those parameters act on the component-wise differences between the classical input and weight vectors. In this way, the parametrized constant-depth quantum neuron (PCDQN) fires with the probability presented in~(\ref{eq:pcdqn_act_func}). By adjusting $\tau$ and $\delta$, the PCDQN can change its activation function shape. Thus, that parametrization generates flexibility. Such flexibility is supposed to allow the PCDQN to fit different underlying patterns.

\begin{equation}
    |\vec{w}^* \cdot \vec{i}|^2 = \Bigg| \frac{1}{N} \prod_{j=0}^{m-1} (1 + e^{i [\tau (\theta_j - \phi_j) + \delta]}) \Bigg| ^2
\label{eq:pcdqn_act_func}
\end{equation}

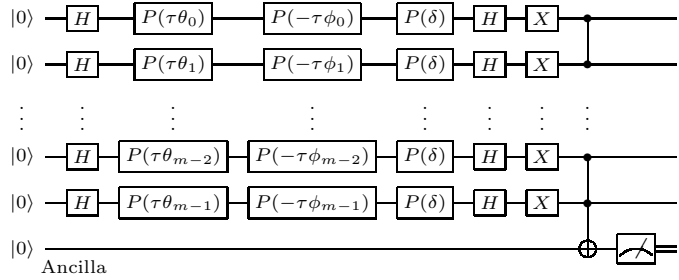
\begin{figure*}[!t]
    \centering
    \leavevmode
    \scriptsize
    \Qcircuit @C=1.0em @R=0.7em @!R {
        \lstick{\ket{0}} & \gate{H} & \gate{P(\tau \theta_0)} & \gate{P(-\tau \phi_0)} & \gate{P(\delta)} & \gate{H} & \gate{X} & \ctrl{1} & \qw & \qw \\
        \lstick{\ket{0}} & \gate{H} & \gate{P(\tau \theta_1)} & \gate{P(-\tau \phi_1)} & \gate{P(\delta)} & \gate{H} & \gate{X} & \control \qw & \qw & \qw \\
        \lstick{\vdots \phantom{0}} & \vdots & \vdots & \vdots & \vdots & \vdots & \vdots & \vdots & & \\
        \lstick{\ket{0}} & \gate{H} & \gate{P(\tau \theta_{m-2})} & \gate{P(-\tau \phi_{m-2})} & \gate{P(\delta)} & \gate{H} & \gate{X} & \ctrl{1} & \qw & \qw \\
        \lstick{\ket{0}} & \gate{H} & \gate{P(\tau \theta_{m-1})} & \gate{P(-\tau \phi_{m-1})} & \gate{P(\delta)} & \gate{H} & \gate{X} & \ctrl{1} & \qw & \qw \\
        \lstick{\ket{0}} & \dstick{\text{{\scriptsize Ancilla \phantom{0}}}} \qw & \qw & \qw & \qw & \qw & \qw & \targ & \meter & \cw \\
    }
    \vspace{0.5em}
    \caption{Circuit implementation of the PCDQN. Qubit encoding provides a constant-depth circuit with a linear number of elementary single-qubit gates. Additionally, the parameters $\tau$ and $\delta$ can change the PCDQN activation function shape.}
    \label{fig:pcdqn_circuit}
\end{figure*}

At the circuit level, that parametrization is achieved by passing $\tau \boldsymbol{\theta}$ as the argument of $E(\cdot)$ and $\tau \boldsymbol{\phi}$ as the argument of $D(\cdot)$ followed by applying $P(\delta)$ to each qubit. Figure~\ref{fig:pcdqn_circuit} shows the circuit implementation of the PCDQN. As can be seen, the PCDQN is implemented by a circuit of constant depth with a linear number of elementary single-qubit gates. The PCDQN is not only efficient but also flexible due to $\tau$ and $\delta$.

The PCDQN can be even more efficient by applying only one phase gate $P(\lambda)$ to each qubit, where $\lambda = \tau (\theta_j - \phi_j) + \delta$. Therefore, the parametrization can be achieved with no additional cost in the circuit and the total circuit depth becomes 5. However, the circuit presented in Figure~\ref{fig:pcdqn_circuit} has a total depth of 7 because 3 phase gates are respectively used to encode the input, decode the weight, and add the parameter $\delta$, where the number of operations is also increased by $m$ due to the parametrization.

\begin{figure}[!t]
    \centering
    \subfloat[]{
        \includegraphics[width=0.47\columnwidth]{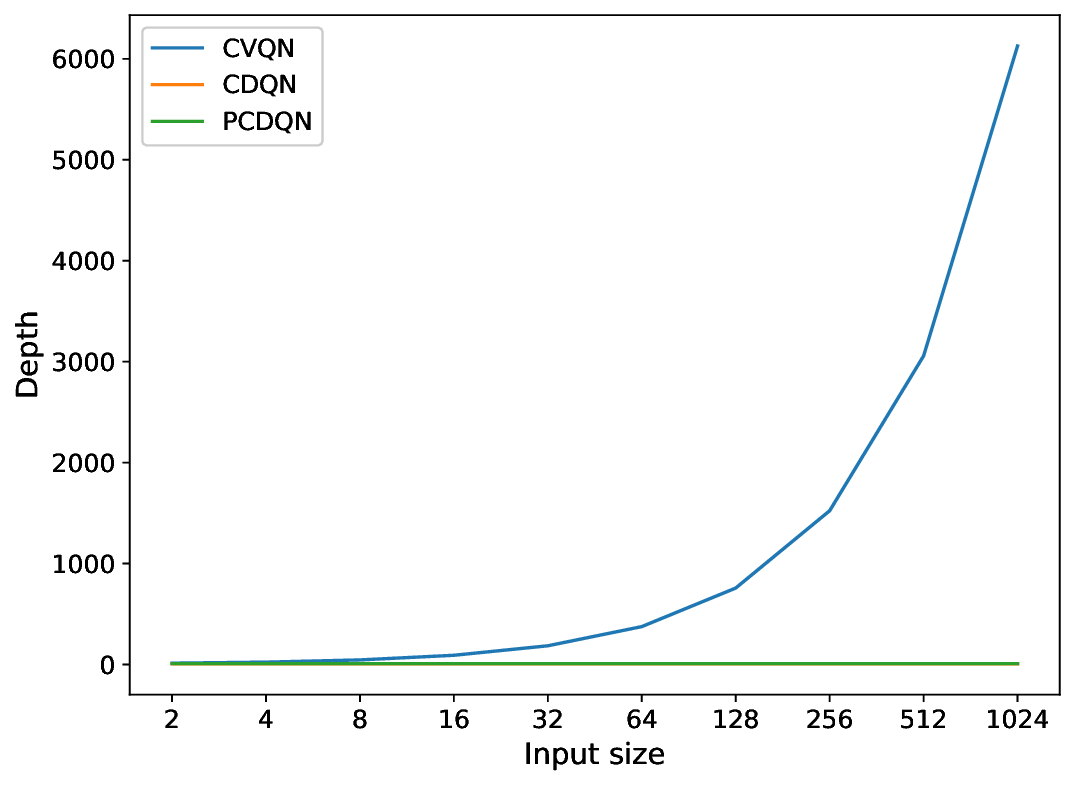}
        \label{fig:depth_growth}
    }
    \subfloat[]{
        \includegraphics[width=0.47\columnwidth]{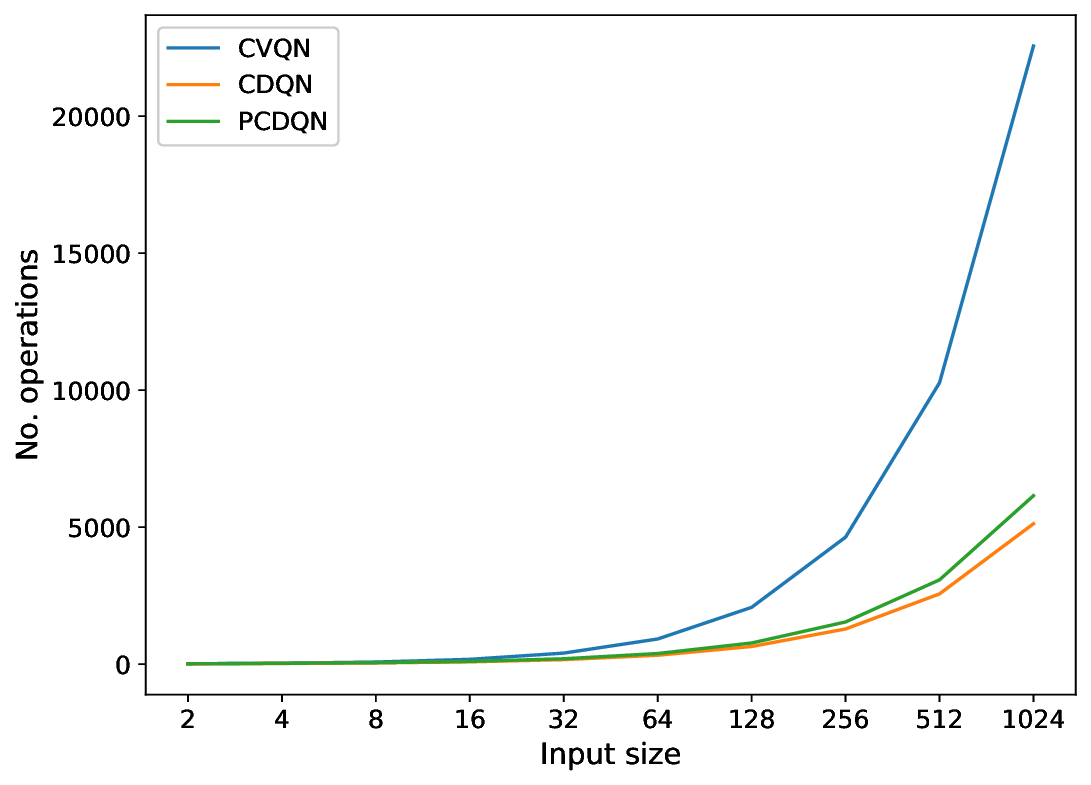}
        \label{fig:op_number_growth}
    }
    \caption{Circuit growth of each quantum neuron as the input size increases. The blue, orange, and green curves represent the CVQN, the CDQN, and the PCDQN growths, respectively. (a) in terms of circuit depth. (b) in terms of the number of operations.}
    \label{fig:circuit_growth}
\end{figure}

To demonstrate the efficiency of the CDQN and the PCDQN over the CVQN, Figure~\ref{fig:circuit_growth} compares the circuit growth of each quantum neuron as the input size increases. Figure~\ref{fig:depth_growth} compares in terms of circuit depth and Figure~\ref{fig:op_number_growth} compares in terms of number of operations by using, respectively, the \textit{depth} and \textit{size} functions provided by Qiskit~\cite{qiskit}. The blue curve represents the CVQN, the orange curve represents the CDQN, and the green curve represents the PCDQN.

As can be seen, the CVQN grows linearly with the input size $m$ in terms of depth and number of operations. Since $m = 2^n$ for the CVQN, that quantum neuron grows exponentially with the number of qubits $n$. On the other hand, the CDQN and the PCDQN have a constant depth and a lower number of operations. Additionally, the CDQN and the PCDQN are efficiently implemented with elementary single-qubit gates only, while the CVQN is expensive, even using complex phase-shift blocks.
\section{Activation Function Shapes}
\label{sec:act_functions}

Classical neuron outputs depend on some interaction between the input and weight vectors. For example, the larger the inner product between the vectors, the larger the output of the sigmoid activation function. As another example, the larger the Euclidean distance between the vectors, the smaller the output of the radial basis activation function. On the other hand, the quantum neuron outputs depend on the inner product between the input and weight vectors in the feature space. In this work, we study the quantum neuron outputs as a function of some relations between the original vectors, which can be seen as an indirect way to visualize quantum activation shapes.

Then, we defined a set of two-dimensional input vectors by splitting the interval $[0, \pi/2]$ into ten equidistant values and taking all possible combinations, which gives an input set of one hundred vectors. The weight vector is fixed here as $(\pi/2, \pi/2)$. Thus, we compute the classical relations and the corresponding quantum neuron outputs for each input vector with respect to that weight vector of reference. Input vectors that give the same measure to that reference vector do not necessarily give the same neuron outputs. Specifically, we considered the following metrics provided by scikit-learn~\cite{scikit-learn}: Manhattan distance, Euclidean distance, linear kernel, polynomial kernel, RBF kernel, and sigmoid kernel.

\begin{figure}[!t]
    \centering
    \subfloat[]{
        \includegraphics[width=0.3\columnwidth]{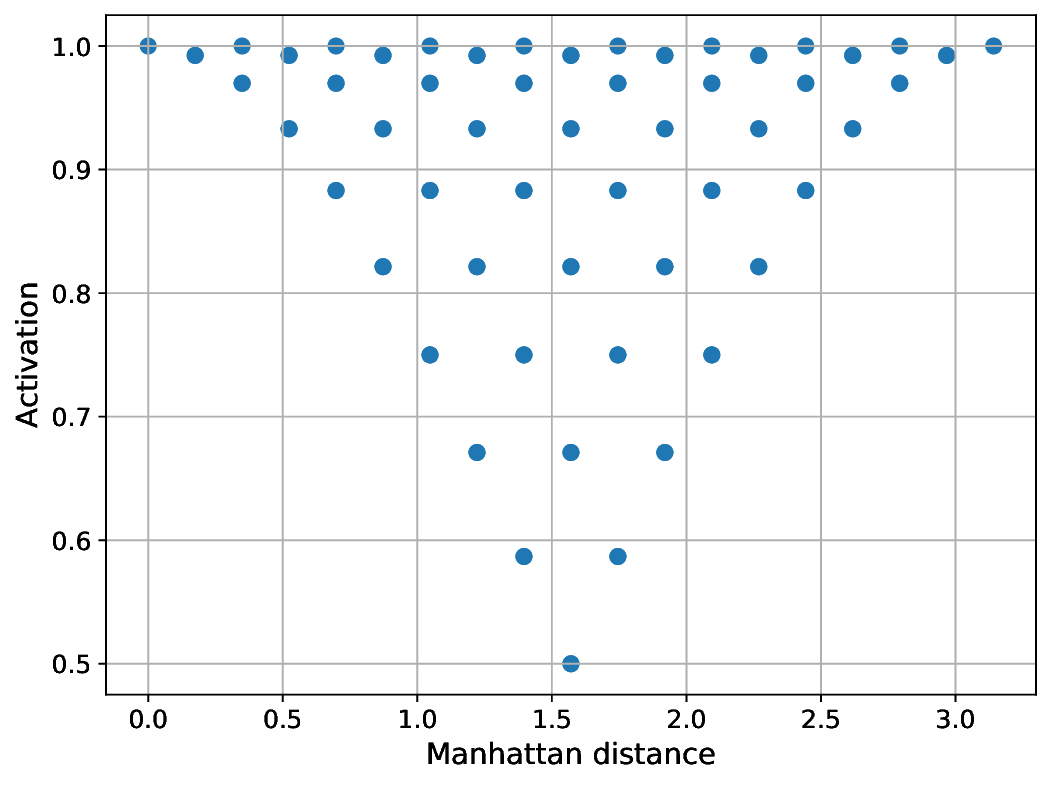}
        \label{fig:cvqn_act_function_for_manhattan}
    }
    \subfloat[]{
        \includegraphics[width=0.3\columnwidth]{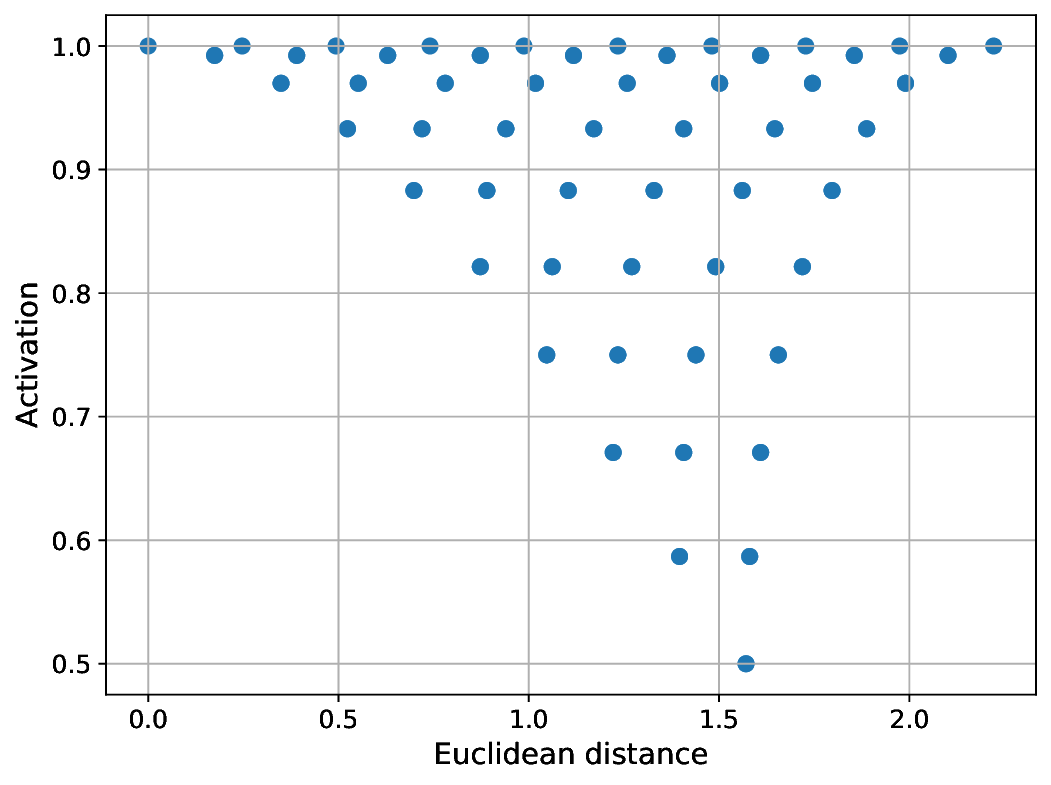}
        \label{fig:cvqn_act_function_for_euclidean}
    }
    \subfloat[]{
        \includegraphics[width=0.3\columnwidth]{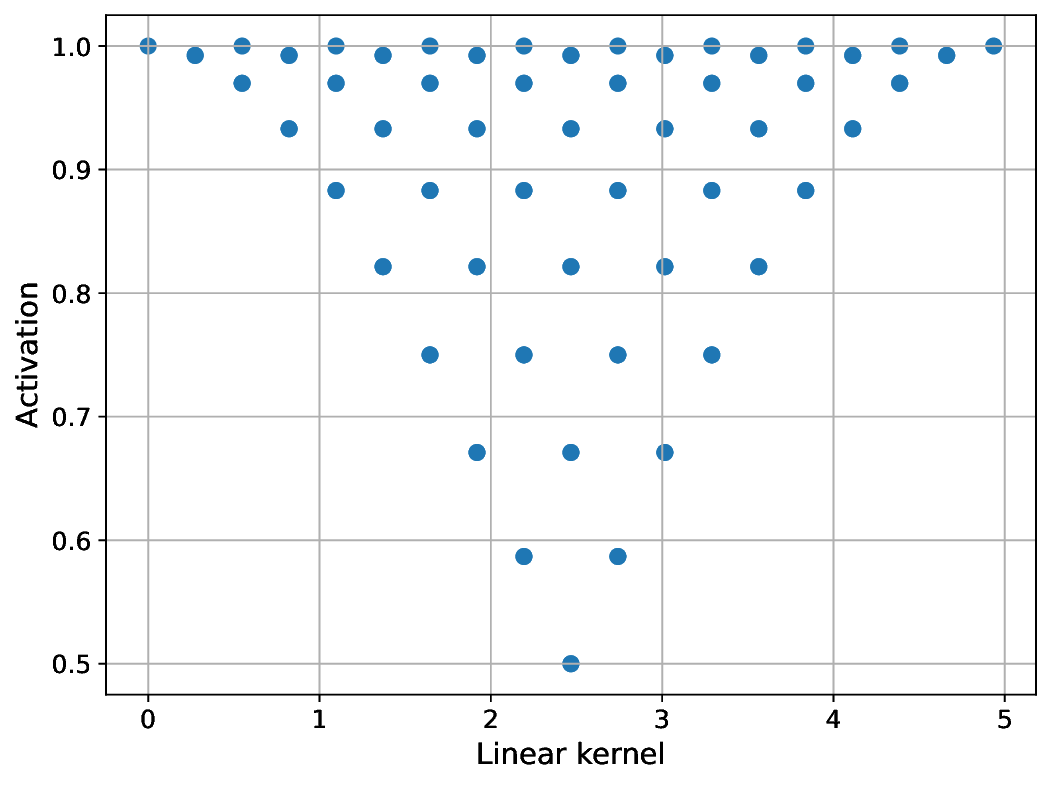}
        \label{fig:cvqn_act_function_for_linear}
    }
    
    \subfloat[]{
        \includegraphics[width=0.3\columnwidth]{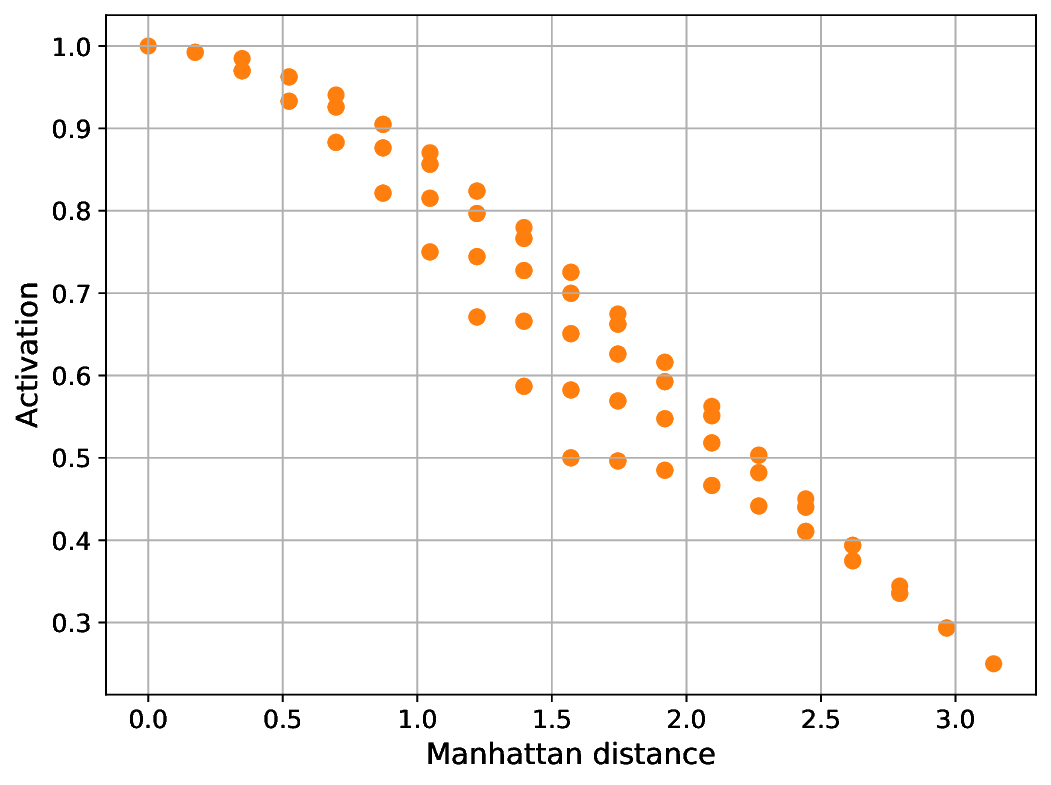}
        \label{fig:cdqn_act_function_for_manhattan}
    }
    \subfloat[]{
        \includegraphics[width=0.3\columnwidth]{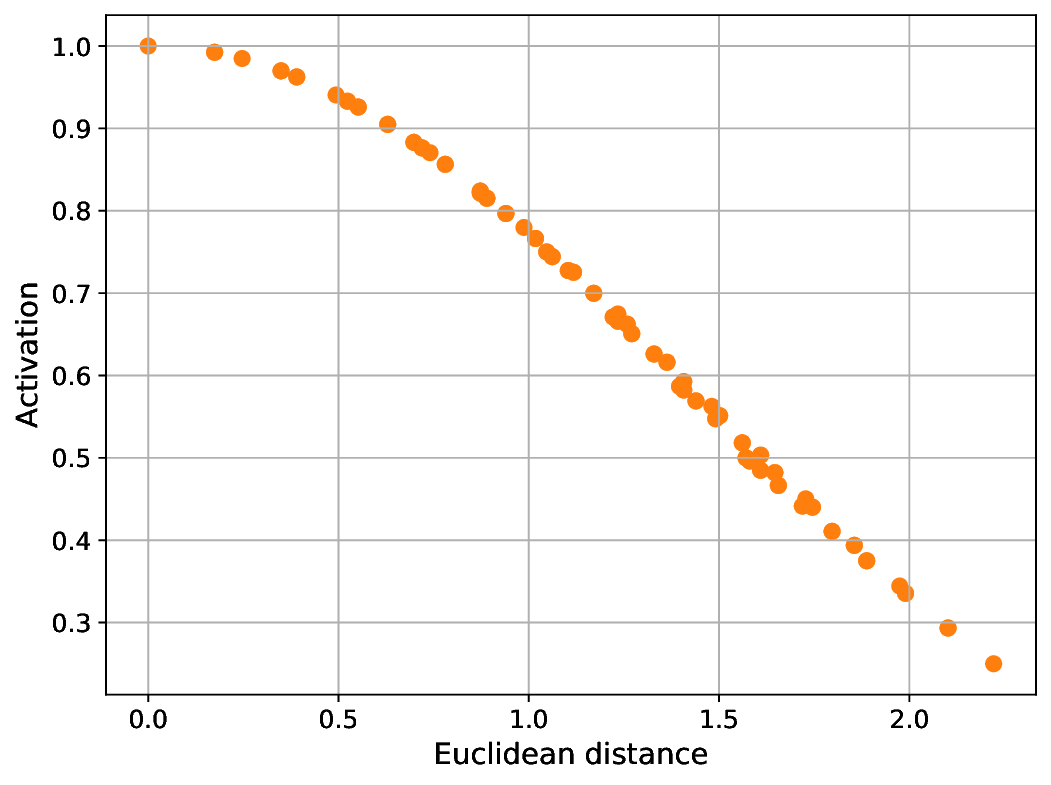}
        \label{fig:cdqn_act_function_for_euclidean}
    }
    \subfloat[]{
        \includegraphics[width=0.3\columnwidth]{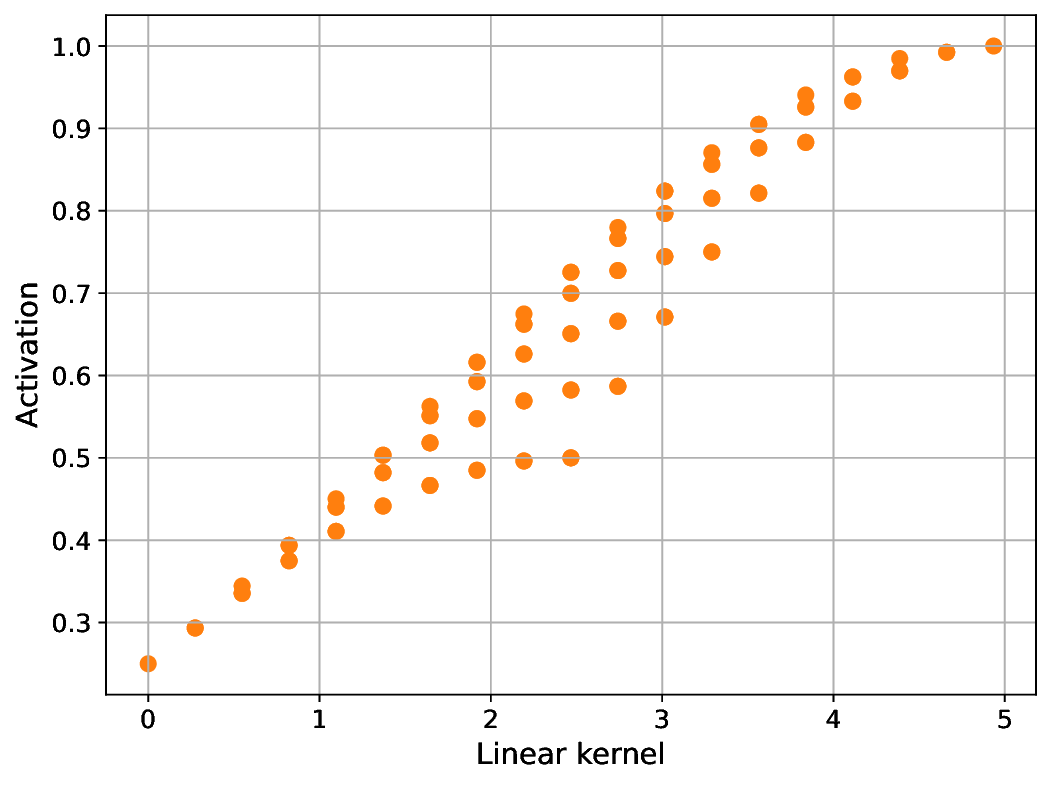}
        \label{fig:cdqn_act_function_for_linear}
    }
    \caption{Quantum neuron activation as a function of the Manhattan distance, the Euclidean distance, and the linear kernel between the original input and weight vectors. (a), (b), and (c) for the CVQN. (d), (e), and (f) for the CDQN.}
    \label{fig:act_functions_for_manhattan-euclidean-linear}
\end{figure}

Figure~\ref{fig:act_functions_for_manhattan-euclidean-linear} contrasts the CVQN activation with the CDQN activation as functions of the Manhattan distance, the Euclidean distance, and the linear kernel. The CVQN activation presents the same qualitative behavior in all cases but with a shift to the right for the Euclidean distance, as can be seen in Figure~\ref{fig:cvqn_act_function_for_manhattan}, Figure~\ref{fig:cvqn_act_function_for_euclidean}, and Figure~\ref{fig:cvqn_act_function_for_linear}. In all cases, high values of activation are generated for the input vectors with extreme measures. Moving away from the extremes, the measures generate smaller and smaller values of activation in addition to the previous high values. Regarding the CDQN, its activation monotonically decreases as the Manhattan and Euclidean distances increase, as shown in Figure~\ref{fig:cdqn_act_function_for_manhattan} and Figure~\ref{fig:cdqn_act_function_for_euclidean}, and monotonically increases as the linear kernel also increases, as shown in Figure~\ref{fig:cdqn_act_function_for_linear}.

\begin{figure}[!t]
    \centering
    \subfloat[]{
        \includegraphics[width=0.3\columnwidth]{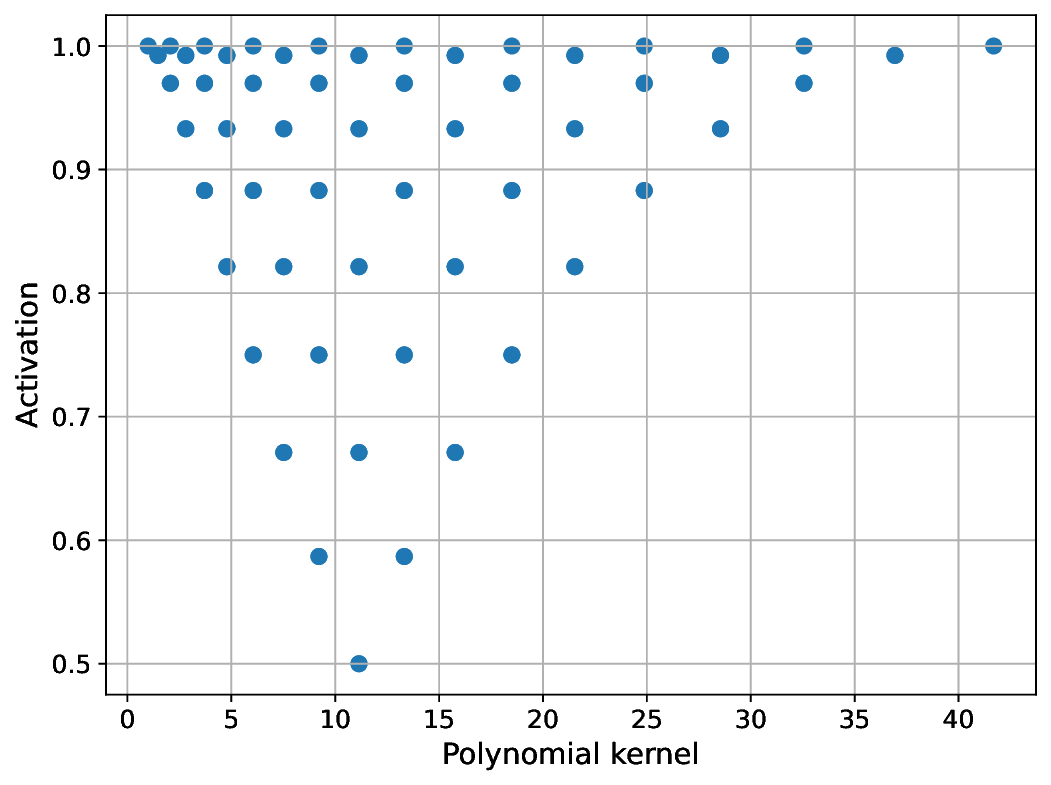}
        \label{fig:cvqn_act_function_for_poly}
    }
    \subfloat[]{
        \includegraphics[width=0.3\columnwidth]{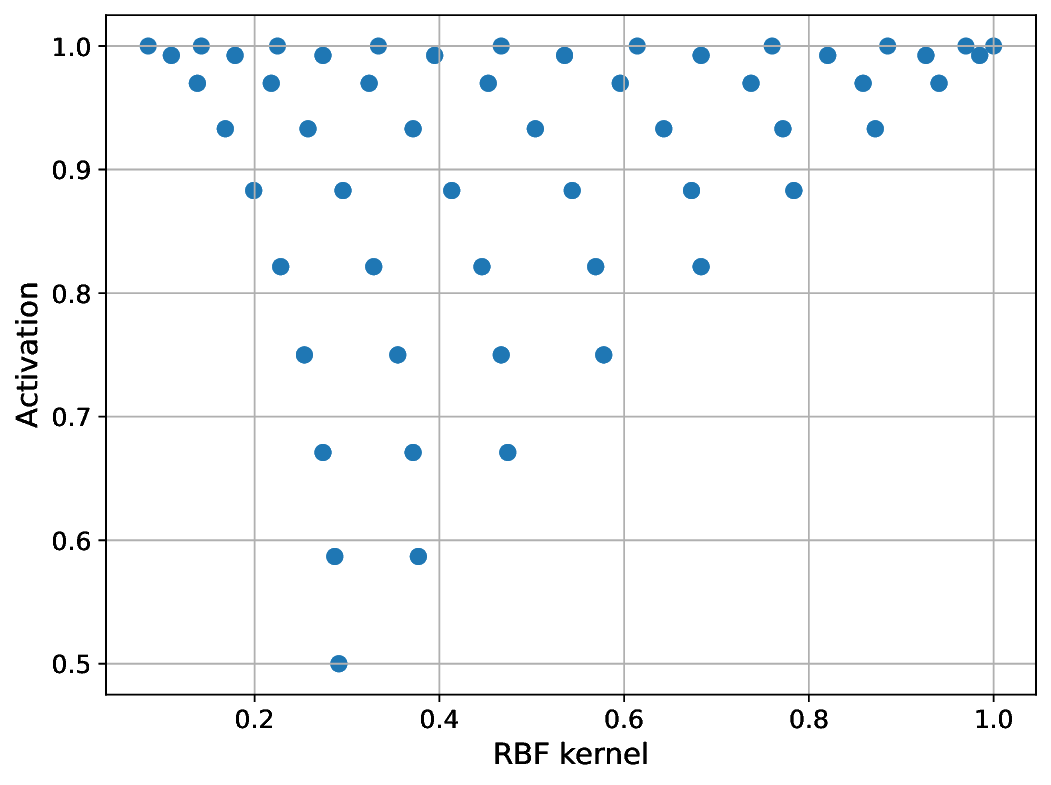}
        \label{fig:cvqn_act_function_for_rbf}
    }
    \subfloat[]{
        \includegraphics[width=0.3\columnwidth]{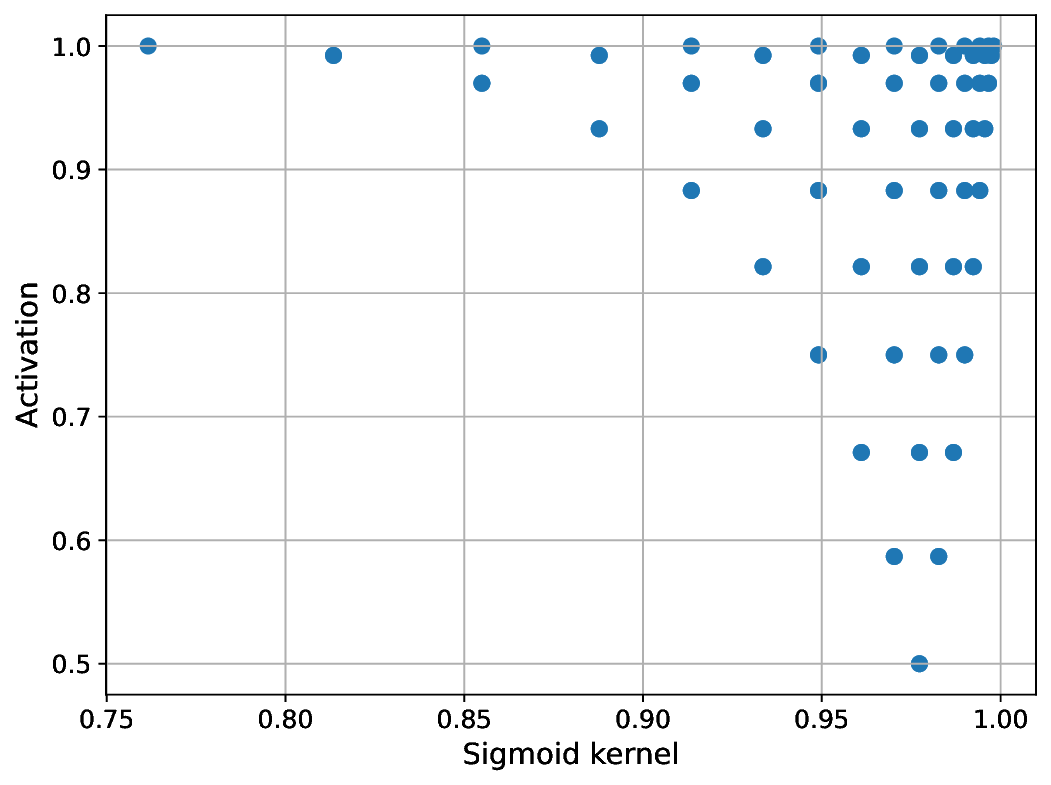}
        \label{fig:cvqn_act_function_for_sig}
    }
    
    \subfloat[]{
        \includegraphics[width=0.3\columnwidth]{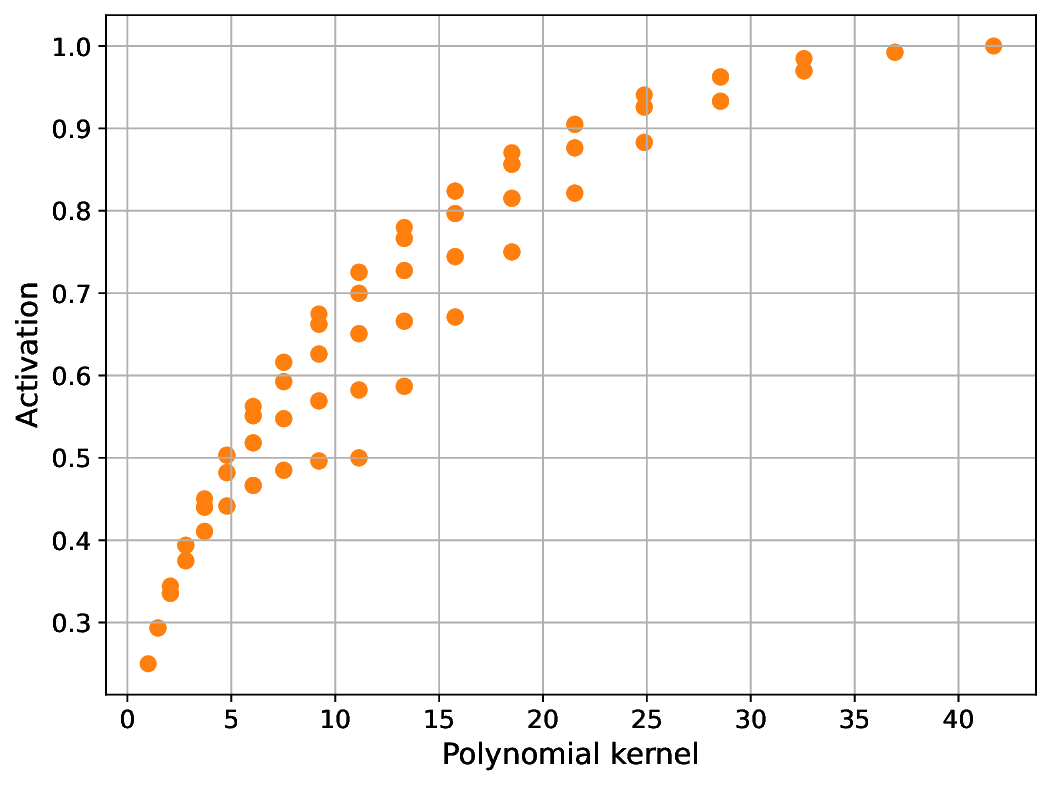}
        \label{fig:cdqn_act_function_for_poly}
    }
    \subfloat[]{
        \includegraphics[width=0.3\columnwidth]{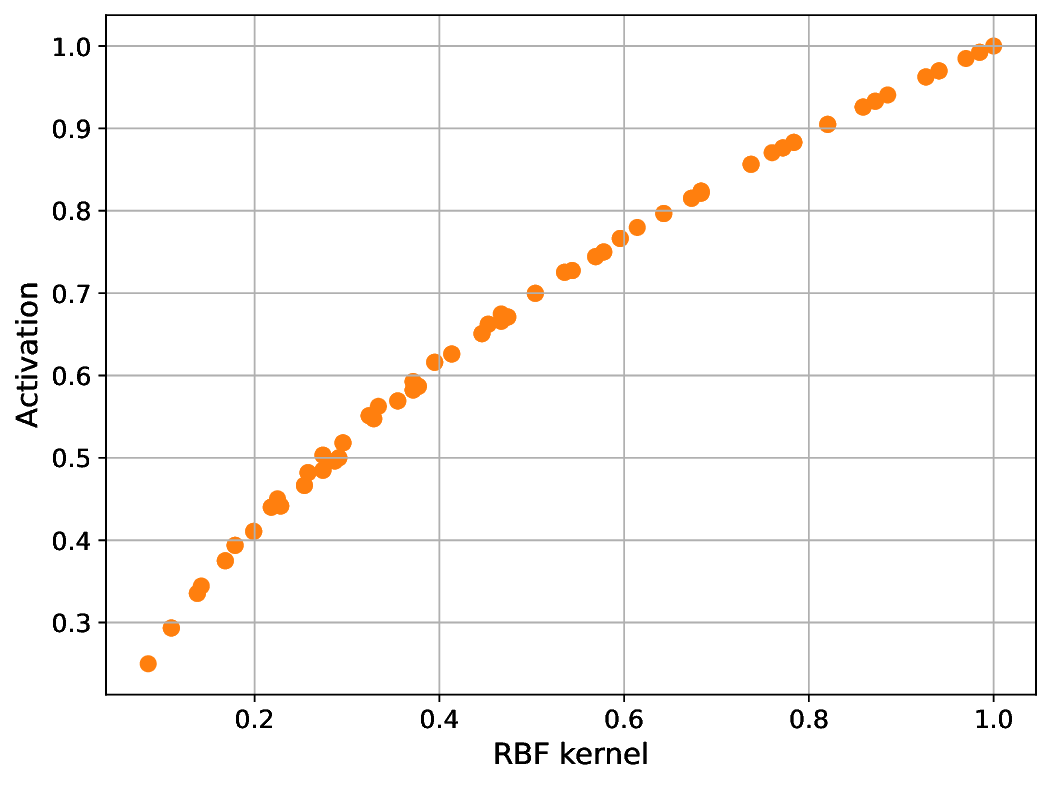}
        \label{fig:cdqn_act_function_for_rbf}
    }
    \subfloat[]{
        \includegraphics[width=0.3\columnwidth]{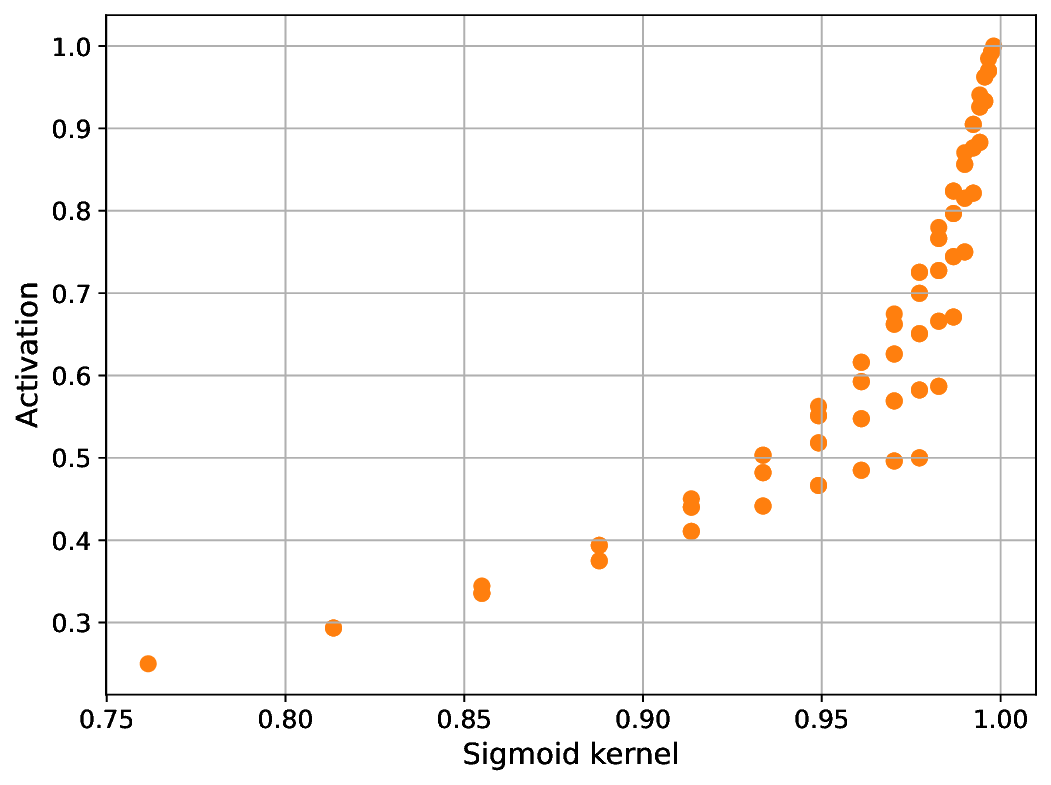}
        \label{fig:cdqn_act_function_for_sig}
    }
    \caption{Quantum neuron activation as a function of the polynomial kernel, the RBF kernel, and the sigmoid kernel between the original input and weight vectors. (a), (b), and (c) for the CVQN. (d), (e), and (f) for the CDQN.}
    \label{fig:act_functions_for_poly-rbf-sig}
\end{figure}

In a similar way, Figure~\ref{fig:act_functions_for_poly-rbf-sig} contrasts the CVQN activation with the CDQN activation as functions of the polynomial kernel, the RBF kernel, and the sigmoid kernel. The CVQN activation presents relatively the same qualitative behavior already presented in Figure~\ref{fig:act_functions_for_manhattan-euclidean-linear} but with a shift to the left for the polynomial and RBF kernels, as shown in Figure~\ref{fig:cvqn_act_function_for_poly} and Figure~\ref{fig:cvqn_act_function_for_rbf}, and a considerable shift to the right for the sigmoid kernel, as shown in Figure~\ref{fig:cvqn_act_function_for_sig}. The CDQN activation presents relatively the same qualitative behavior already presented in Figure~\ref{fig:cdqn_act_function_for_linear} but with shapes that suggest a logarithmic growth in Figure~\ref{fig:cdqn_act_function_for_poly}, a linear growth in Figure~\ref{fig:cdqn_act_function_for_rbf}, and an exponential growth in Figure~\ref{fig:cdqn_act_function_for_sig}.

Therefore, the CVQN and the CDQN can solve classification problems where those activation shapes fit correctly. It turns out the nature of each quantum neuron is preserved for any of those metrics. Thus, we follow the analysis in this work using only the Euclidean distance due to its intuitiveness.

\begin{figure}[!t]
    \centering
    \subfloat[]{
        \includegraphics[width=0.3\columnwidth]{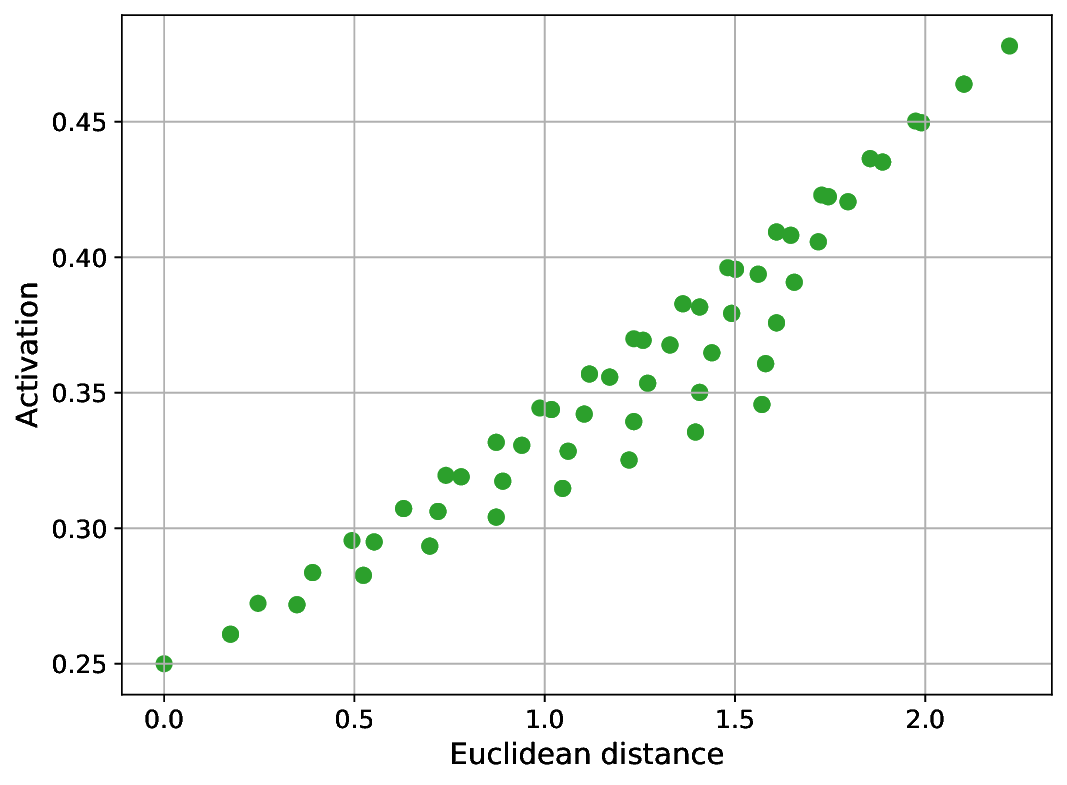}
        \label{fig:pcdqn_act_function_for_1over4_piover2}
    }
    \subfloat[]{
        \includegraphics[width=0.3\columnwidth]{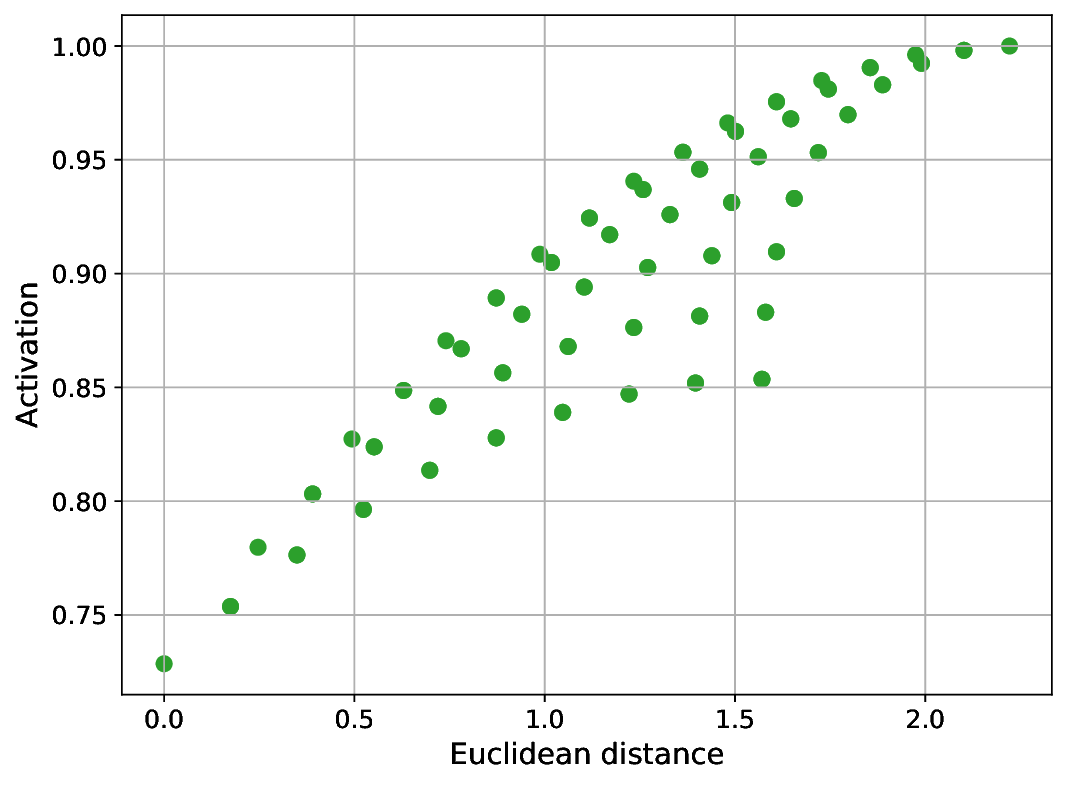}
        \label{fig:pcdqn_act_function_for_1over2_piover4}
    }
    \subfloat[]{
        \includegraphics[width=0.3\columnwidth]{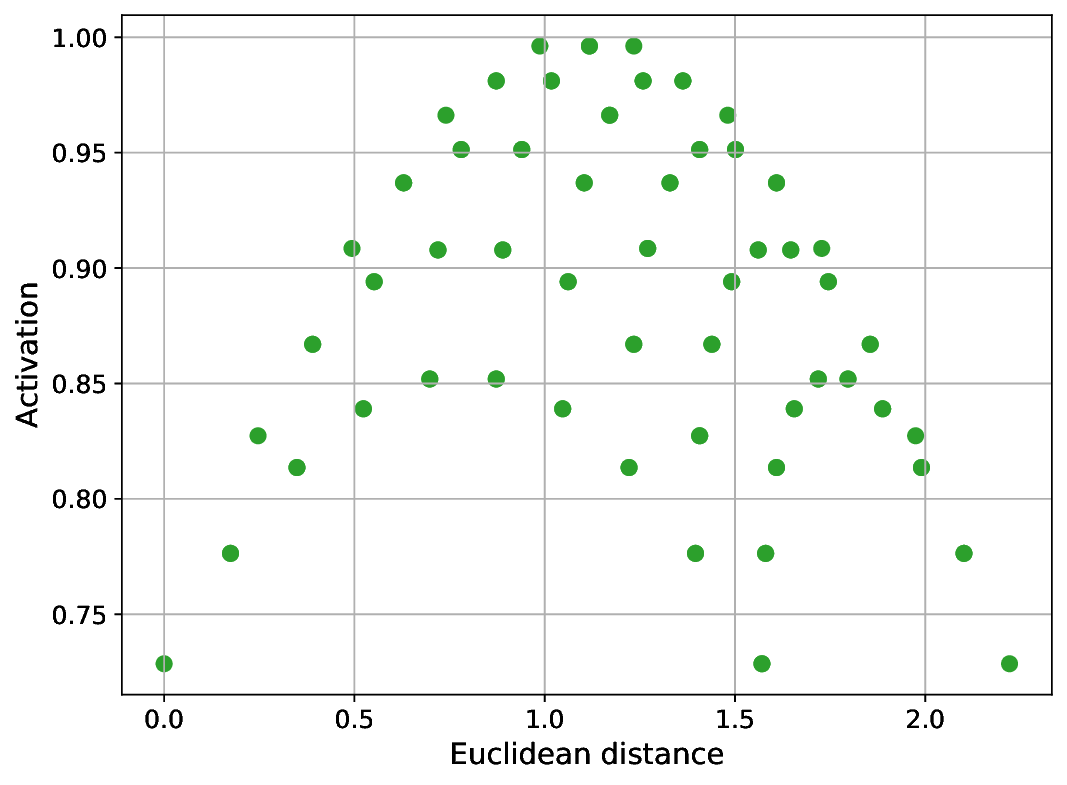}
        \label{fig:pcdqn_act_function_for_1_piover4}
    }
    
    \subfloat[]{
        \includegraphics[width=0.3\columnwidth]{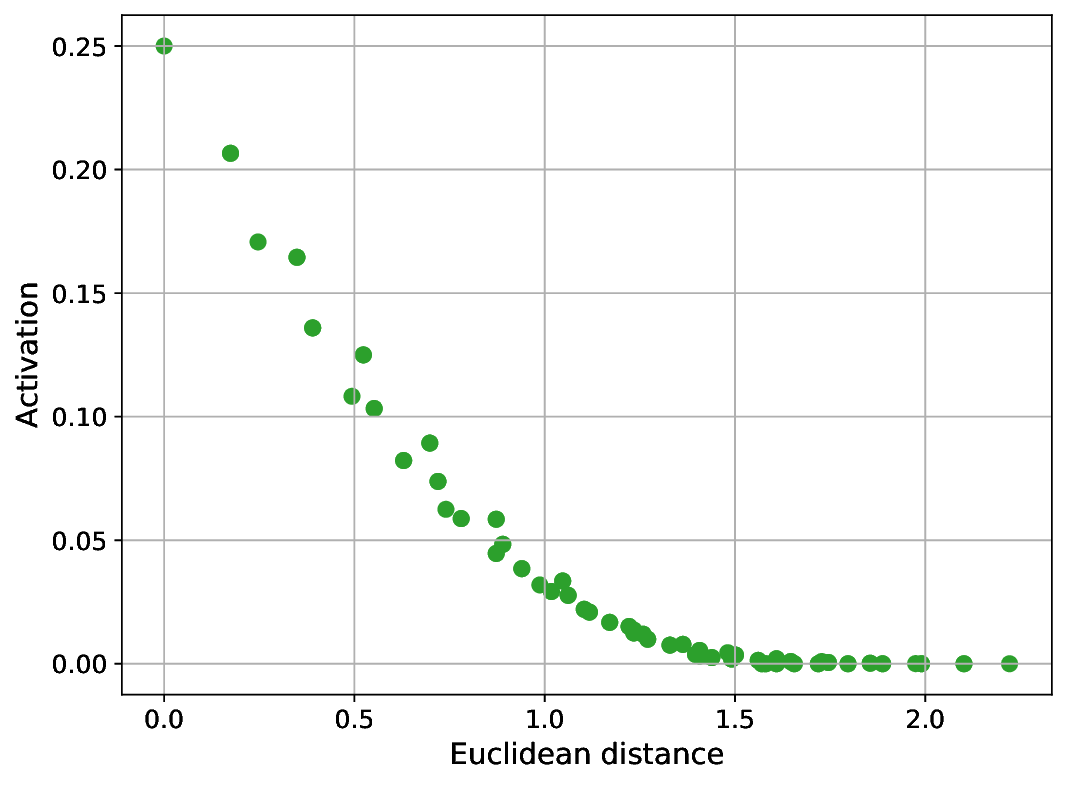}
        \label{fig:pcdqn_act_function_for_1_3piover2}
    }
    \subfloat[]{
        \includegraphics[width=0.3\columnwidth]{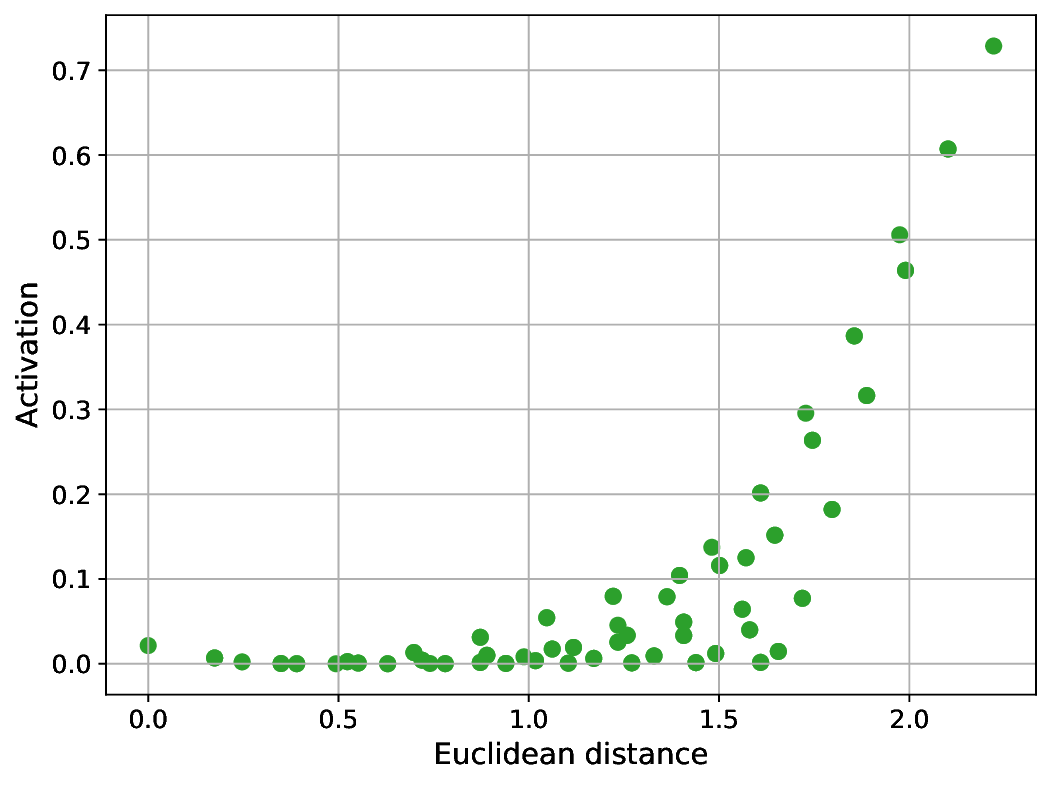}
        \label{fig:pcdqn_act_function_for_2_5piover4}
    }
    \subfloat[]{
        \includegraphics[width=0.3\columnwidth]{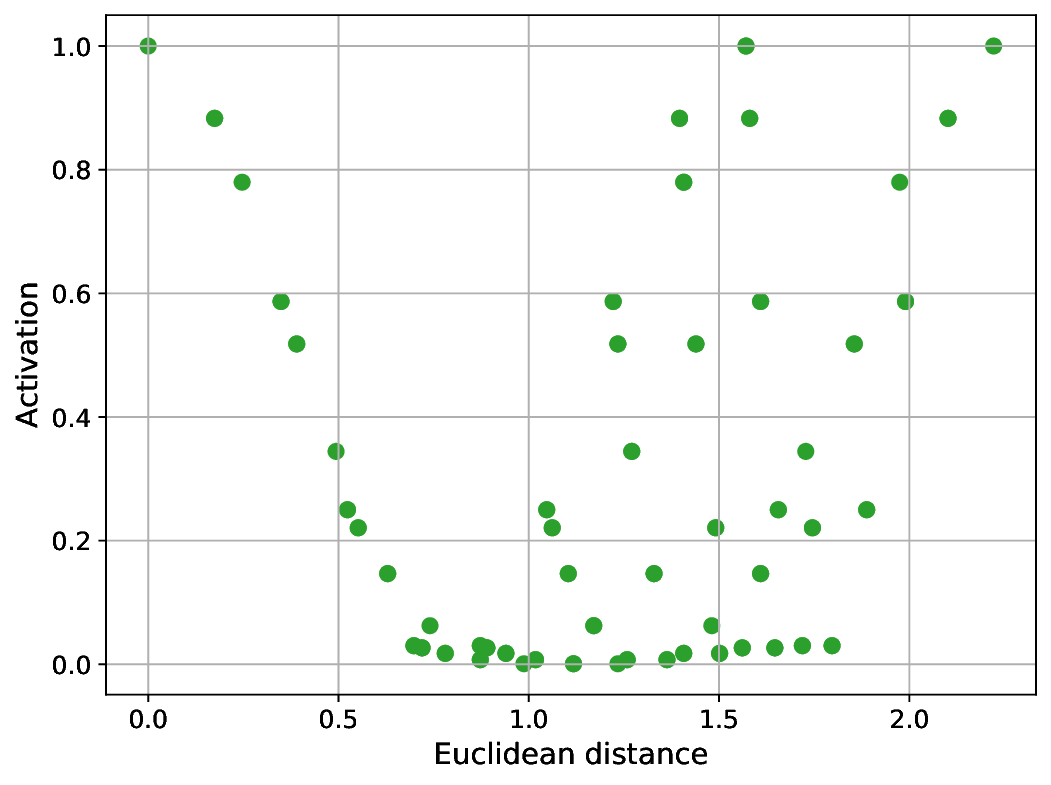}
        \label{fig:pcdqn_act_function_for_4_0}
    }
    \caption{PCDQN activation as a function of the Euclidean distance between the original input and weight vectors for some combinations of $\tau$ and $\delta$. (a) for $\tau=1/4$ and $\delta=\pi/2$. (b) for $\tau=1/2$ and $\delta=\pi/4$. (c) for $\tau=1$ and $\delta=\pi/4$. (d) for $\tau=1$ and $\delta=3\pi/2$. (e) for $\tau=2$ and $\delta=5\pi/4$. (f) for $\tau=4$ and $\delta=0$.}
    \label{fig:pcdqn_act_functions_for_euclidean-dist}
\end{figure}

Figure~\ref{fig:pcdqn_act_functions_for_euclidean-dist} shows the PCDQN activation as a function of the Euclidean distance for some combinations of $\tau$ and $\delta$, which gives examples of substantial changes that are achieved by means of parametrization on that linear decay already presented in Figure~\ref{fig:cdqn_act_function_for_euclidean}. For example, $(\tau=1/4, \delta=\pi/2)$ and $(\tau=1/2, \delta=\pi/4)$ give approximately a linear growth and a logarithmic growth, as shown in Figure~\ref{fig:pcdqn_act_function_for_1over4_piover2} and Figure~\ref{fig:pcdqn_act_function_for_1over2_piover4}. Figure~\ref{fig:pcdqn_act_function_for_1_piover4} shows a chain of concave-down parabolas achieved by $(\tau=1, \delta=\pi/4)$. An exponential decay and an exponential growth are achieved by $(\tau=1, \delta=3\pi/2)$ and $(\tau=2, \delta=5\pi/4)$ respectively, as shown in Figure~\ref{fig:pcdqn_act_function_for_1_3piover2} and Figure~\ref{fig:pcdqn_act_function_for_2_5piover4}. Figure~\ref{fig:pcdqn_act_function_for_4_0} approaches multiple concave-up parabolas with $(\tau=4, \delta=0)$. In summary, the PCDQN is flexible. In practice, such flexibility is expected to enable the PCDQN to fit problems that the other neurons cannot fit correctly.

\section{Demonstration on Toy Classification Problems}
\label{sec:qn_demo}

Here, we subject those kernel-based quantum neurons to a first demonstration of their classification abilities. The demonstration consists of solving toy problems, followed by quantitative and qualitative comparisons of the best solutions of the quantum neurons. On one hand, quantitatively, the solutions are compared here by a classification metric. On the other hand, qualitatively, the solutions are compared here by the effectiveness of the activation function shapes applied to the problems. Finally, we demonstrate the feasibility of those numerical solutions by executing the quantum neuron circuits on a quantum simulator.

\subsection{Datasets}

\begin{figure}[!t]
    \centering
    \subfloat[]{
        \includegraphics[width=0.3\columnwidth]{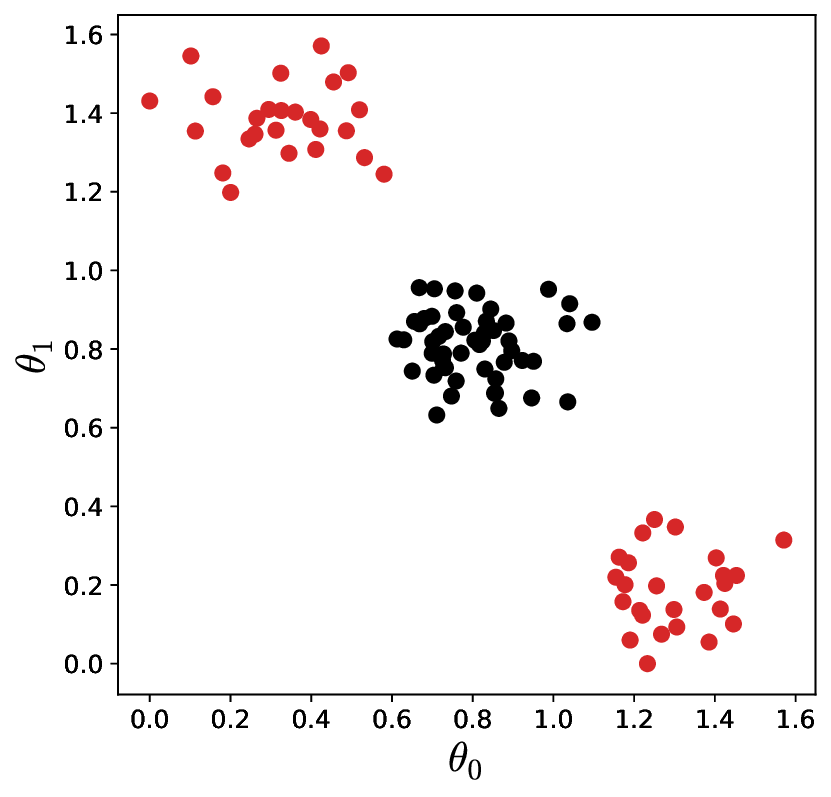}
        \label{fig:diagonal_blobs-target_center}
    }
    \subfloat[]{
        \includegraphics[width=0.3\columnwidth]{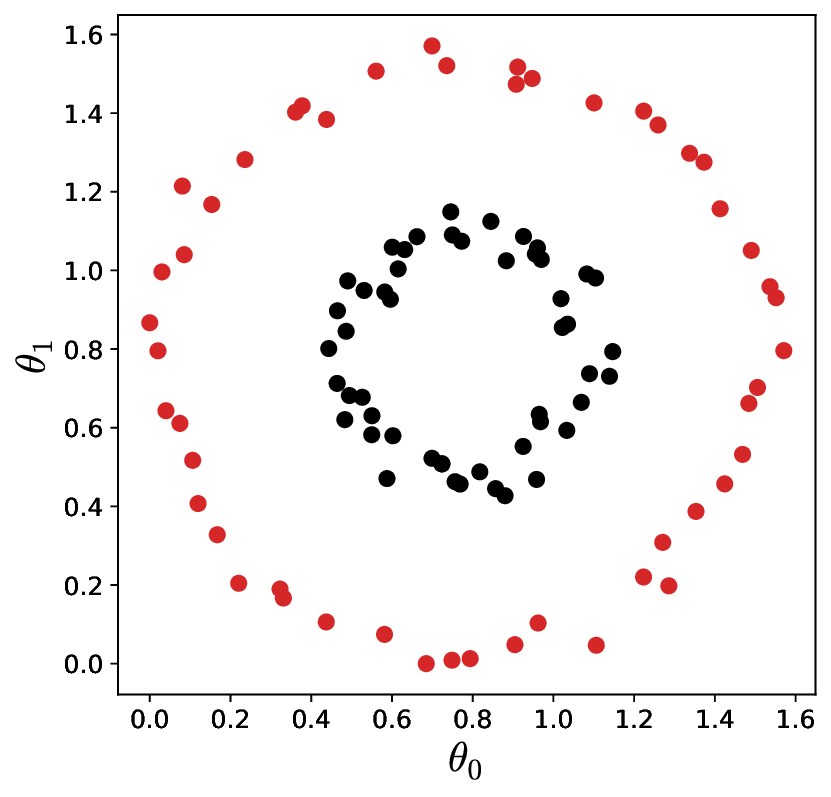}
        \label{fig:concentric_circles-target_inner}
    }
    \subfloat[]{
        \includegraphics[width=0.3\columnwidth]{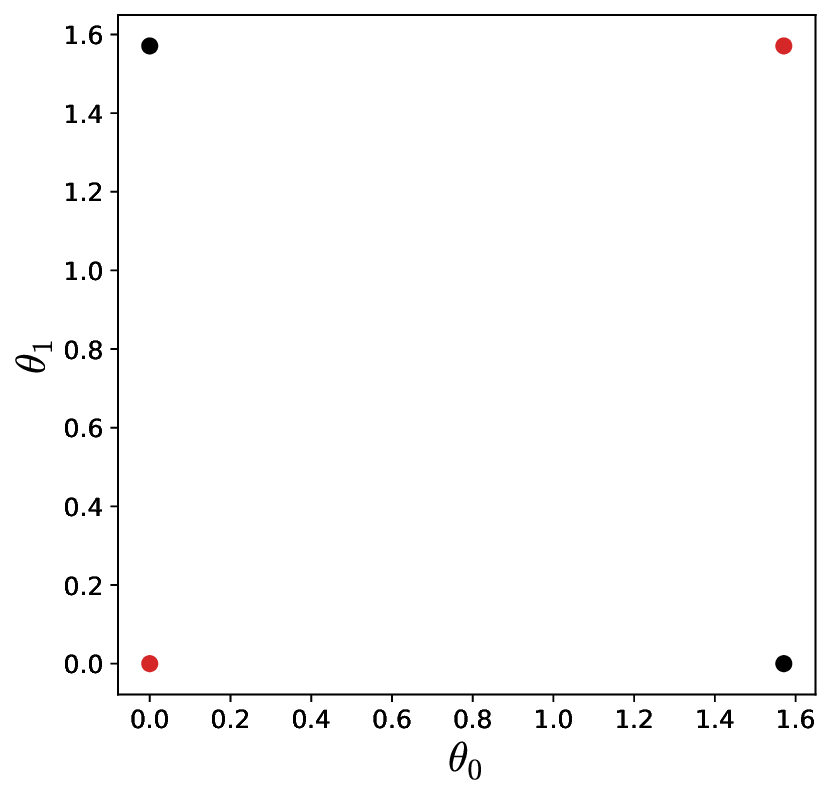}
        \label{fig:square_blobs-target_XORlike}
    }
    
    \subfloat[]{
        \includegraphics[width=0.3\columnwidth]{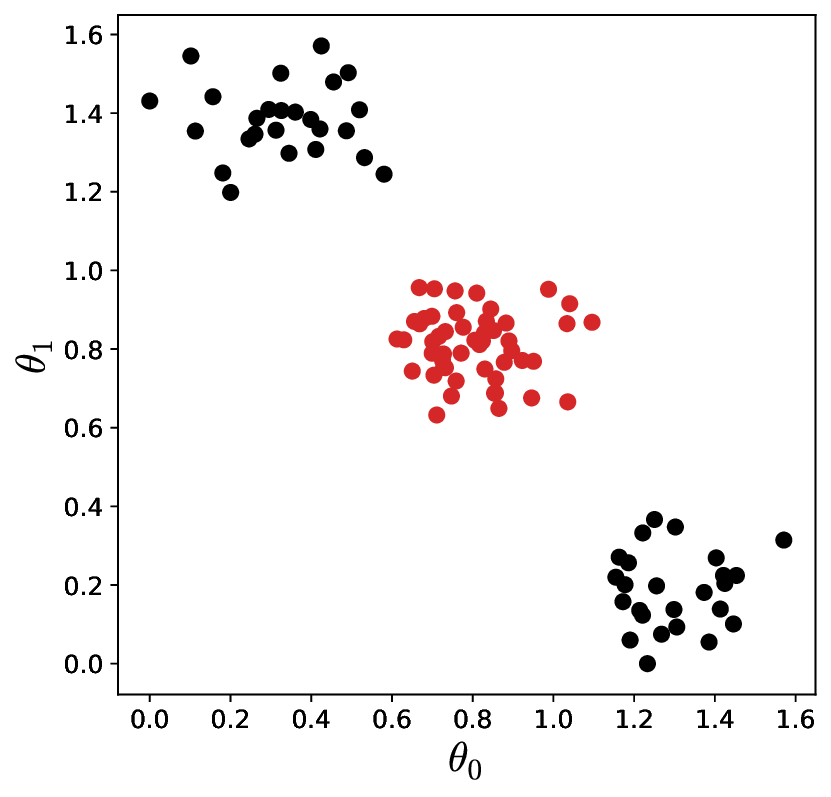}
        \label{fig:diagonal_blobs-target_corner}
    }
    \subfloat[]{
        \includegraphics[width=0.3\columnwidth]{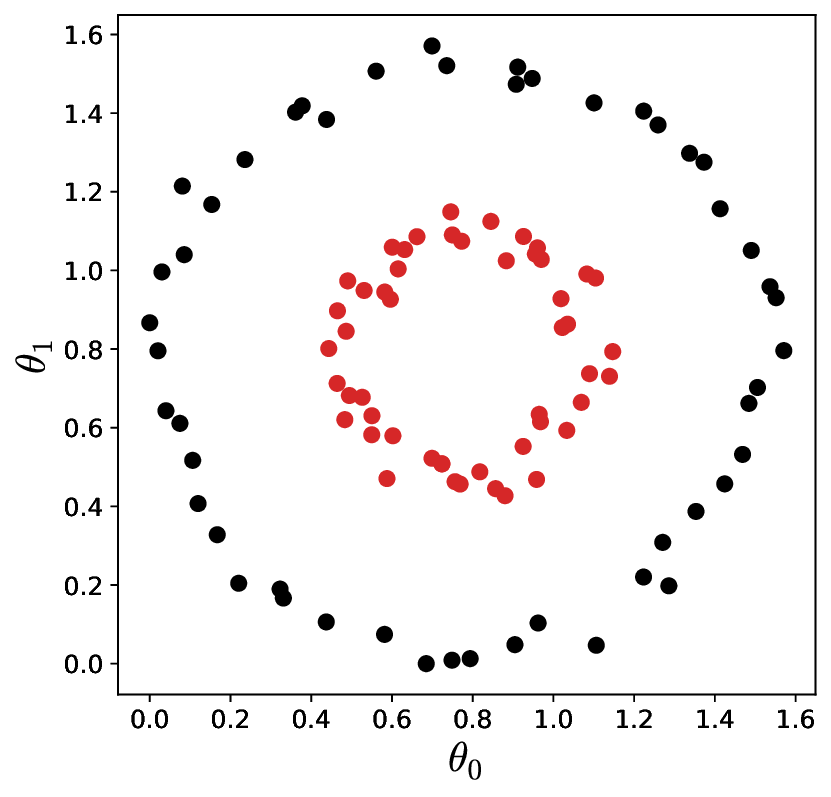}
        \label{fig:concentric_circles-target_outer}
    }
    \subfloat[]{
        \includegraphics[width=0.3\columnwidth]{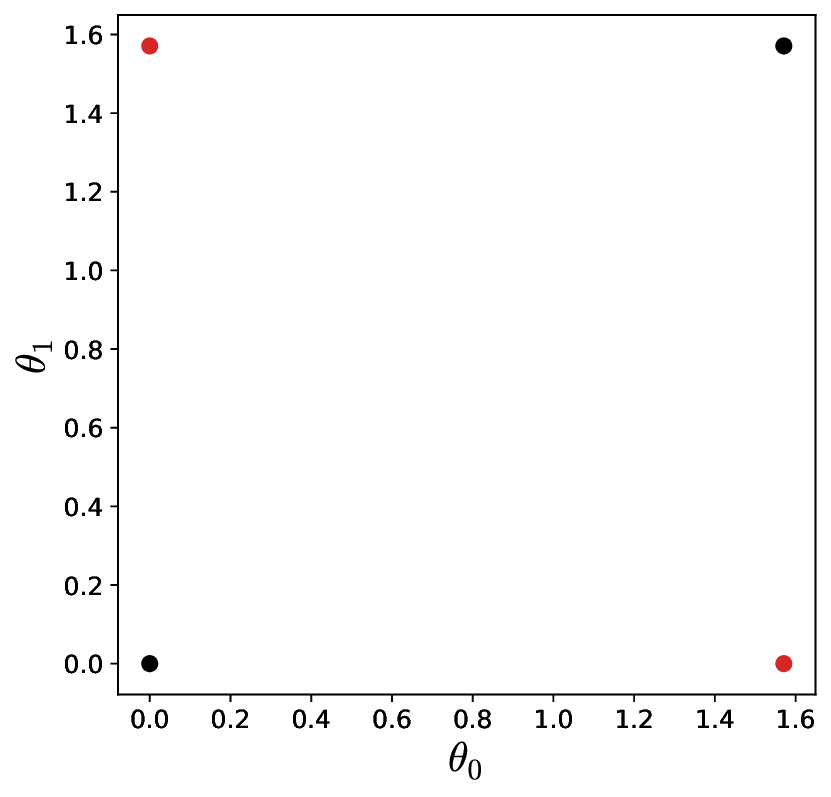}
        \label{fig:square_blobs-target_NXORlike}
    }
    \caption{Toy classification problems where the quantum neurons are applied in this demonstration. The positive class is represented by the black blobs, while the negative class is represented by the red ones. All inputs are in the interval $[0, \pi/2]$.}
    \label{fig:toy_problems}
\end{figure}

The three datasets used in this demonstration are depicted in Figure~\ref{fig:diagonal_blobs-target_center}, Figure~\ref{fig:concentric_circles-target_inner}, and Figure~\ref{fig:square_blobs-target_XORlike}. Here, the positive class is represented by the black blobs, while the negative class is represented by the red blobs, although there is no previous indication of that labeling. Thus, we decide to swap the labels, deriving three more classification problems where the old negative classes become the new positive classes and vice versa, as depicted in Figure~\ref{fig:diagonal_blobs-target_corner}, Figure~\ref{fig:concentric_circles-target_outer}, and Figure~\ref{fig:square_blobs-target_NXORlike}. Note that a line is not able to separate the classes of any of those two-dimensional binary classification problems. The two attributes of each of those artificial and toy problems are merely coordinates, $\theta_0$ and $\theta_1$, that are defined in a way that visually organizes the data in a diagonal, in circles, or in a square, respectively. It is worth mentioning that $\theta_0$ and $\theta_1$ are scaled to the interval $[0, \pi/2]$. We discuss those classification problems and how to generate them in the following.

We refer to Figure~\ref{fig:diagonal_blobs-target_center} and Figure~\ref{fig:diagonal_blobs-target_corner} as the diagonal blobs due to the formed structure. The difference is that the target is the center blobs in Figure~\ref{fig:diagonal_blobs-target_center}, while the target is the corner blobs in Figure~\ref{fig:diagonal_blobs-target_corner}. That dataset is provided by the scikit-learn \textit{make\_blobs} function, letting \textit{centers} be [(-5,8), (0,0), (5,-8)], \textit{cluster\_std} be 1.2, and \textit{random\_state} be 0. Each class has 50 samples. Alternatively, the target is the inner circle in Figure~\ref{fig:concentric_circles-target_inner}, while the target is the outer circle in Figure~\ref{fig:concentric_circles-target_outer}. Those concentric circles are provided by the scikit-learn \textit{make\_circles} function, letting \textit{noise} be 0.05, \textit{factor} be 0.4, and \textit{random\_state} be 0. Each circle has 50 samples. Finally, the blobs in the coordinates $[(0, 0), (0, \pi/2), (\pi/2, 0), (\pi/2, \pi/2)]$ form a square, where the target is like the Exclusive OR (XOR) problem in Figure~\ref{fig:square_blobs-target_XORlike}, while the target is like the Not Exclusive OR (NXOR) problem in Figure~\ref{fig:square_blobs-target_NXORlike}.

\subsection{Search for Weights and Parameters}

In this demonstration, we conduct a grid search for the best weight vector of each quantum neuron for each classification problem. The space of weight vectors is formed here by all vectors $(\phi_0, \phi_1)$ where each component assumes one-hundred equidistant values in the interval $[0, \pi/2]$, which gives a ten-thousand-vector weight space. Then, we choose the first weight vector in the search that maximizes the Area Under the Receiver Operating Characteristic Curve (AUC ROC). The AUC ROC allows measuring the classification quality in a threshold-free manner. We compute that metric by means of the scikit-learn \textit{roc\_auc\_score} function.

Specifically for the PCDQN, we conduct a nested grid search for the best parameter combination and weight vector for each classification problem. The space of parameters is formed here by all combinations $(\tau, \delta)$, where

\begin{equation*}
    \tau \in \bigg\{ \frac{1}{4}, \frac{1}{2}, 1, 2, 4 \bigg\}
\end{equation*}
and

\begin{equation*}
    \delta \in \bigg\{0, \frac{\pi}{4}, \frac{\pi}{2}, \frac{3\pi}{4}, \pi, \frac{5\pi}{4}, \frac{3\pi}{2} \bigg\}.
\end{equation*}

Actually, the combination (1, 0) is not allowed since it produces the CDQN. Consequently, there are 34 allowed combinations to search. Those combinations explore a variety of lengths and shifts along the activation function period, which is $2\pi$. For each combination of $\tau$ and $\delta$, we search that ten-thousand-vector weight space. Then, we choose the first values of $(\tau, \delta)$ and $(\phi_0, \phi_1)$ that maximize the AUC ROC together in the nested search.

\subsection{Results and Discussion}

Table~\ref{tab:toy_max_aucroc} shows the maximum AUC ROC that each quantum neuron achieved for each classification problem. The CVQN and the CDQN perfectly solved the Diagonal blobs when targeting the class Center. The CVQN also solved the NXOR-like Square blobs, while the CDQN performed like a random model. On the other hand, while the CVQN did not solve the Concentric circles when targeting the class Inner, the CDQN did. In summary, the CVQN and the CDQN only solved two problems each. The PCDQN produced optimal solutions for all problems, even for those problems where the other neurons performed like merely random models. Those results confirm the hypothesis that parametrization gives flexibility.

\begin{table*}
    \centering
    \caption{Maximum AUC ROC of Each Quantum Neuron for Each Toy Problem.}
    \label{tab:toy_max_aucroc}
    \begin{tabular}{cc|c|c|c}
        \hline
        Dataset & Target & CVQN & CDQN & PCDQN \\
        \hline
        \multirow{2}{*}{Diagonal blobs} & Center & \textbf{1.0} & \textbf{1.0} & \textbf{1.0} \\
        & Corner & 0.5 & 0.5 & \textbf{1.0} \\
        \multirow{2}{*}{Concentric circles} & Inner & 0.8244 & \textbf{1.0} & \textbf{1.0} \\
        & Outer & 0.502 & 0.4088 & \textbf{0.9524} \\
        \multirow{2}{*}{Square blobs} & XOR-like & 0.5 & 0.5 & \textbf{1.0} \\
        & NXOR-like & \textbf{1.0} & 0.5 & \textbf{1.0} \\
        \hline
    \end{tabular}
\end{table*}

The best parameter combinations and weight vectors of each quantum neuron for each classification problem are reported in Table~\ref{tab:best_params_weights}. Actually, the weights have about sixteen decimal places. We report here only the first two decimal places, which is enough to retrieve the original weights.

\begin{table*}
    \centering
    \caption{Best Parameter Combinations and Weight Vectors of Each Quantum Neuron for Each Toy Problem.}
    \label{tab:best_params_weights}
    \begin{tabular}{cc|c|c|cc}
        \hline
        \multirow{2}{*}{Dataset} & \multirow{2}{*}{Target} & CVQN & CDQN & \multicolumn{2}{c}{PCDQN} \\
        & & ($\phi_0$, $\phi_1$) & ($\phi_0$, $\phi_1$) & ($\tau$, $\delta$) & ($\phi_0$, $\phi_1$) \\
        \hline
        \multirow{2}{*}{Diagonal blobs} & Center & (0, 0) & (0.46, 0.49) & (1/4, 0) & (0.46, 0.50) \\
        & Corner & (0, 0.87) & (0, 1.07) & (1/4, $\pi$) & (0.69, 0.76) \\
        \multirow{2}{*}{Concentric circles} & Inner & (0, 0.03) & (0.60, 0.76) & (1/4, 0) & (0.60, 0.76) \\
        & Outer & (0.79, 0) & (1.57, 1.45) & (1/4, $\pi$) & (0.82, 0.74) \\
        \multirow{2}{*}{Square blobs} & XOR-like & (0, 0.79) & (0, 0) & (4, 0) & (0.15, 1.20) \\
        & NXOR-like & (0, 0) & (0, 0) & (4, 0) & (0.04, 0.04) \\
        \hline
    \end{tabular}
\end{table*}

In the following, we use those best parameters and weights to study the quantum neuron solutions. We then plot the activation shape of each quantum neuron for each classification problem, which depends on the best parameters for the PCDQN. Those shapes are drawn as functions of the Euclidean distance from the input data to the corresponding best weight vector. A problem is solved if the neuron outputs for the black blobs are higher than the ones for the red blobs, regardless of the Euclidean distances to the best weight vector. In this way, there will be a threshold that separates the classes perfectly.

\begin{figure}[!t]
    \centering
    \subfloat[]{
        \includegraphics[width=0.3\columnwidth]{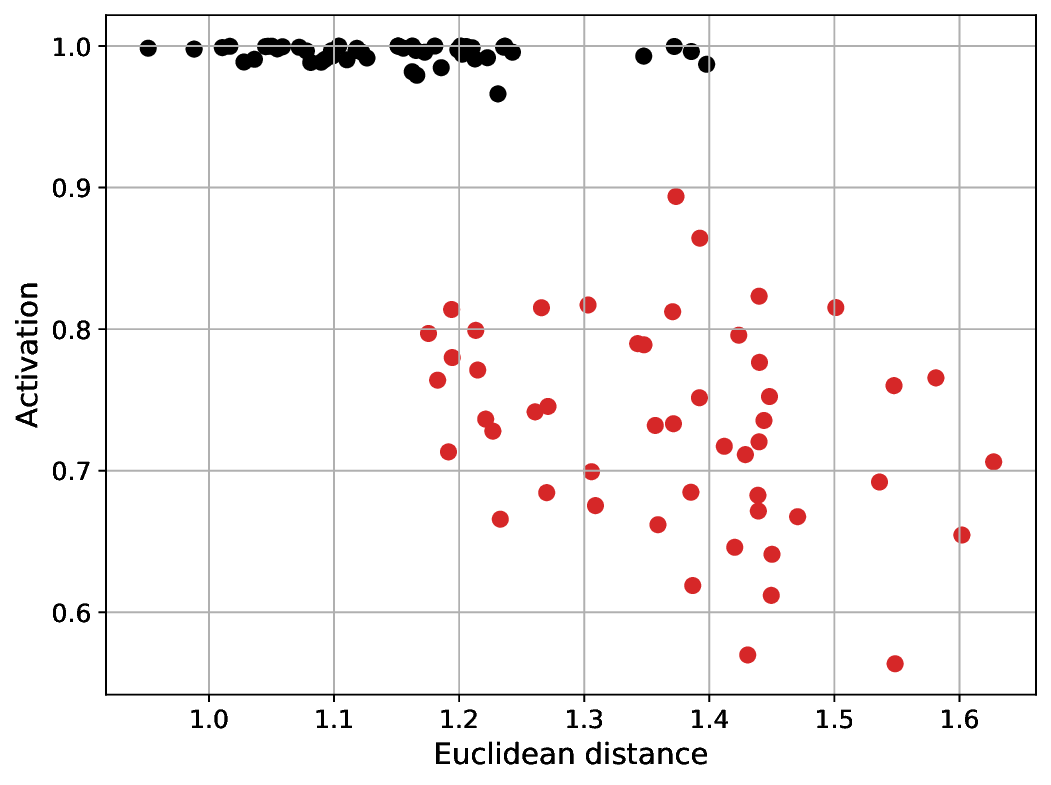}
        \label{fig:cvqn_act_func-db_center}
    }
    \subfloat[]{
        \includegraphics[width=0.3\columnwidth]{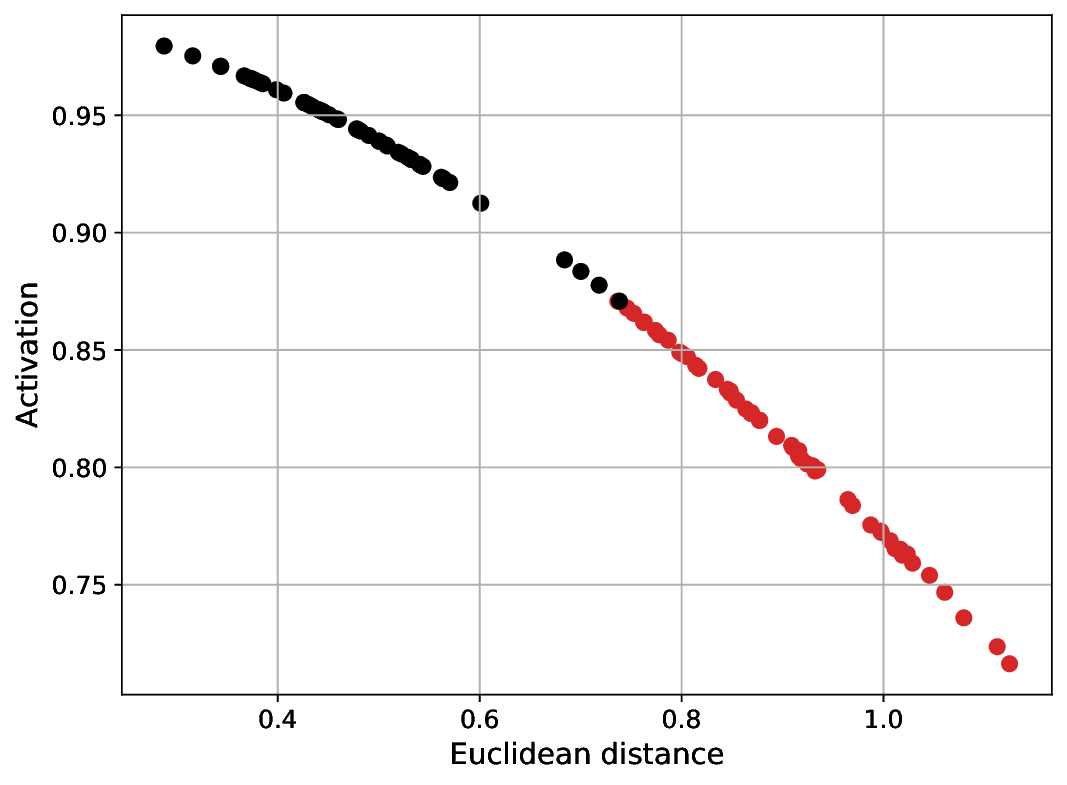}
        \label{fig:cdqn_act_func-db_center}
    }
    \subfloat[]{
        \includegraphics[width=0.3\columnwidth]{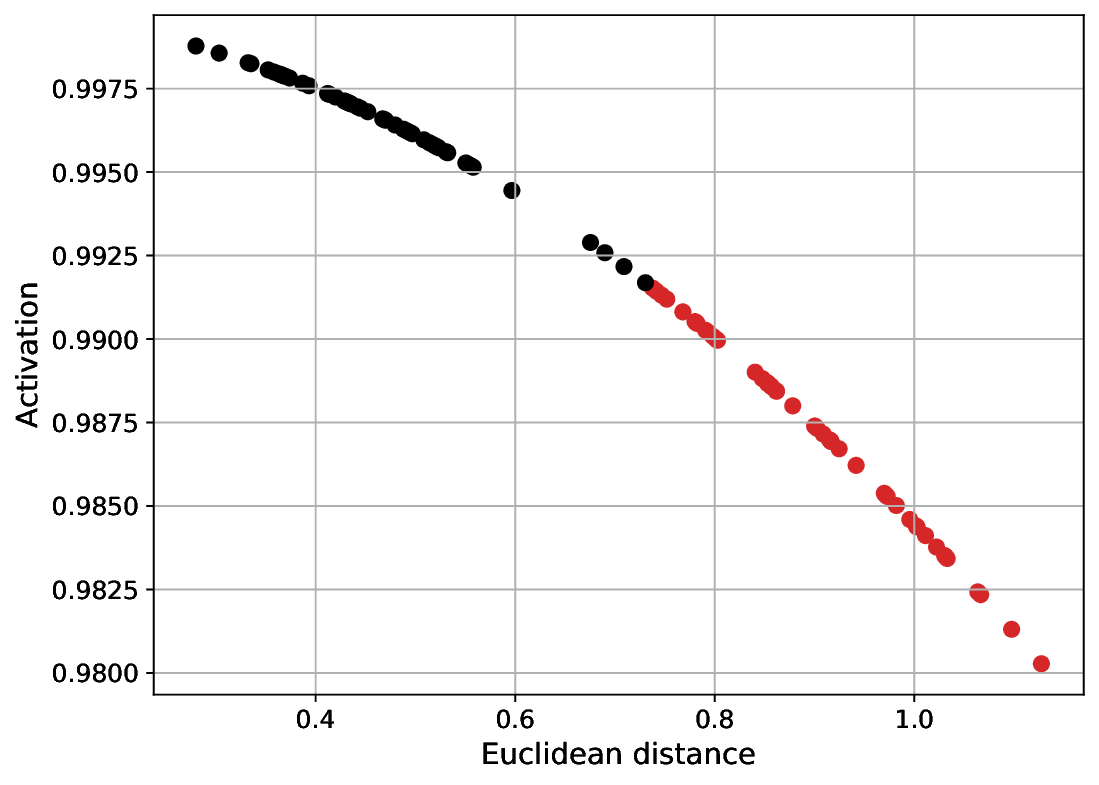}
        \label{fig:pcdqn_act_func-db_center}
    }
    
    \subfloat[]{
        \includegraphics[width=0.3\columnwidth]{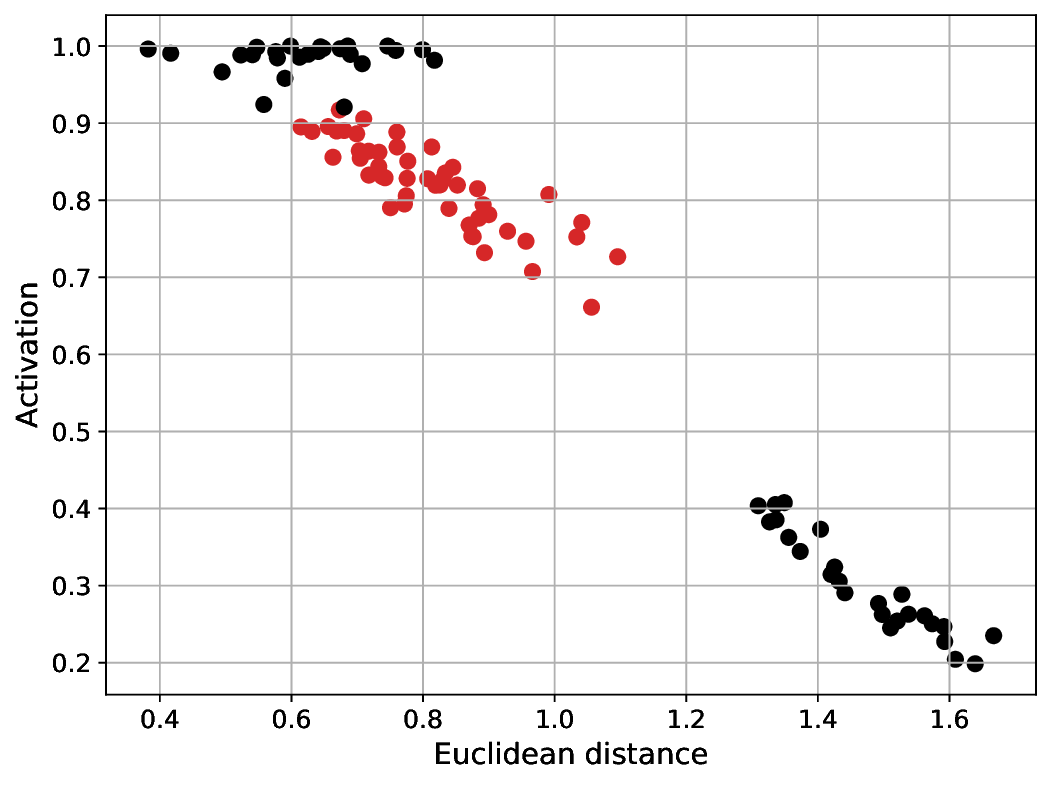}
        \label{fig:cvqn_act_func-db_corner}
    }
    \subfloat[]{
        \includegraphics[width=0.3\columnwidth]{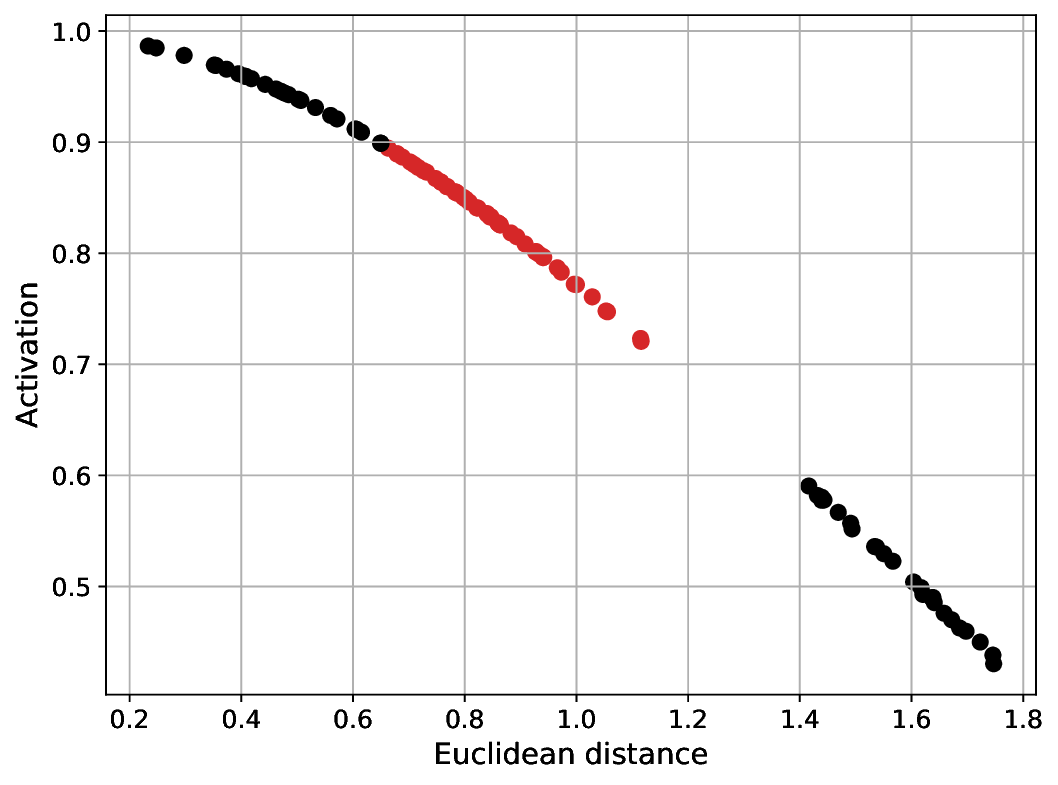}
        \label{fig:cdqn_act_func-db_corner}
    }
    \subfloat[]{
        \includegraphics[width=0.3\columnwidth]{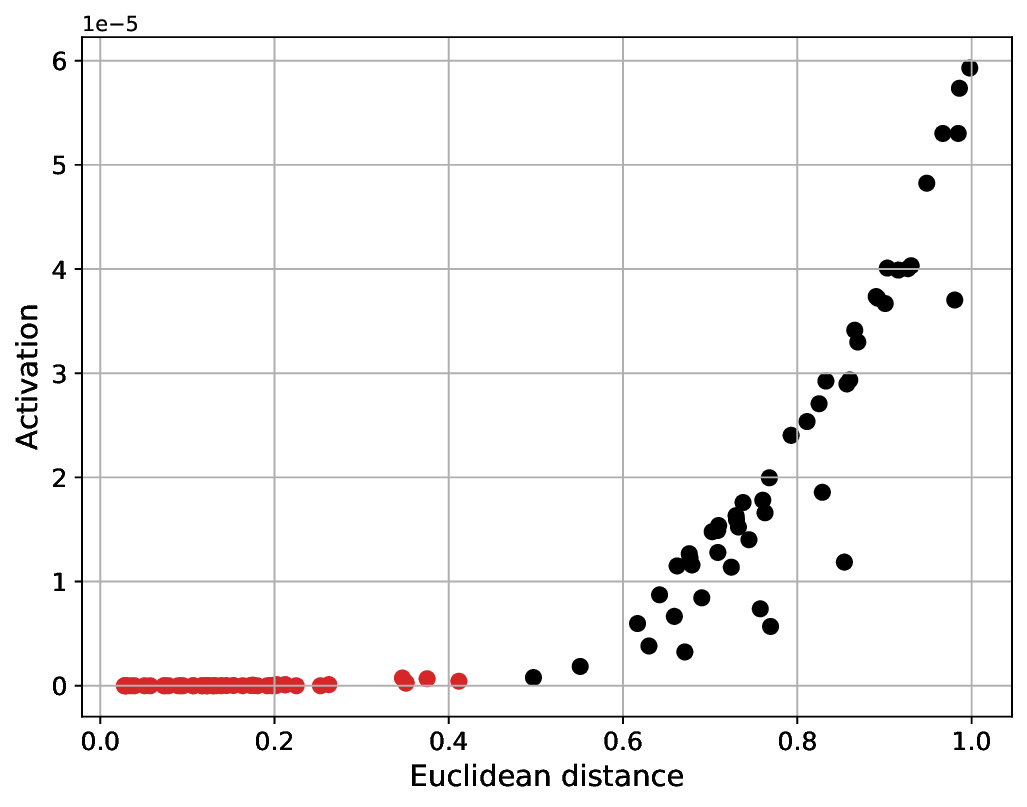}
        \label{fig:pcdqn_act_func-db_corner}
    }
    \caption{Solutions obtained by the CVQN, the CDQN, and the PCDQN, respectively, for the two problems derived from the Diagonal blobs. (a), (b), and (c) when targeting the class Center. (d), (e), and (f) when targeting the class Corner.}
    \label{fig:db_act_functions}
\end{figure}

Figure~\ref{fig:db_act_functions} contrasts the quantum neuron solutions for the two problems derived from the Diagonal blobs. The CVQN solves the problem when targeting the class Center by positioning the weight vector closer to the center cluster, which gives values of activation for that cluster higher than the ones for the corner clusters, as shown in Figure~\ref{fig:cvqn_act_func-db_center}. That CVQN solution required the ability to also distinguish the black blobs from the red blobs at the intersection of Euclidean distances between the clusters. A similar CVQN solution correctly separated the upper left cluster from the center cluster when targeting the class Corner, although the bottom right cluster has incorrectly obtained the lowest values, as shown in Figure~\ref{fig:cvqn_act_func-db_corner}. The CDQN and the PCDQN solve the problem when targeting the class Center by positioning the weight vector close to the target cluster, and then implementing a monotonic decay as the Euclidean distance increases in any direction, as shown in Figure~\ref{fig:cdqn_act_func-db_center} and Figure~\ref{fig:pcdqn_act_func-db_center}. A monotonic decay does not solve the problem when targeting the class Corner, as shown in Figure~\ref{fig:cdqn_act_func-db_corner} for the CDQN. Actually, a monotonic growth from a weight vector within the center cluster solves that problem, as shown in Figure~\ref{fig:pcdqn_act_func-db_corner} for the PCDQN, which can be achieved by means of parametrization.

\begin{figure}[!t]
    \centering
    \subfloat[]{
        \includegraphics[width=0.3\columnwidth]{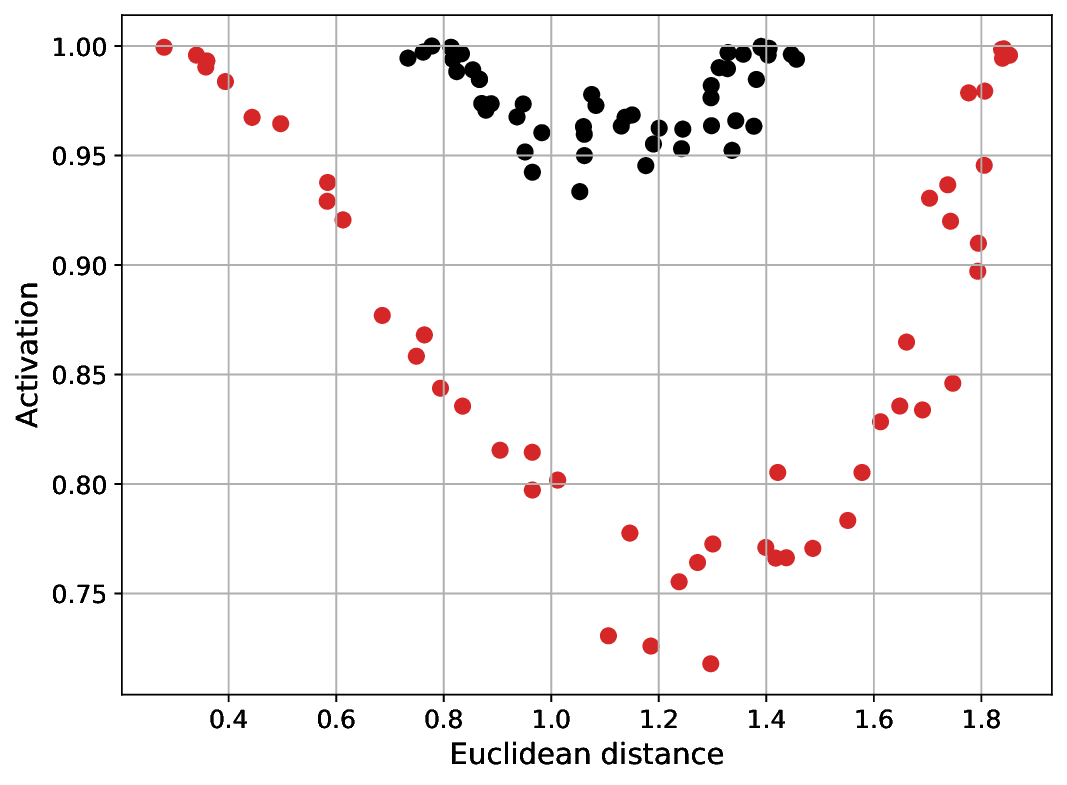}
        \label{fig:cvqn_act_func-cc_inner}
    }
    \subfloat[]{
        \includegraphics[width=0.3\columnwidth]{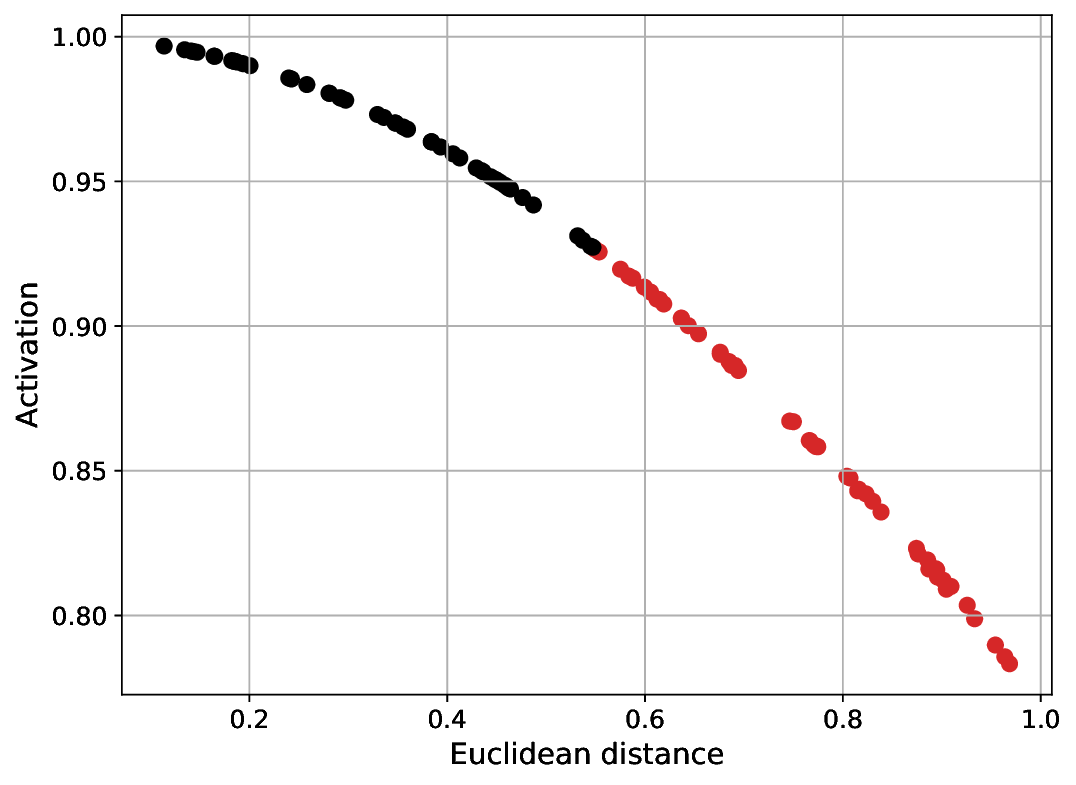}
        \label{fig:cdqn_act_func-cc_inner}
    }
    \subfloat[]{
        \includegraphics[width=0.3\columnwidth]{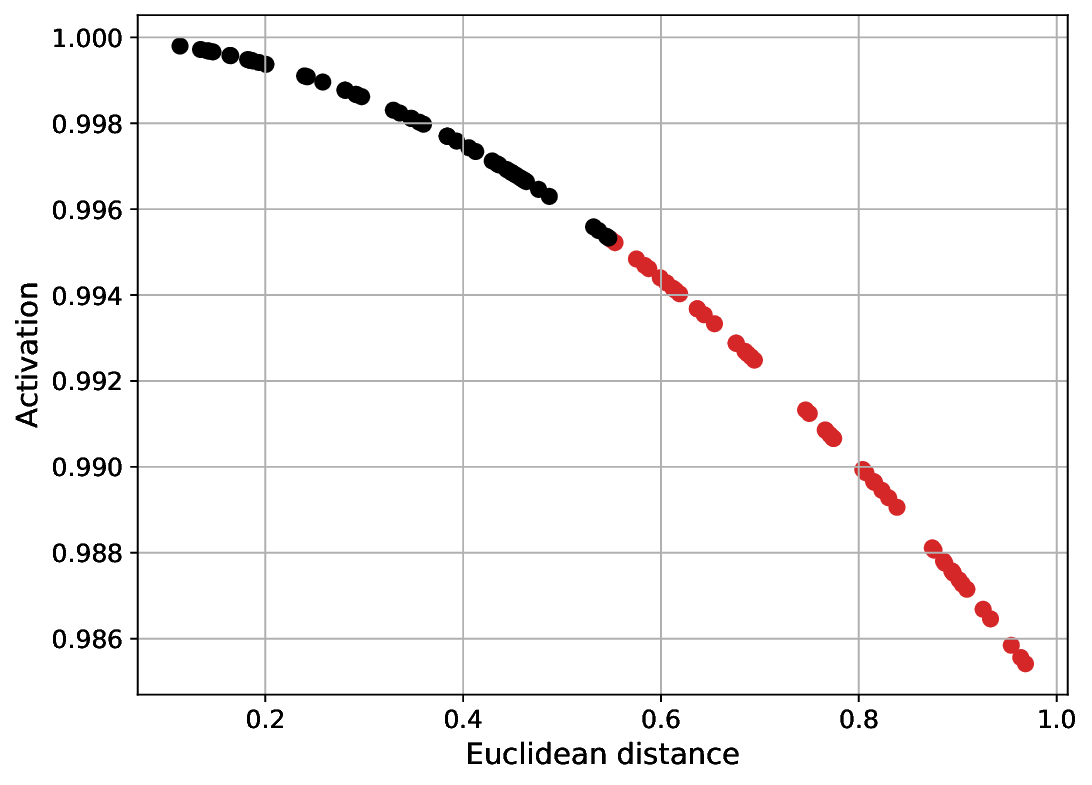}
        \label{fig:pcdqn_act_func-cc_inner}
    }
    
    \subfloat[]{
        \includegraphics[width=0.3\columnwidth]{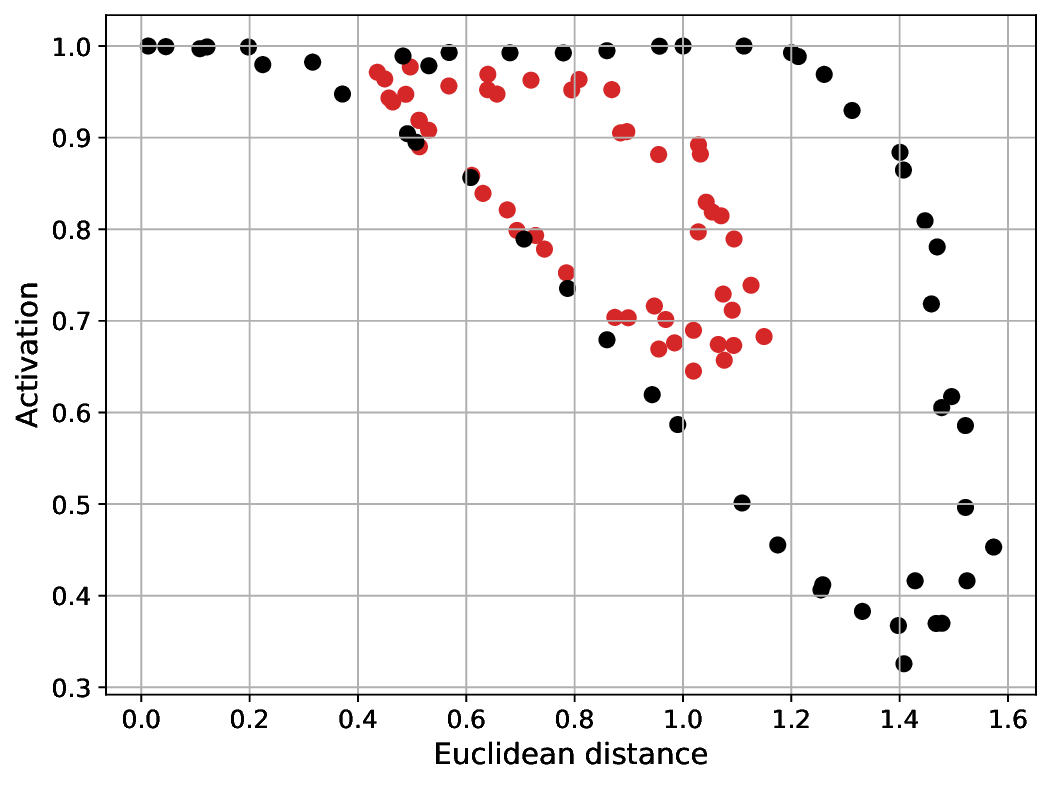}
        \label{fig:cvqn_act_func-cc_outer}
    }
    \subfloat[]{
        \includegraphics[width=0.3\columnwidth]{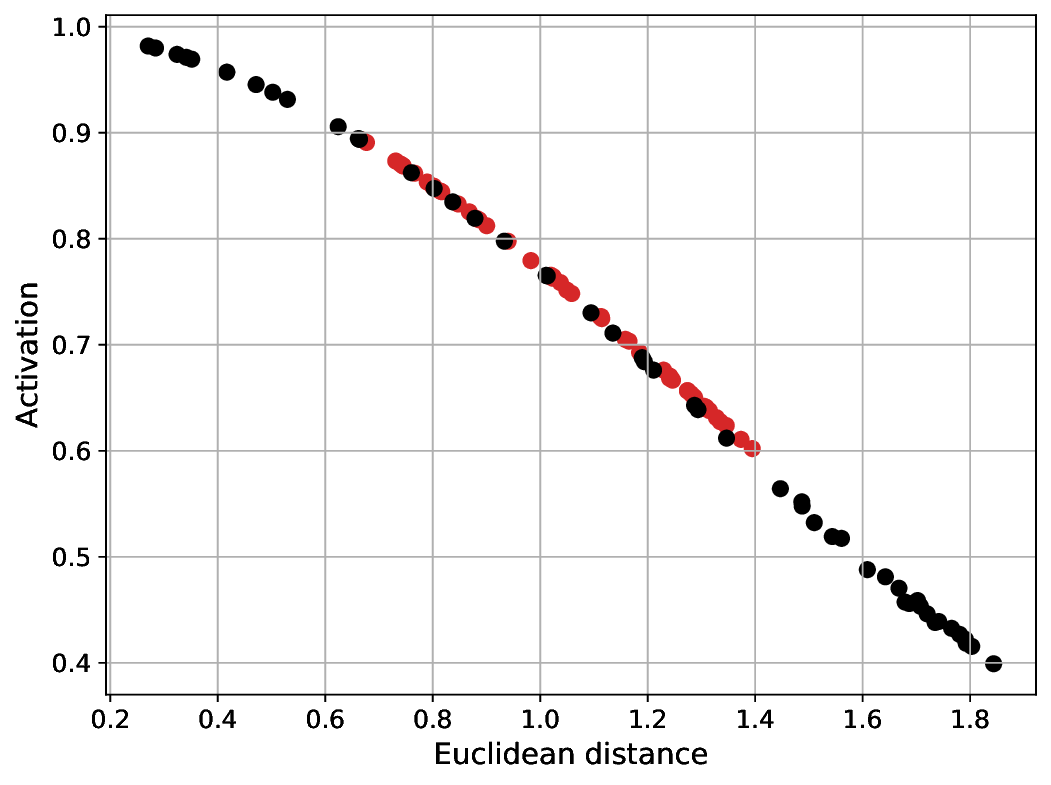}
        \label{fig:cdqn_act_func-cc_outer}
    }
    \subfloat[]{
        \includegraphics[width=0.3\columnwidth]{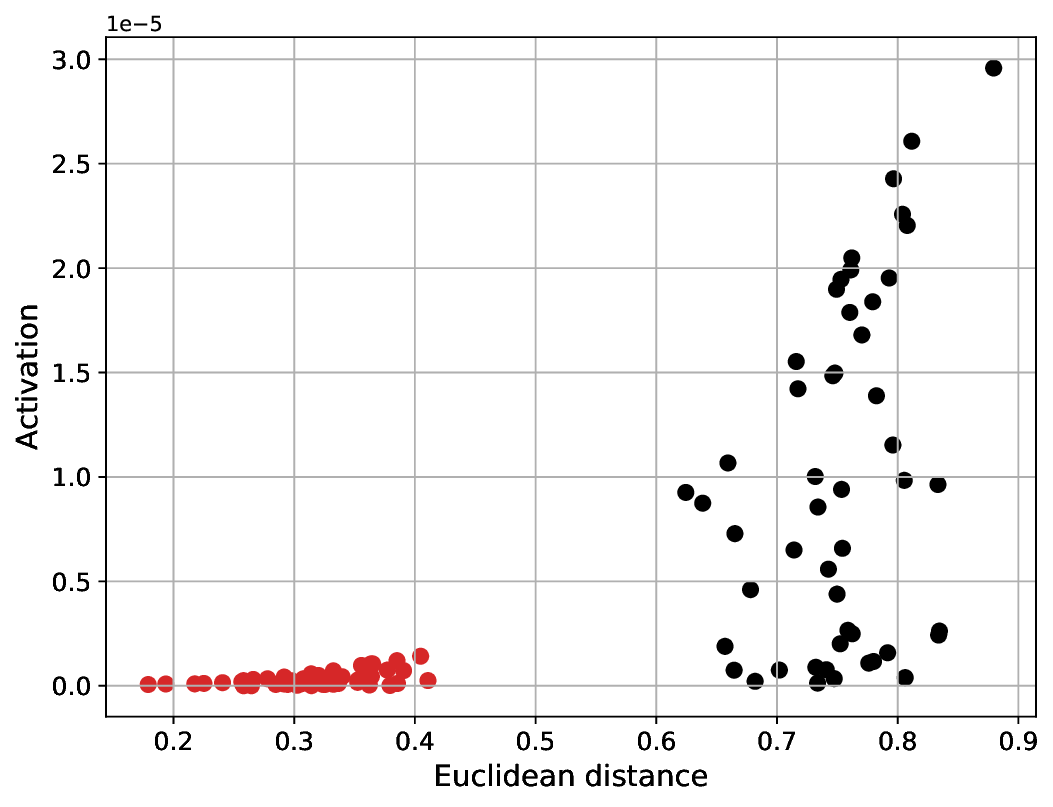}
        \label{fig:pcdqn_act_func-cc_outer}
    }
    \caption{Solutions obtained by the CVQN, the CDQN, and the PCDQN, respectively, for the two problems derived from the Concentric circles. (a), (b), and (c) when targeting the class Inner. (d), (e), and (f) when targeting the class Outer.}
    \label{fig:cc_act_functions}
\end{figure}

The quantum neuron solutions for the two problems derived from the Concentric circles are contrasted in Figure~\ref{fig:cc_act_functions}. The CVQN correctly separates the classes at the intersection of Euclidean distances between them when targeting the class Inner, as shown in Figure~\ref{fig:cvqn_act_func-cc_inner}. The CVQN does not solve the problem because the closest and the most distant blobs, which are of the negative class, obtained high values of activation. Figure~\ref{fig:cvqn_act_func-cc_outer} shows that the CVQN performed like a random model when targeting the class Outer because the solution basically implements a decay. However, the circle shapes inside that solution suggest the multiple neuron outputs for the same Euclidean distance depend on the distance direction. The CDQN and the PCDQN solve the problem when targeting the class Inner by implementing a monotonic decay from that target circle, as shown in Figure~\ref{fig:cdqn_act_func-cc_inner} and Figure~\ref{fig:pcdqn_act_func-cc_inner}. That shape does not fit the problem when targeting the class Outer, as shown in Figure~\ref{fig:cdqn_act_func-cc_outer} for the CDQN. A monotonic growth from the center can really fit that problem, which can be achieved by parametrization, as shown in Figure~\ref{fig:pcdqn_act_func-cc_outer} for the PCDQN. That solution does not fit perfectly because far blobs obtained low values of activation. Thus, the distance direction really matters.

\begin{figure}[!t]
    \centering
    \subfloat[]{
        \includegraphics[width=0.3\columnwidth]{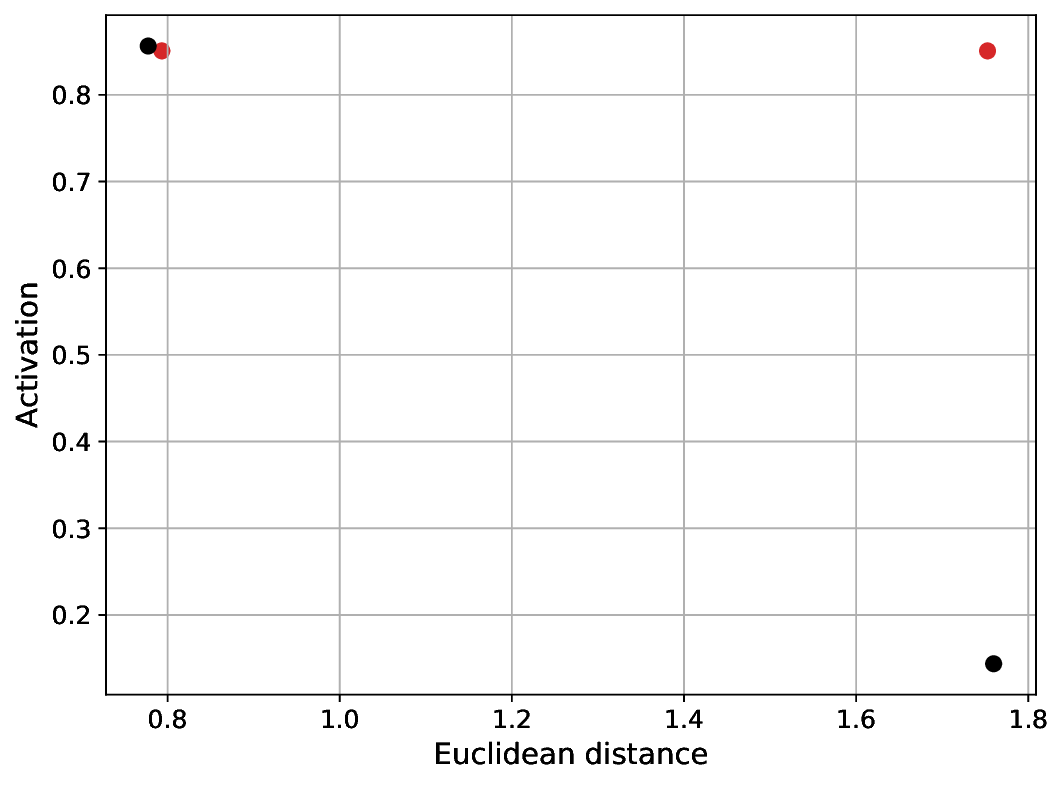}
        \label{fig:cvqn_act_func-sb_xor}
    }
    \subfloat[]{
        \includegraphics[width=0.3\columnwidth]{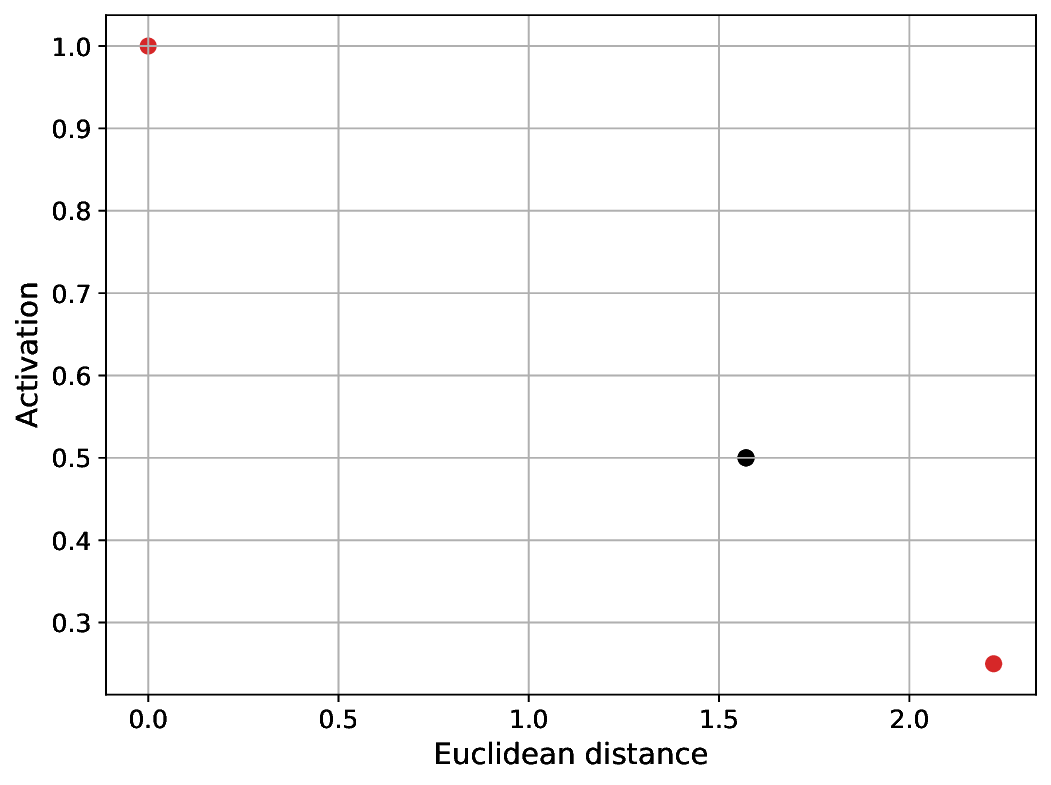}
        \label{fig:cdqn_act_func-sb_xor}
    }
    \subfloat[]{
        \includegraphics[width=0.3\columnwidth]{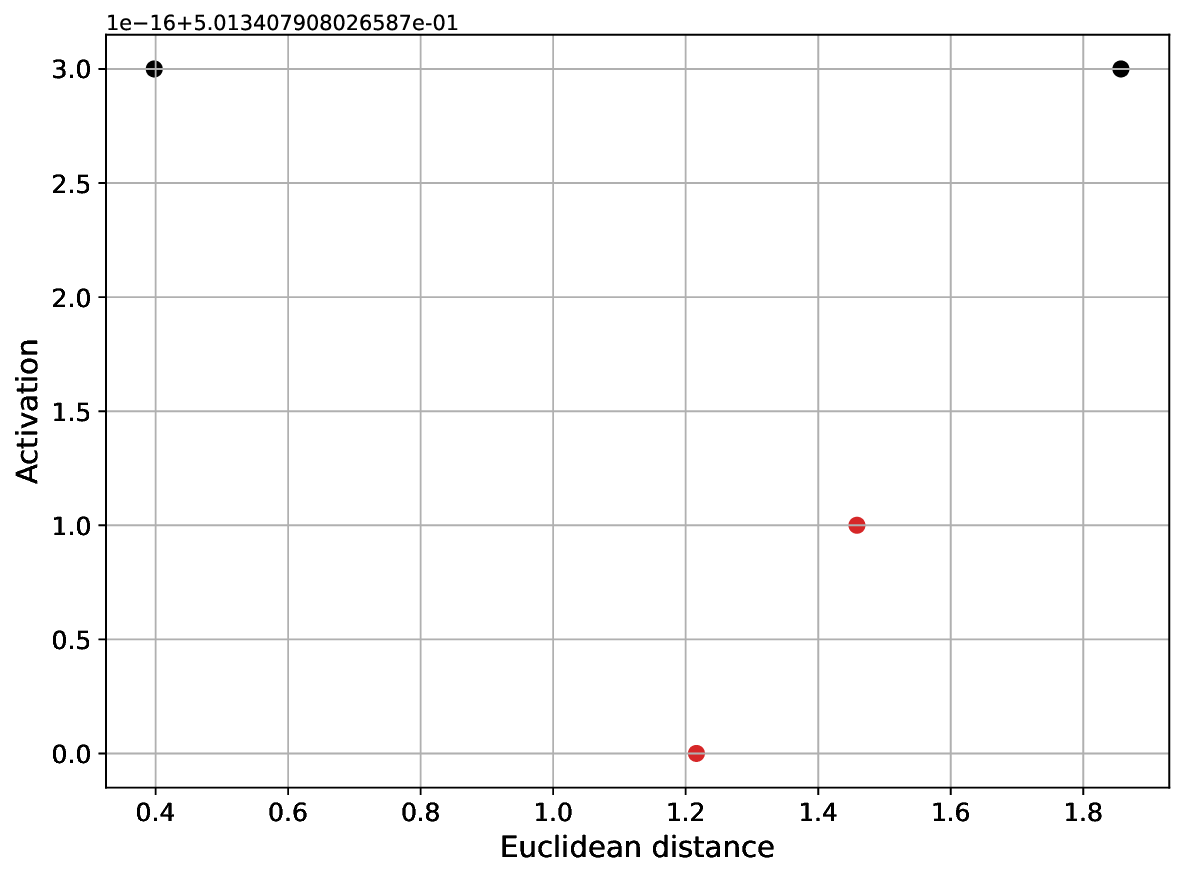}
        \label{fig:pcdqn_act_func-sb_xor}
    }
    
    \subfloat[]{
        \includegraphics[width=0.3\columnwidth]{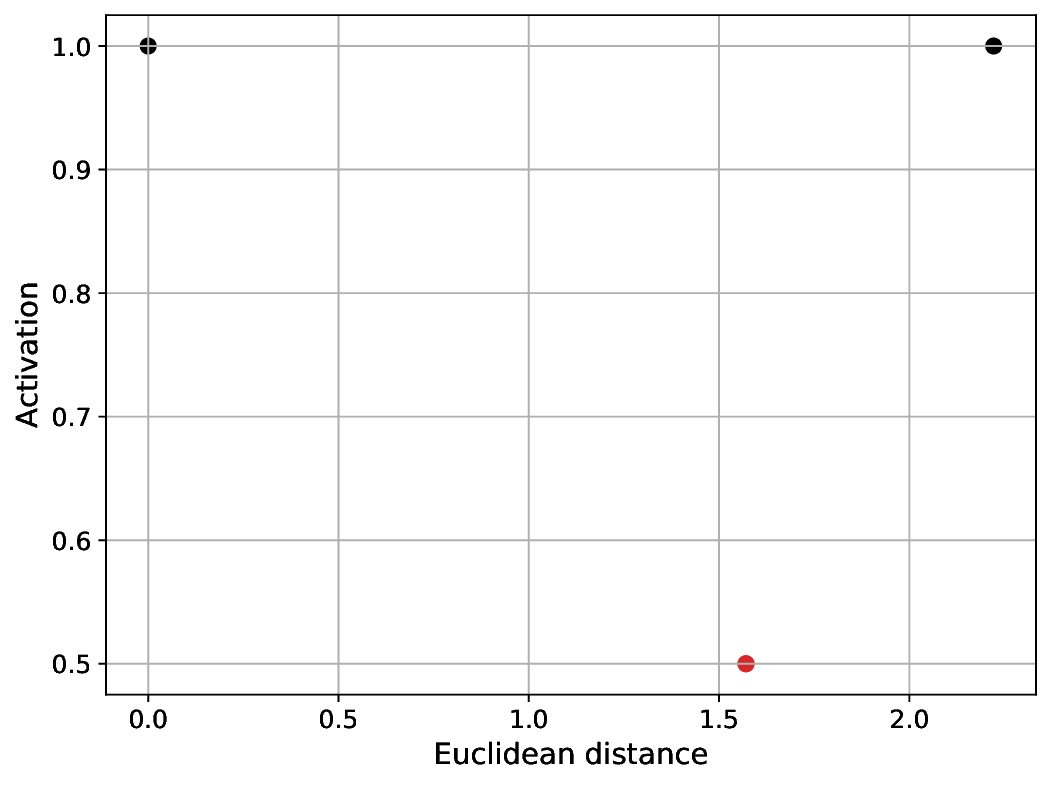}
        \label{fig:cvqn_act_func-sb_nxor}
    }
    \subfloat[]{
        \includegraphics[width=0.3\columnwidth]{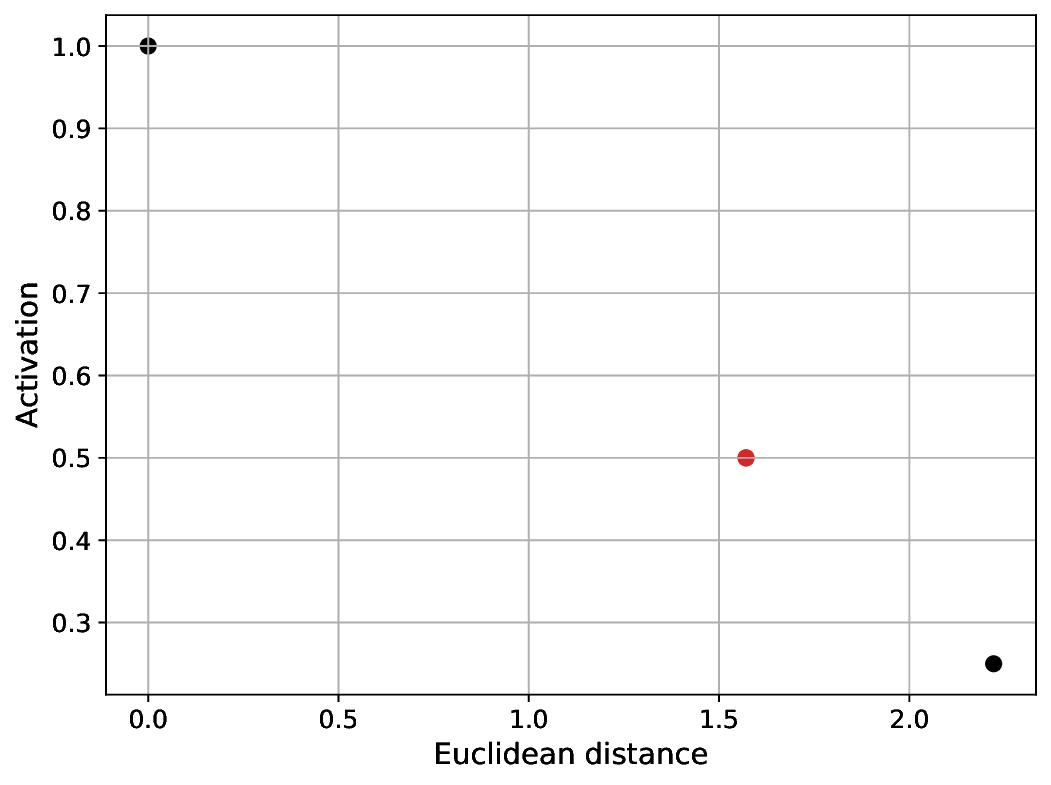}
        \label{fig:cdqn_act_func-sb_nxor}
    }
    \subfloat[]{
        \includegraphics[width=0.3\columnwidth]{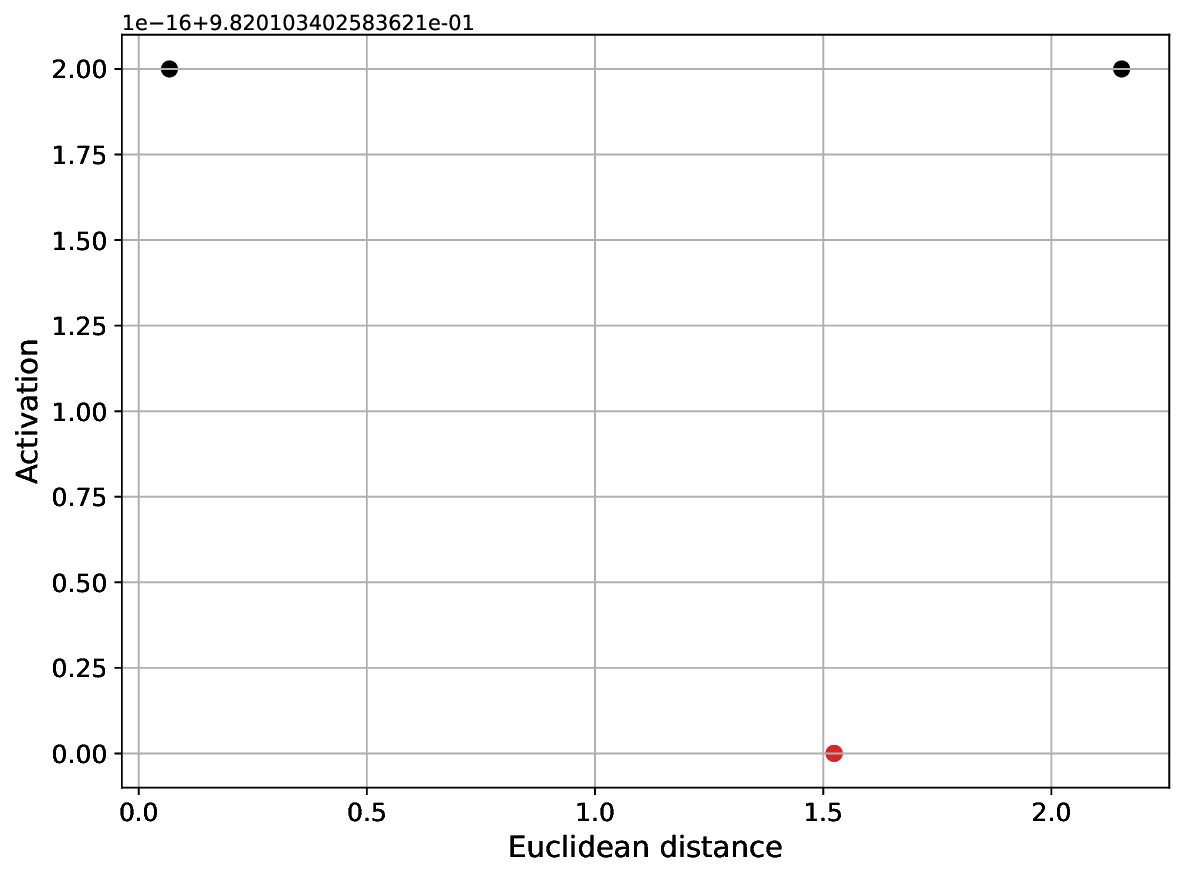}
        \label{fig:pcdqn_act_func-sb_nxor}
    }
    \caption{Solutions obtained by the CVQN, the CDQN, and the PCDQN, respectively, for the two problems derived from the Square blobs. (a), (b), and (c) when targeting the XOR problem. (d), (e), and (f) when targeting the NXOR problem.}
    \label{fig:sb_act_functions}
\end{figure}

Finally, Figure~\ref{fig:sb_act_functions} contrasts the quantum neuron solutions for the two problems derived from the Square blobs. Since those problems contain only four instances, we try to figure out the activation shape that is being formed in each case based on the way that each neuron works. In some of those cases, two instances that are at the same Euclidean distance from the best weight vector also give the same neuron output, which means that the activation function shape will be formed by three points only. When targeting the XOR-like problem, the CVQN performed as a random model because the solution basically implements a decay, as shown in Figure~\ref{fig:cvqn_act_func-sb_xor}. A distant point obtained a high value of activation due to its distance direction. The CVQN solves the problem perfectly when targeting the NXOR-like problem by taking advantage of all spectrum of its activation shape to generate high values of activation in the extremes and low values in the intermediate distances, as shown in Figure~\ref{fig:cvqn_act_func-sb_nxor}. A monotonic decay does not fit any of those two problems, as shown in Figure~\ref{fig:cdqn_act_func-sb_xor} and Figure~\ref{fig:cdqn_act_func-sb_nxor}, both for the CDQN. In contrast, a concave-up parabola can fit those two problems. By the way, the PCDQN can implement such a shape, as already shown in Figure~\ref{fig:pcdqn_act_function_for_4_0}. Thus, the PCDQN can solve those two problems perfectly, as shown in Figure~\ref{fig:pcdqn_act_func-sb_xor} and Figure~\ref{fig:pcdqn_act_func-sb_nxor}.

Those results demonstrate limitations in the CVQN and in the CDQN since their activation shapes can fit only particular problems. The PCDQN, in turn, can dynamically change its activation shape to a monotonic decay, a monotonic growth, and a concave-up parabola, for example. That list of shapes is not exhaustive. The PCDQN can implement even more shapes by exploring its real parameters, depending on the underlying pattern of the problem. Thus, the PCDQN is flexible. Such flexibility allows the PCDQN to fit problems that are beyond the capabilities of the other neurons.

\subsection{Execution on a Quantum Simulator}

Finally, we demonstrate the physical feasibility of the previous numerical solutions through a proof-of-concept experiment on a quantum simulator. We address here the two classification problems used in~\cite{mangini_CVQN}: Diagonal blobs targeting the class Center and Concentric circles targeting the class Inner. The quantum neuron solutions for those two problems are then implemented in quantum circuits that are executed on the Qiskit QASM simulator.

To estimate the quantum neuron activation for a given pair of vectors, we took the proportion of 1's in the circuit outputs after 20,000 executions of the circuit. Thus, each quantum neuron is executed 20,000 times for each input vector of each classification problem to estimate the activation with respect to the corresponding best weight vector. The amount of circuit executions for 3 quantum neurons applied in 2 classification problems of 100 samples each is 12,000,000. Nevertheless, finite numbers of execution can still cause small errors in the activation estimates.

\begin{figure}[!t]
    \centering
    \includegraphics[width=0.6\columnwidth]{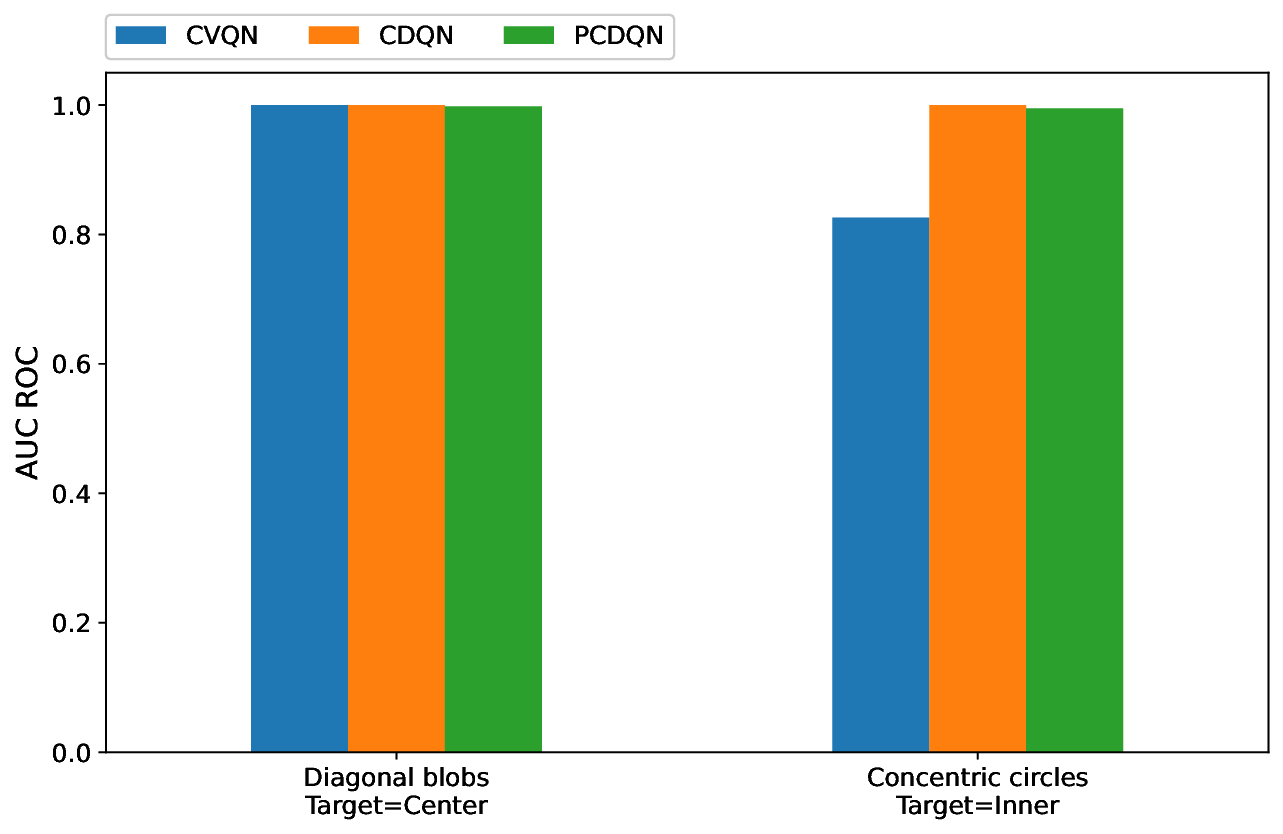}
    \caption{AUC ROC generated by the quantum neuron circuits on a quantum simulator for the Diagonal blobs when targeting the class Center and the Concentric circles when targeting the class Inner. The blue, orange, and green bars represent the CVQN, the CDQN, and the PCDQN, respectively.}
    \label{fig:aucroc_on_simulator}
\end{figure}

After estimating the activation, we can compute the circuit-generated AUC ROC of each quantum neuron for each classification problem on the quantum simulator. Those circuit-generated values of AUC ROC are contrasted in Figure~\ref{fig:aucroc_on_simulator}. The blue bars represent the CVQN, the orange bars represent the CDQN, and the green bars represent the PCDQN. In fact, the quantum neurons reproduced the reference values reported in Table~\ref{tab:toy_max_aucroc} for those two problems. The CDQN and the PCDQN generated optimal results, while the CVQN was optimal in the first problem but generated an AUC ROC of about 0.82 in the second problem. Therefore, the quantum neuron solutions are validated by the circuit realizations in the quantum simulator.

\section{Recognizing Handwritten Digits}
\label{sec:mnist_exp}

\begin{figure}[!t]
    \centering
    \subfloat[]{
        \includegraphics[width=0.17\columnwidth]{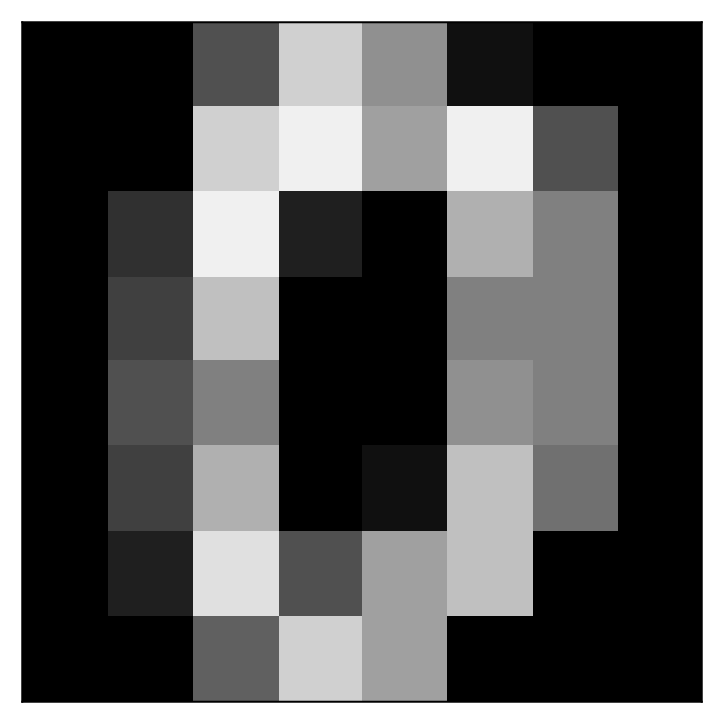}
        \label{fig:digit0_example}
    }
    \subfloat[]{
        \includegraphics[width=0.17\columnwidth]{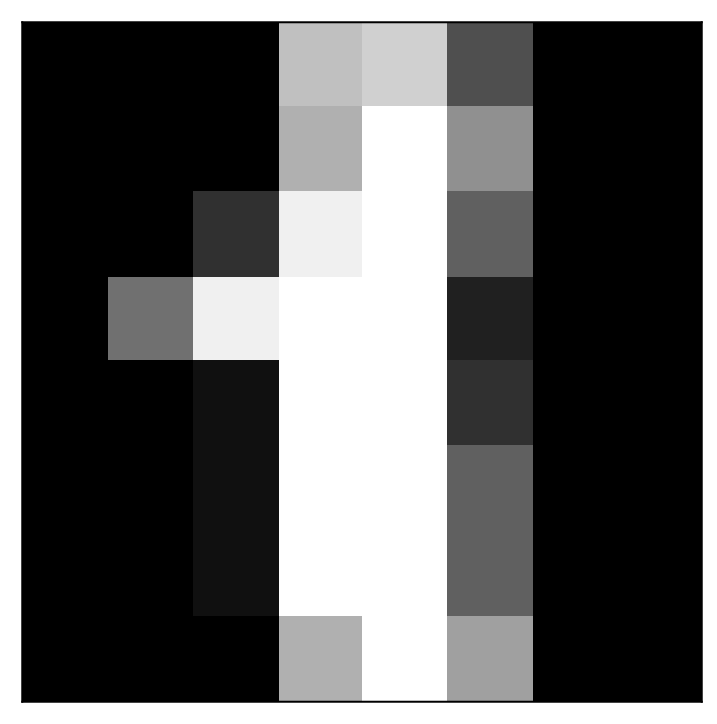}
        \label{fig:digit1_example}
    }
    \subfloat[]{
        \includegraphics[width=0.17\columnwidth]{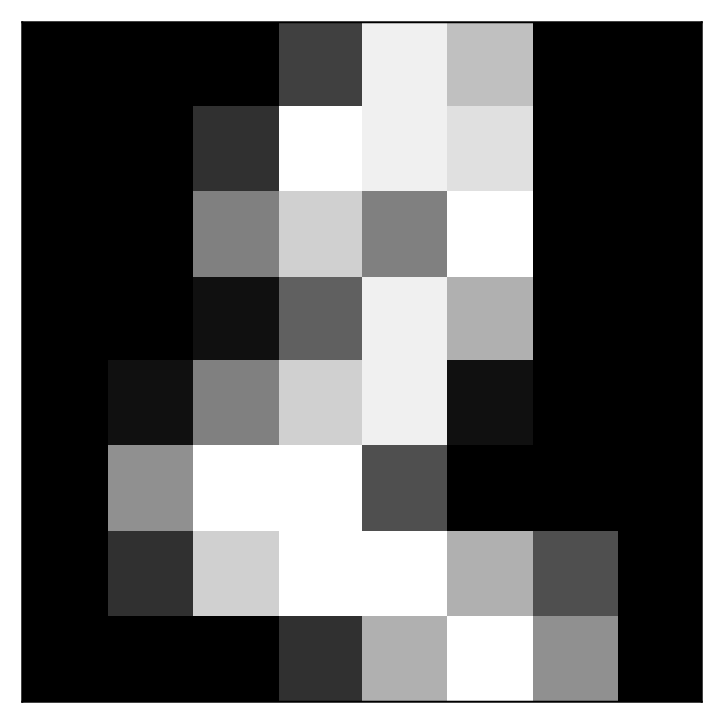}
        \label{fig:digit2_example}
    }
    \subfloat[]{
        \includegraphics[width=0.17\columnwidth]{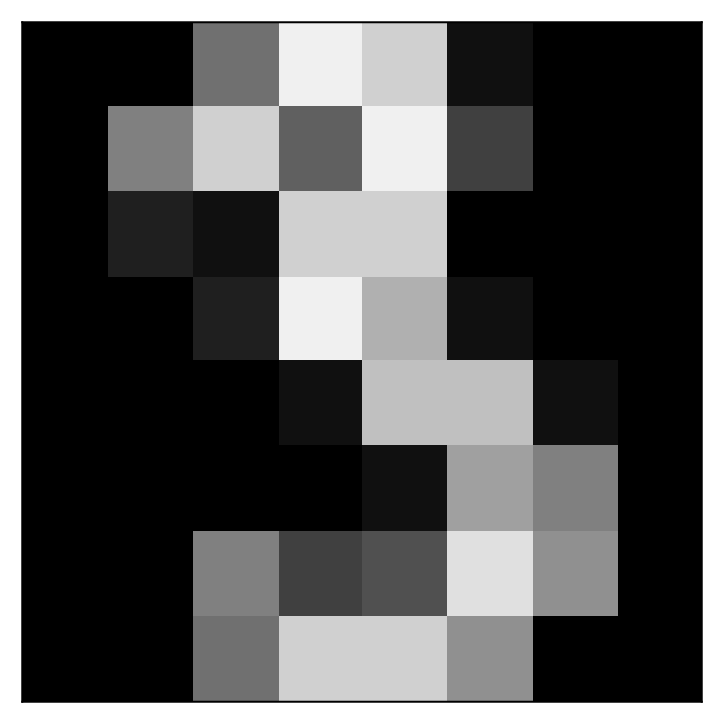}
        \label{fig:digit3_example}
    }
    \subfloat[]{
        \includegraphics[width=0.17\columnwidth]{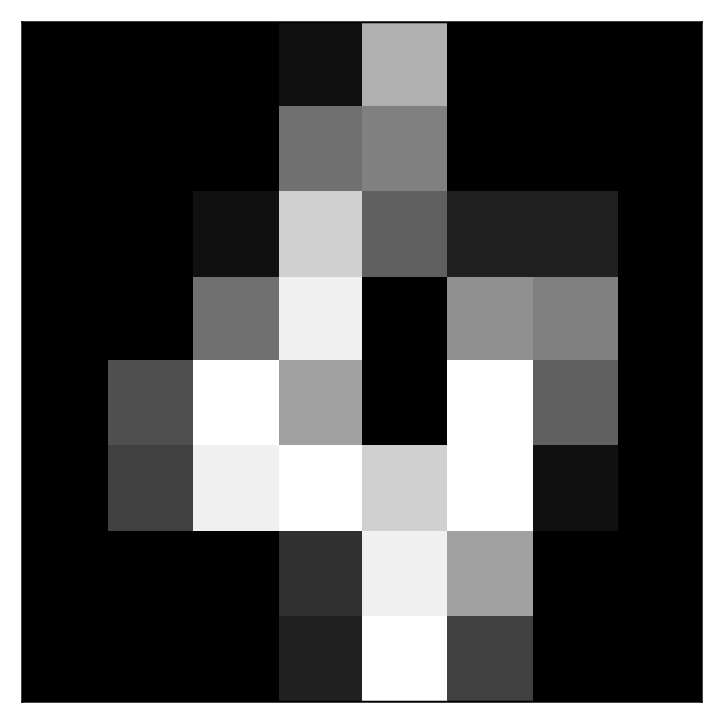}
        \label{fig:digit4_example}
    }
    
    \subfloat[]{
        \includegraphics[width=0.17\columnwidth]{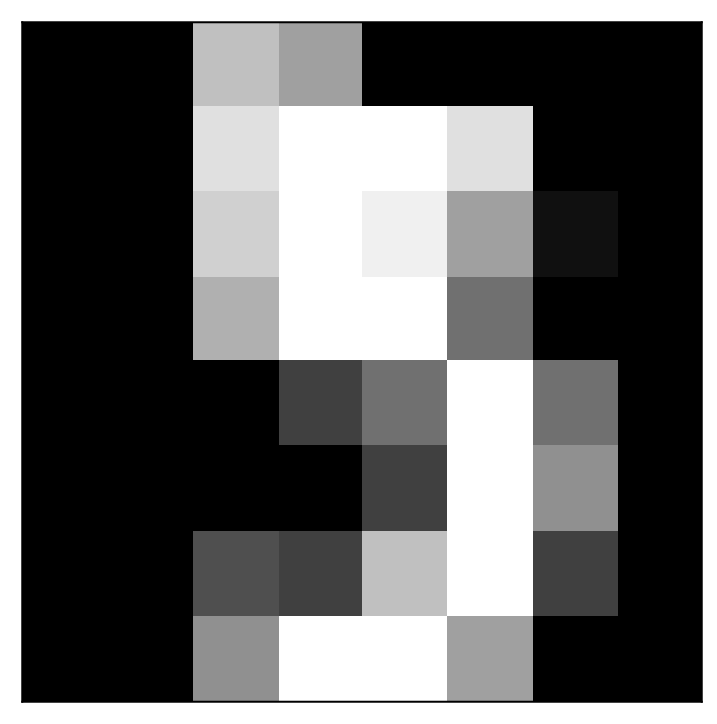}
        \label{fig:digit5_example}
    }
    \subfloat[]{
        \includegraphics[width=0.17\columnwidth]{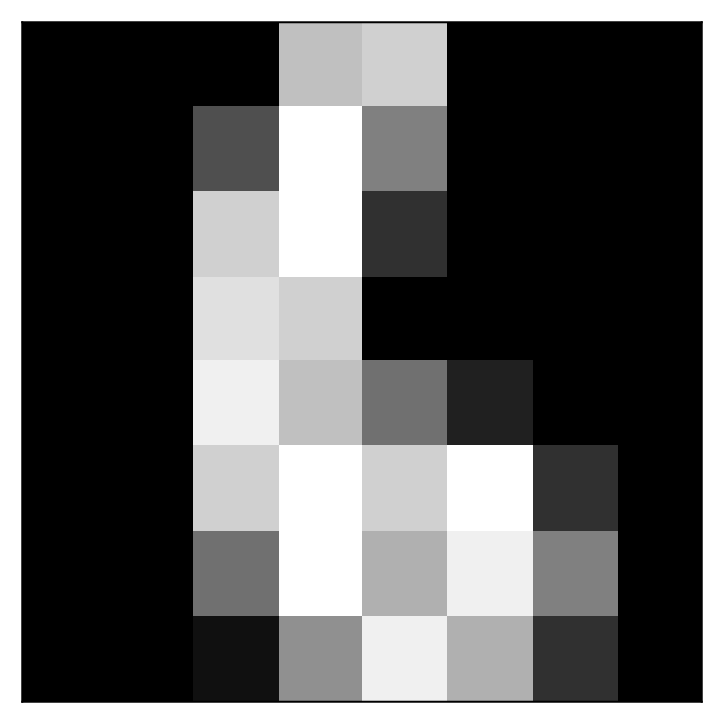}
        \label{fig:digit6_example}
    }
    \subfloat[]{
        \includegraphics[width=0.17\columnwidth]{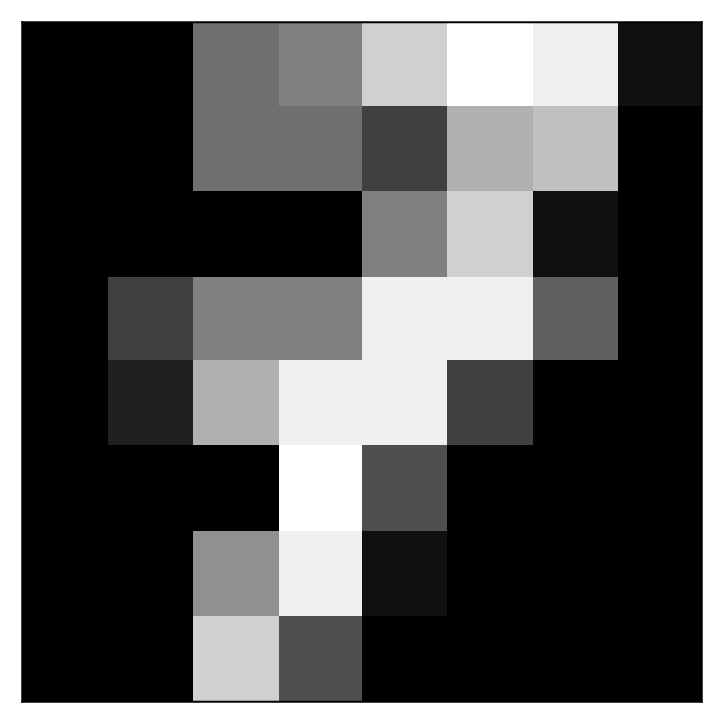}
        \label{fig:digit7_example}
    }
    \subfloat[]{
        \includegraphics[width=0.17\columnwidth]{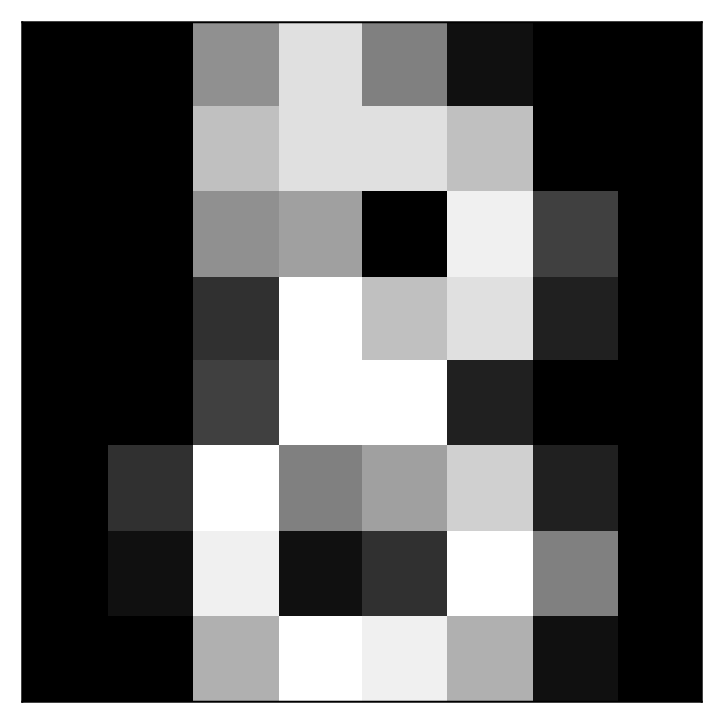}
        \label{fig:digit8_example}
    }
    \subfloat[]{
        \includegraphics[width=0.17\columnwidth]{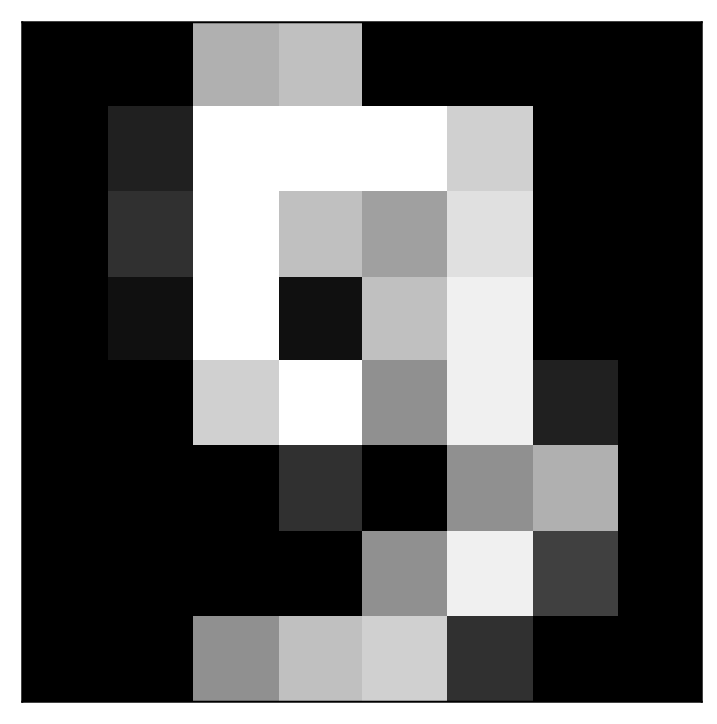}
        \label{fig:digit9_example}
    }
    \caption{First occurrence of each digit in the dataset of handwritten digits addressed here. The 8x8 image matrices are displayed from 64-dimensional vectors with integer entries between 0 and 16.}
    \label{fig:digit_examples}
\end{figure}

Since we validated the quantum neurons in toy datasets, the next step is to verify their performances in a real dataset in pursuit of more conclusive evidence. In this way, we addressed the handwritten digits dataset provided by the scikit-learn \textit{load\_digits} function. Each one of the 10 digits has approximately 180 samples in the dataset. For illustrative purposes, the first occurrence of each digit is shown in Figure~\ref{fig:digit_examples}, where Figure~\ref{fig:digit0_example} shows the digit 0, Figure~\ref{fig:digit1_example} shows the digit 1, Figure~\ref{fig:digit2_example} shows the digit 2, and so on. These 8x8 images are represented by vectors of 64 dimensions, where each entry is an integer in the interval [0, 16], although the entries are scaled to [0, $\pi$/2] before being given as inputs to the quantum neurons. By using a one-vs-all approach for each digit, the objective of the quantum neurons here is to recognize the samples of a digit by producing values of activation higher than the ones for all other digits, which can be measured by the AUC ROC. To maximize the AUC ROC, each quantum neuron tried ten thousand weight vectors $\boldsymbol{\phi}$ randomly generated by NumPy~\cite{numpy} with a seed equal to 0, where $\boldsymbol{\phi} \in [0, \pi/2]^{64}$. As before, we tried those 34 parameter combinations for the PCDQN.

Conclusive evidence can be obtained not only by addressing more complex problems but also by contrasting quantum neurons based on different architectures. Specifically, we incorporated in the experiments a quantum model capable of implementing classical activation functions in a discrete manner given some precision~\cite{paula-neto_any_nonlinear-QN, yan_any_nonlinear-QN_efficiently}. Inspired in~\cite{paula-neto_any_nonlinear-QN}, we compared against the linear, sigmoid, and radial-basis activation functions, which are defined respectively as $\varphi_1(\vec{i} \cdot \vec{w}) = \vec{i} \cdot \vec{w}$, $\varphi_2(\vec{i} \cdot \vec{w}) = (1 + e^{-(\vec{i} \cdot \vec{w})})^{-1}$, and $\varphi_3(\lVert \vec{i} - \vec{w} \rVert) = e^{-\frac{1}{2} \lVert \vec{i} - \vec{w} \rVert^2}$, all of them with a precision of 4 qubits, i.e., subdividing the output interval into $2^4=16$ discrete parts. The implementations of these 3 discrete activation functions are named here as linear discrete quantum neuron (LDQN), sigmoid discrete quantum neuron (SDQN), and radial-basis discrete quantum neuron (RBDQN). Since this quantum architecture encodes information in the basis states as bit strings, we can use the data in the original integer interval [0, 16] for each entry as well as the entries for the ten thousand random weight vectors sampled from NumPy with a seed of 0. However, due to exploding inner products and Euclidean distances, we scaled the input and weight entries to [0, 0.5] in the activation function computations. Finally, for each classification problem, each discrete quantum neuron can search for the AUC ROC maximization by trying all weight vectors.

As a natural consequence, it is possible to see that the quantum neurons that approximate existing activation functions do not represent any advantage over their classical counterparts in terms of discriminative power. Nonetheless, those discrete quantum neurons provide a quantum baseline for the kernel-based quantum neurons with no classical analog that we support here, mainly the ones of constant circuit depth. Comparing against the discrete quantum neurons is an opportunity to investigate the advantages that can emerge by using strictly quantum activation functions in classification tasks. Table~\ref{tab:mnist_max_aucroc} shows the maximum AUC ROC achieved by the quantum neurons when targeting each one of the 10 digits in a one-vs-all approach. The simplest of the discrete quantum neurons, which is the LDQN, obtained the worst values of AUC ROC in all cases, except when targeting the digit 8 in which the SDQN was the worst. Targeting the digit 1 was the only case where the SDQN outperformed a neuron other than the LDQN. In that case, the RBDQN obtained a value of AUC ROC smaller. Actually, the RBDQN was the discrete neuron to beat here due to its significant improvements over the LDQN and the SDQN in all other cases. It turns out that kernel-based quantum neurons could surpass the performances of such a baseline. For example, the CVQN was better than the RBDQN in all cases, excluding the case of targeting the digit 8. Particularly, targeting the digit 8 was the case where the CDQN surpassed the CVQN the most, which also implied a reasonable improvement over the RBDQN. As the CDQN was better than the CVQN also in the other cases, the neuron that we proposed, even in its non-parametrized form, could better recognize handwritten digits than the previous proposals of quantum neurons. Except for digit 6, the PCDQN improved the results even more, especially for digit 1 where a boost in the AUC ROC really exhibited the potential of the activation function parametrization.

\begin{table*}
    \centering
    \caption{Maximum AUC ROC of Discrete Quantum Neurons and Kernel-Based Quantum Neurons in the Recognition of Handwritten Digits.}
    \label{tab:mnist_max_aucroc}
    \begin{tabular}{c|c|c|c|c|c|c}
        \hline
        Target & LDQN & SDQN & RBDQN & CVQN & CDQN & PCDQN \\
        \hline
        0 & 0.7584 & 0.7744 & 0.9717 & 0.9852 & 0.9968 & \textbf{0.9977} \\
        1 & 0.6916 & 0.7139 & 0.7039 & 0.7857 & 0.8244 & \textbf{0.9444} \\
        2 & 0.7346 & 0.7488 & 0.8819 & 0.9237 & 0.9485 & \textbf{0.9585} \\
        3 & 0.724 & 0.728 & 0.8944 & 0.9325 & 0.9355 & \textbf{0.9367} \\
        4 & 0.7639 & 0.7646 & 0.888 & 0.9572 & 0.9715 & \textbf{0.9717} \\
        5 & 0.6836 & 0.7164 & 0.8618 & 0.9499 & 0.977 & \textbf{0.9774} \\
        6 & 0.7373 & 0.8086 & 0.9373 & 0.9518 & \textbf{0.9699} & 0.9694 \\
        7 & 0.6875 & 0.7456 & 0.9026 & 0.9527 & 0.9638 & \textbf{0.9642} \\
        8 & 0.7345 & 0.7296 & 0.8361 & 0.8161 & 0.9065 & \textbf{0.9092} \\
        9 & 0.6878 & 0.736 & 0.8708 & 0.9003 & 0.9302 & \textbf{0.9316} \\
        \hline
    \end{tabular}
\end{table*}

\begin{figure*}[!t]
    \centering
    \subfloat[]{
        \includegraphics[width=0.31\columnwidth]{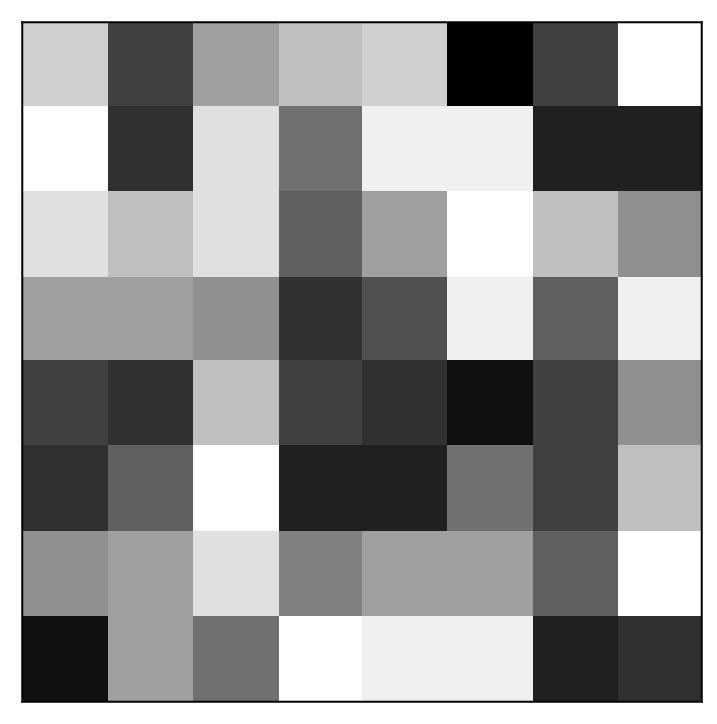}
        \label{fig:wvector_linear_digit0}
    }
    \subfloat[]{
        \includegraphics[width=0.31\columnwidth]{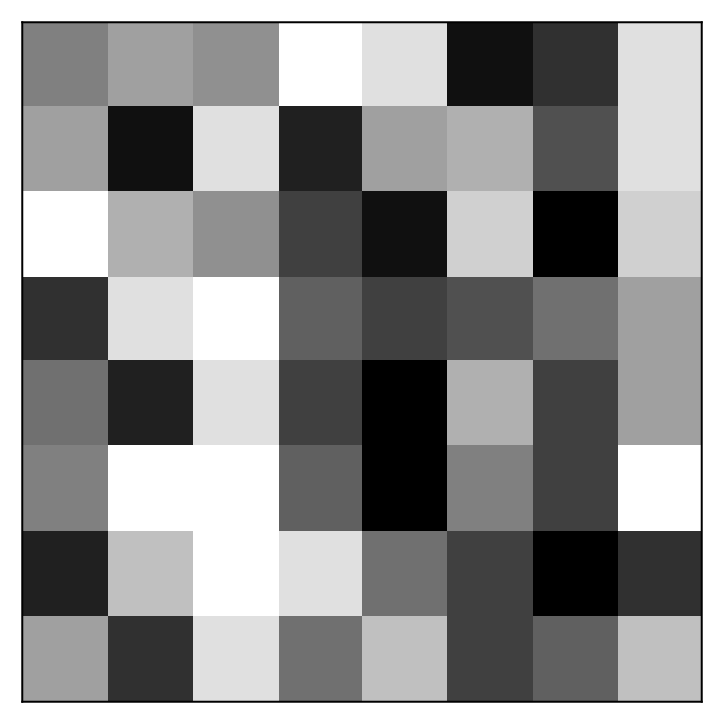}
        \label{fig:wvector_sigmoid_digit0}
    }
    \subfloat[]{
        \includegraphics[width=0.31\columnwidth]{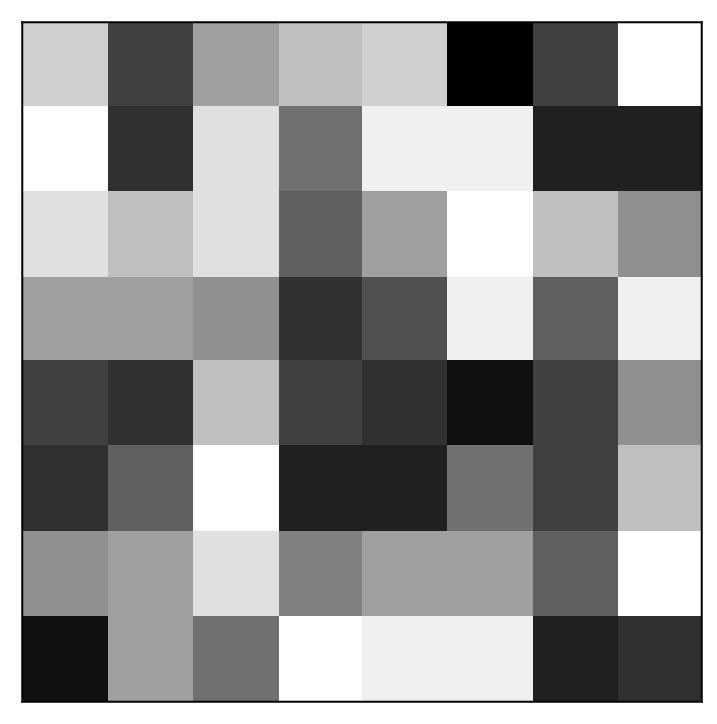}
        \label{fig:wvector_rbf_digit0}
    }
    \subfloat[]{
        \includegraphics[width=0.31\columnwidth]{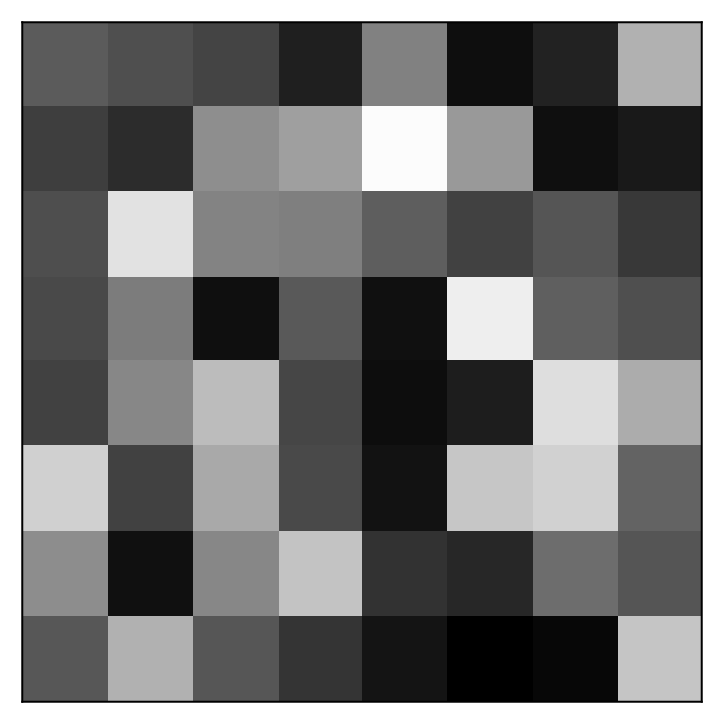}
        \label{fig:wvector_cvqn_digit0}
    }
    \subfloat[]{
        \includegraphics[width=0.31\columnwidth]{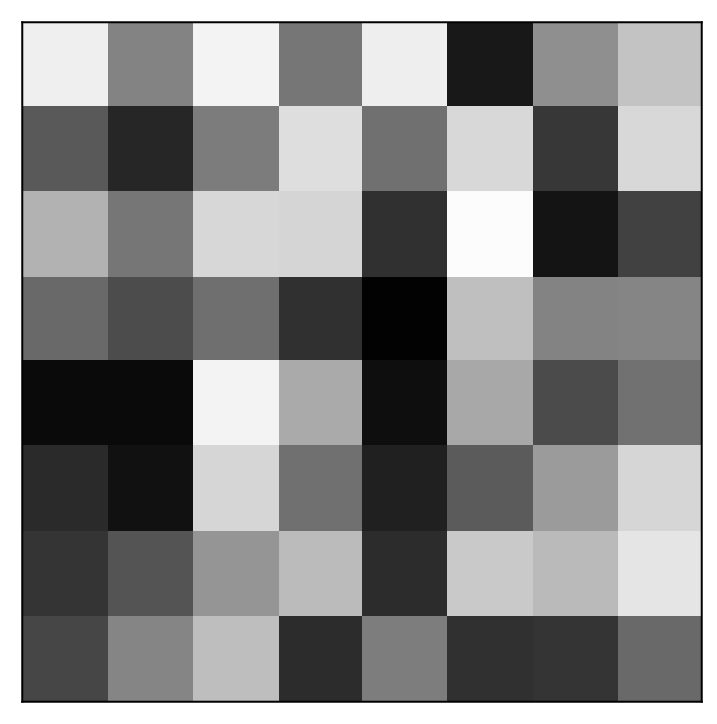}
        \label{fig:wvector_cdqn_digit0}
    }
    \subfloat[]{
        \includegraphics[width=0.31\columnwidth]{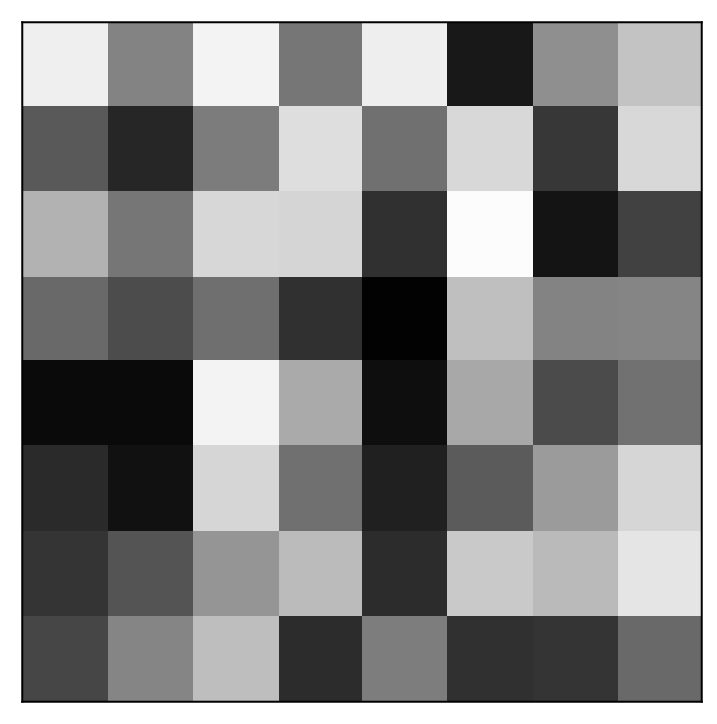}
        \label{fig:wvector_pcdqn_digit0}
    }
    
    \subfloat[]{
        \includegraphics[width=0.31\columnwidth]{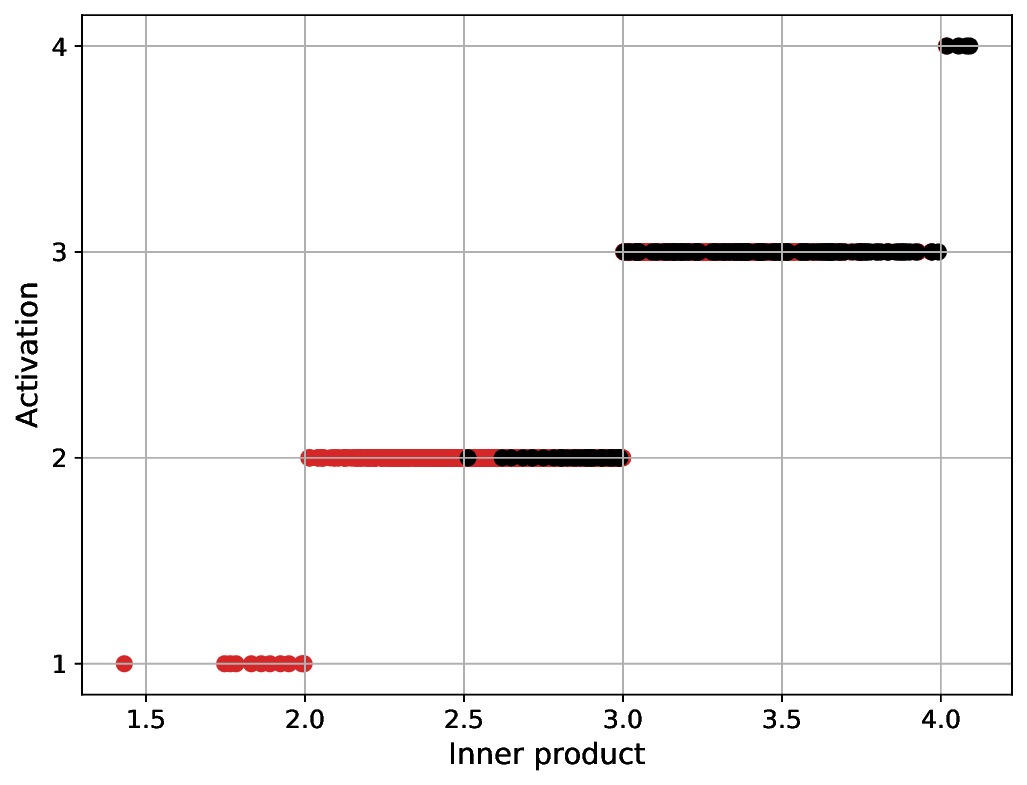}
        \label{fig:act-function_linear_digit0}
    }
    \subfloat[]{
        \includegraphics[width=0.31\columnwidth]{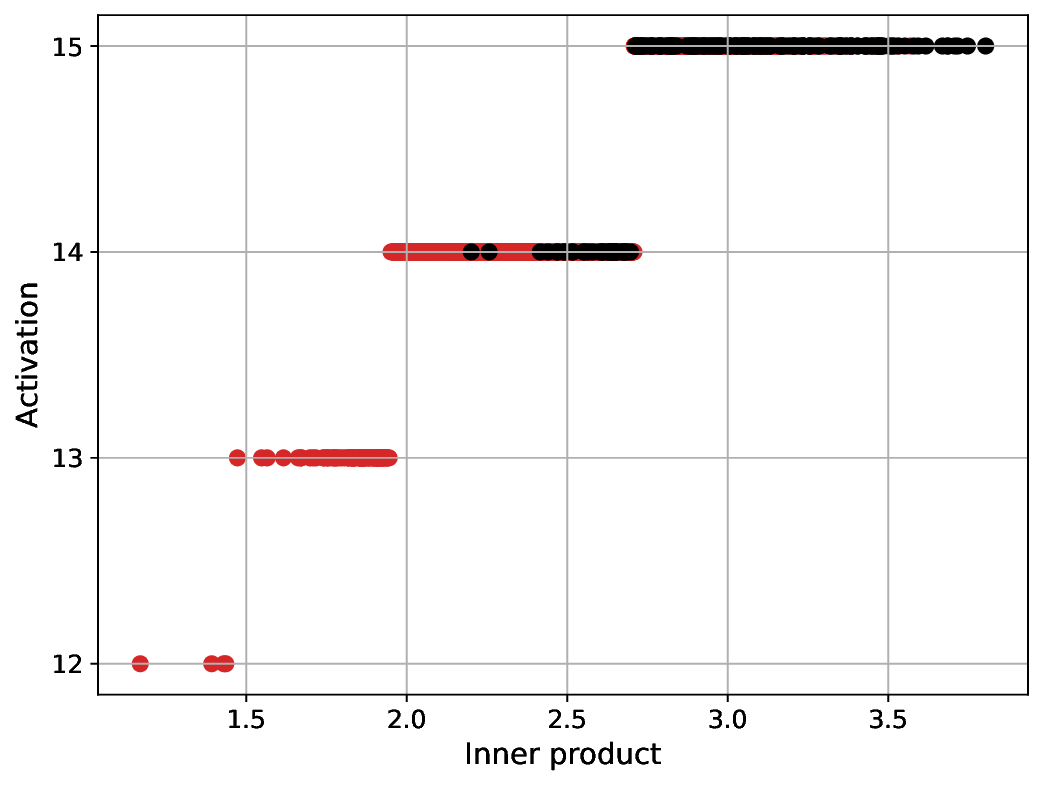}
        \label{fig:act-function_sigmoid_digit0}
    }
    \subfloat[]{
        \includegraphics[width=0.31\columnwidth]{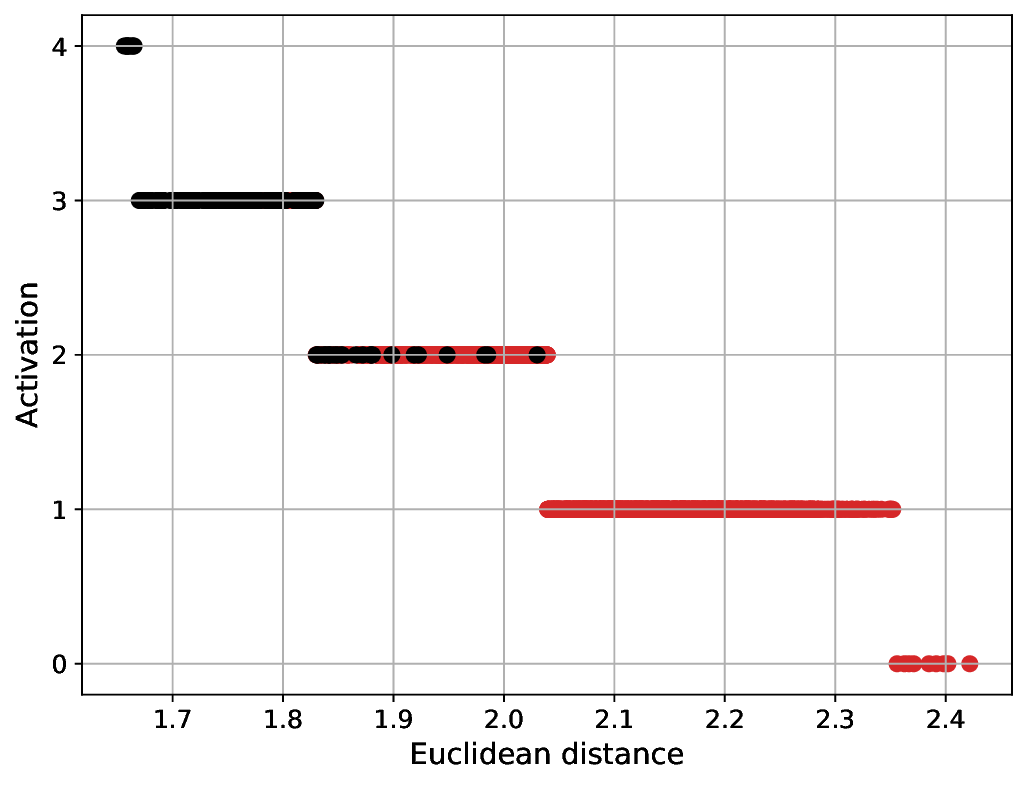}
        \label{fig:act-function_rbf_digit0}
    }
    \subfloat[]{
        \includegraphics[width=0.31\columnwidth]{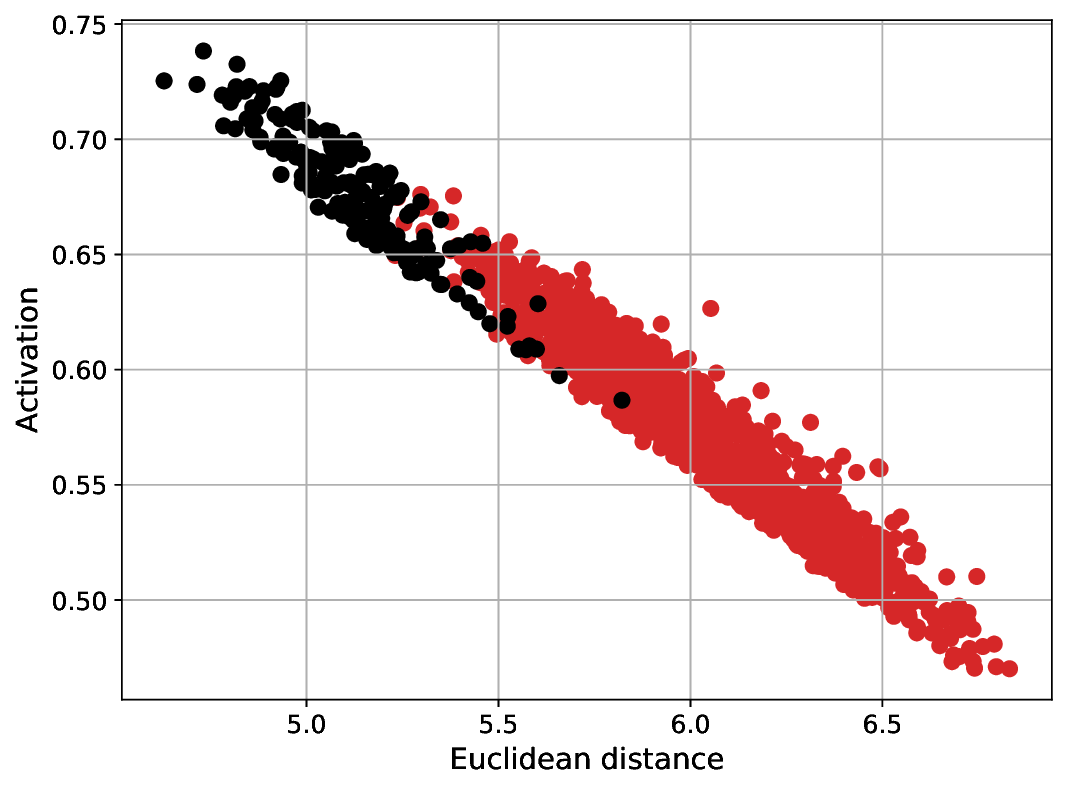}
        \label{fig:act-function_cvqn_digit0}
    }
    \subfloat[]{
        \includegraphics[width=0.31\columnwidth]{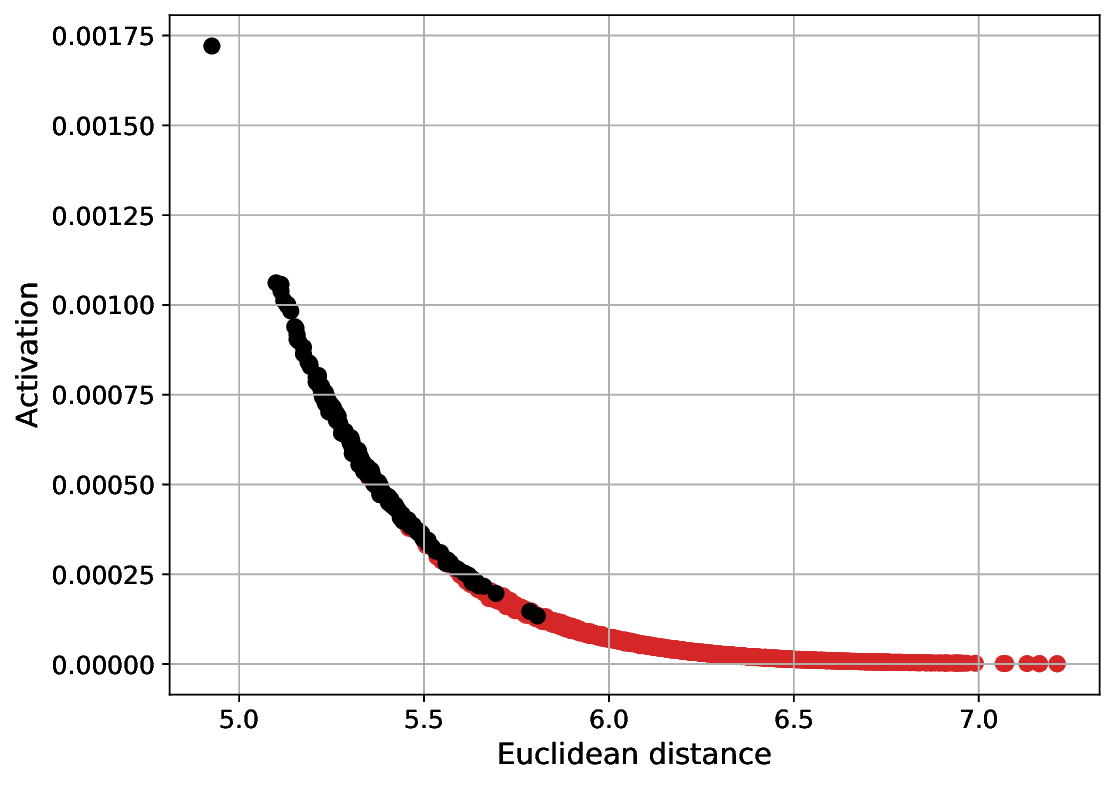}
        \label{fig:act-function_cdqn_digit0}
    }
    \subfloat[]{
        \includegraphics[width=0.31\columnwidth]{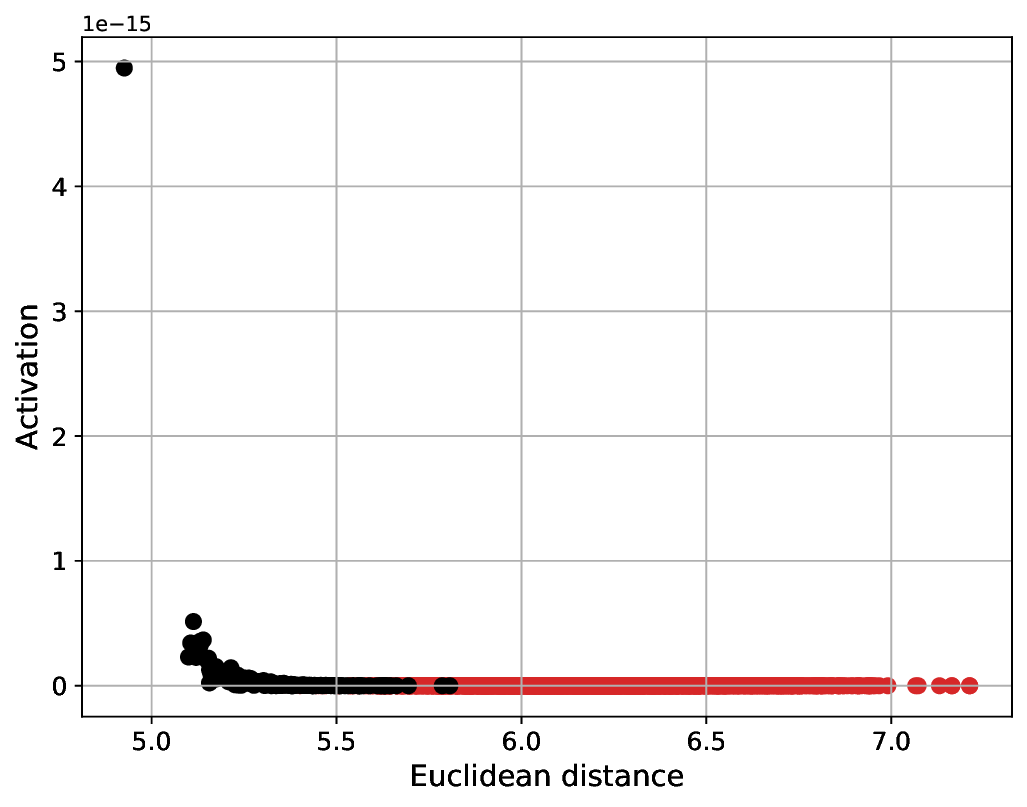}
        \label{fig:act-function_pcdqn_digit0}
    }
    \caption{Best weight vector and activation function concerning such weight vector for each quantum neuron in the task of recognizing the digit 0. (a) and (g) for the LDQN. (b) and (h) for the SDQN. (c) and (i) for the RBDQN. (d) and (j) for the CVQN. (e) and (k) for the CDQN. (f) and (l) for the PCDQN.}
    \label{fig:wvectors&act-functions_digit0}
\end{figure*}

A deeper analysis of the results can be made by examining both the best weight vectors found in the random search and the activation functions with respect to such vectors, as evidenced in Figure~\ref{fig:wvectors&act-functions_digit0} for the digit 0. Figure~\ref{fig:wvector_linear_digit0} – Figure~\ref{fig:wvector_pcdqn_digit0} show that the best weight vector of each quantum neuron has traces of the target digit. Such a similarity is confirmed in Figure~\ref{fig:act-function_linear_digit0} for the LDQN and in Figure~\ref{fig:act-function_sigmoid_digit0} for the SDQN because the highest inner products were obtained by the instances of the digit 0. In the same way, the lowest Euclidean distances were obtained by the black blobs in Figure~\ref{fig:act-function_rbf_digit0} – Figure~\ref{fig:act-function_pcdqn_digit0} for the RBDQN, the CVQN, the CDQN, and the PCDQN respectively. Since the LDQN and the SDQN implement monotonic increments in a discrete manner as the inner product increases, the digit-0 instances obtained the highest values of activation predominantly. However, for the task of recognizing the handwritten digit 0, as supported by Table~\ref{tab:mnist_max_aucroc}, the activation function shapes implemented by the RBDQN and the kernel-based quantum neurons fit even better. Those shapes are similar to exponential decrements as the Euclidean distance increases for the RBDQN, the CDQN, and the PCDQN, while the CVQN implements a linear decay. For the other digits, the neurons also searched for weight vectors similar to the target digit and implemented these neuron activation shapes. Table~\ref{tab:mnist_max_aucroc} already showed how well the neurons performed in fitting their activation shapes with respect to their best weight vectors and the target digits.

\begin{figure}[!t]
    \centering
    \subfloat[]{
        \includegraphics[width=0.38\columnwidth]{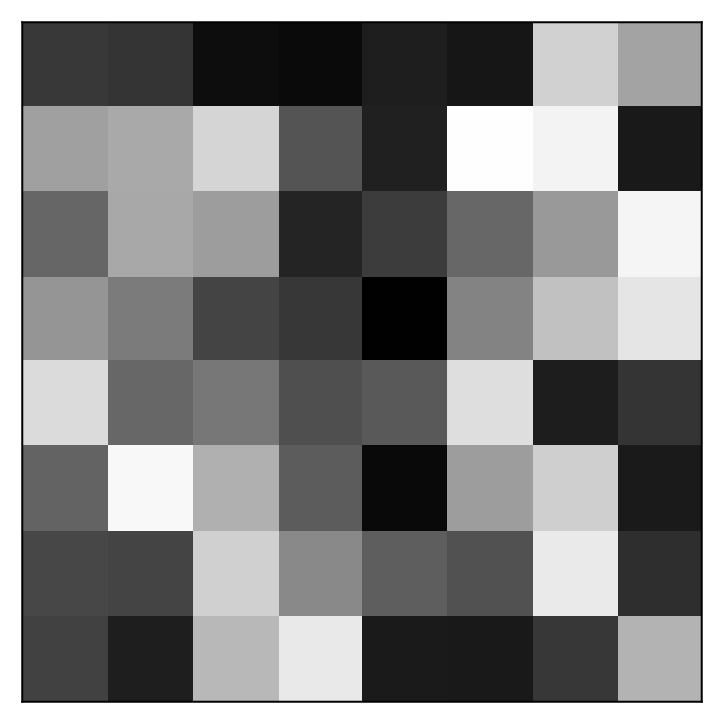}
        \label{fig:wvector_pcdqn_digit1}
    }
    \subfloat[]{
        \includegraphics[width=0.47\columnwidth]{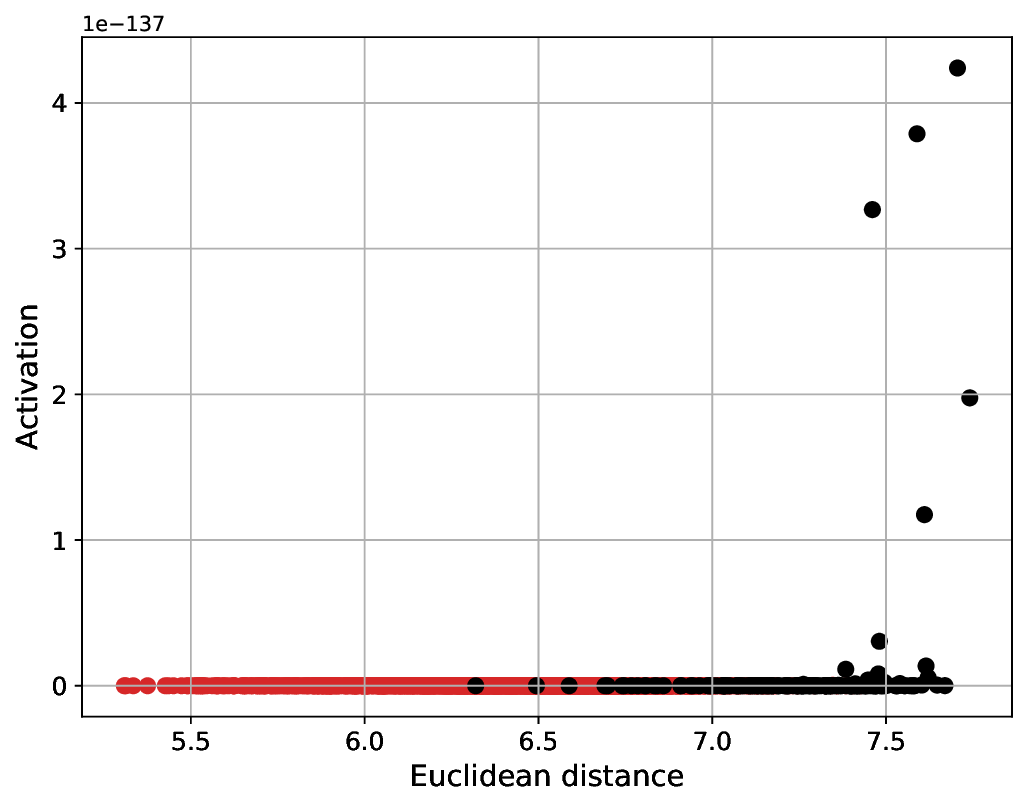}
        \label{fig:act-function_pcdqn_digit1}
    }
    \caption{PCDQN solution for the task of recognizing the digit 1. (a) best weight vector. (b) activation function concerning the best weight vector with $\tau=1/4$ and $\delta=\pi$.}
    \label{fig:pcdqn_digit1}
\end{figure}

Actually, the PCDQN could implement different fitting strategies between target digits by changing the values of $\tau$ and $\delta$. However, the best value of $\delta$ was 0 in all cases, except for the digit 1 where $\delta$ equaled $\pi$. That is, the PCDQN took advantage of the parameter $\tau$ only, which did not provide significant gains in comparison to the CDQN, as shown in Table~\ref{tab:mnist_max_aucroc}. Changing $\tau$ solely controls the decay smooth from linear to exponential shapes as the parameter value increases. The best values of $\tau$ were 2, 1/4, 2, 2, 1/4, 1/4, 1/2, 1/4, 2, and 2, respectively. The actual potential emerges when $\tau$ and $\delta$ are exploited simultaneously, which occurred in the digit 1 where $\tau$ equaled 1/4 and $\delta$ equaled $\pi$. The PCDQN solution for recognizing digit 1 is exhibited in Figure~\ref{fig:pcdqn_digit1}. As can be observed in Figure~\ref{fig:wvector_pcdqn_digit1}, the best weight vector does not have traces of the digit 1 at all, which is confirmed in Figure~\ref{fig:act-function_pcdqn_digit1} because those target instances have the highest Euclidean distances. Figure~\ref{fig:act-function_pcdqn_digit1} also shows that the PCDQN implemented a shape where the activation increases as the Euclidean distance increases instead of the decay that such a neuron implemented in the other cases. In this case, the parametrization changed the activation shape in a way that substantially enhanced the value of AUC ROC. Those results in a real-life problem support the solid conclusion that neurons under the proposed framework can leverage the power of strictly quantum activation functions in classification tasks. As discussed here, we proposed a quantum neuron with better discriminative abilities than the existing ones, especially by parametrizing the activation function.

\section{Conclusions}
\label{sec:conclusions}

In this paper, we proposed a generalized framework to build quantum neurons that apply the kernel trick. Countless quantum neurons can be defined under that framework, including those for actual quantum devices as long as their feature mappings can be implemented in quantum circuits with a limited number of gates. Here, we presented a quantum neuron with a parametrized activation function and an efficient circuit implementation. The parametrization gives flexibility to the neuron since its activation function shape can be changed. That efficient circuit implementation further complies with the constraints of actual quantum devices compared to the existing kernel-based quantum neuron. The first reason is that a circuit with constant depth and a linear number of operations mitigates the errors by decoherence times and gate infidelities. Additionally, qubit encoding generates a straightforward state preparation, where connectivity is not a problem since that encoding strategy generates separable states. Finally, circuit rewriting is facilitated since the circuit uses elementary single-qubit gates.

Each kernel-based quantum neuron implements a fundamentally different activation function. Here, we showed those neuron activation shapes as functions of some relations between the input and weight vectors. While the existing kernel-based neuron is restricted, the parametrized neuron has the flexibility to fit different problem structures. As a first demonstration, the parametrized neuron produced optimal solutions for all nonlinear toy classification problems, including the ones where both the existing kernel-based neuron and the proposed neuron without parametrization performed like random models or worse. We also reproduced those numerical solutions in a simulation environment of quantum circuits as a means to finally validate the neuron implementations. As a last and conclusive test, we contrasted the kernel-based quantum neurons and the discrete quantum neurons in the real problem of recognizing handwritten digits. As a result, the proposed quantum neuron produced the best results in all cases. This repeated evidence of better neuron capabilities in real-life problems, mainly due to the parametrization potential, solidly exhibits the proposed quantum neuron and the proposed framework as effective models for machine learning and its applications.

Future works can discover other quantum neurons under the generalized framework proposed here since the ones used in this work are all special cases. Subsequent works should not limit themselves to studying, extending, and comparing the same kernel-based quantum neurons that we addressed. By using other gates, combining gates, applying heterogeneous gates, or even evolving some evolutionary approach, one can discover better feature mappings for classification problems. A theoretical study can define the bounds of that framework or even prove that it is universal. Delimiting the functions that the framework can approximate or the decision boundaries that can be drawn is fundamental to understanding the class of problems that one can solve as the framework is explored. That class of problems will reveal whether or not the framework can truly lead to practical quantum advantage.

Another example of future research direction is to extend the applications to more classification problems or even other learning tasks. The community can search for problems, also through experimental studies, where those quantum neurons can stand out from the classical models when the input size increases. However, executions in real quantum devices will be required to efficiently explore neuron computations, which also requires that technological advances occur. Additionally, the neuron weights and parameters can be adjusted by variational quantum algorithms~\cite{cerezo_VQA-overview, huang_meta-VQA}. Grid search and random search are certainly not as effective as updating the weights and parameters based on quantum gradients. Finally, a network of quantum neurons under the proposed framework can be constructed. Since a single neuron demonstrated improved abilities here, the connection of multiple of them should be investigated in order to obtain further impressive results.
\section*{Acknowledgments}

This work was financially supported by the Fundação de Amparo à Ciência e Tecnologia do Estado de Pernambuco (FACEPE) under Grant Number IBPG-0084-1.03/20 and Grant Number APQ-1110-1.03/21.



\begin{thebibliography}{10}

\bibitem{nielsen_QC_QI}
M.~A. Nielsen and I.~L. Chuang, {\em Quantum Computation and Quantum
  Information}.
\newblock New York, USA: Cambridge University Press, 2010.

\bibitem{preskill_quantum_supremacy}
J.~Preskill, ``Quantum computing and the entanglement frontier,'' {\em arXiv
  preprint arXiv:1203.5813}, 2012.

\bibitem{nguyen_benchmarking_NNs}
N.~H. Nguyen, E.~C. Behrman, M.~A. Moustafa, and J.~E. Steck, ``Benchmarking
  neural networks for quantum computations,'' {\em IEEE Transactions on Neural
  Networks and Learning Systems}, vol.~31, no.~7, pp.~2522--2531, 2020.

\bibitem{paula-neto_any_nonlinear-QN}
F.~M. de~Paula~Neto, T.~B. Ludermir, W.~R. de~Oliveira, and A.~J. da~Silva,
  ``Implementing any nonlinear quantum neuron,'' {\em IEEE Transactions on
  Neural Networks and Learning Systems}, vol.~31, no.~9, pp.~3741--3746, 2020.

\bibitem{devitt_QEC_review}
S.~J. Devitt, W.~J. Munro, and K.~Nemoto, ``Quantum error correction for
  beginners,'' {\em Reports on Progress in Physics}, vol.~76, no.~7, p.~076001,
  2013.

\bibitem{preskill_NISQ_era}
J.~Preskill, ``Quantum computing in the nisq era and beyond,'' {\em Quantum},
  vol.~2, p.~79, 2018.

\bibitem{leymann_bitter_NISQ-era}
F.~Leymann and J.~Barzen, ``The bitter truth about gate-based quantum
  algorithms in the nisq era,'' {\em Quantum Science and Technology}, vol.~5,
  no.~4, p.~044007, 2020.

\bibitem{zulehner_mapping_to_IBMQX}
A.~Zulehner, A.~Paler, and R.~Wille, ``An efficient methodology for mapping
  quantum circuits to the ibm qx architectures,'' {\em IEEE Transactions on
  Computer-Aided Design of Integrated Circuits and Systems}, vol.~38, no.~7,
  pp.~1226--1236, 2019.

\bibitem{larose_qplatforms_overview}
R.~LaRose, ``Overview and comparison of gate level quantum software
  platforms,'' {\em Quantum}, vol.~3, p.~130, 2019.

\bibitem{acasiete_QW_on_qdev}
F.~Acasiete, F.~P. Agostini, J.~K. Moqadam, and R.~Portugal, ``Implementation
  of quantum walks on ibm quantum computers,'' {\em Quantum Information
  Processing}, vol.~19, no.~426, 2020.

\bibitem{acampora_GA_on_qdev}
G.~Acampora and A.~Vitiello, ``Implementing evolutionary optimization on actual
  quantum processors,'' {\em Information Sciences}, vol.~575, pp.~542--562,
  2021.

\bibitem{tacchino_qneuron_on_qdev}
F.~Tacchino, C.~Macchiavello, D.~Gerace, and D.~Bajoni, ``An artificial neuron
  implemented on an actual quantum processor,'' {\em npj Quantum Information},
  vol.~5, no.~26, 2019.

\bibitem{mangini_CVQN}
S.~Mangini, F.~Tacchino, D.~Gerace, C.~Macchiavello, and D.~Bajoni, ``Quantum
  computing model of an artificial neuron with continuously valued input
  data,'' {\em Machine Learning: Science and Technology}, vol.~1, no.~045008,
  2020.

\bibitem{tacchino_QNN_on_qdev}
F.~Tacchino, P.~Barkoutsos, C.~Macchiavello, I.~Tavernelli, D.~Gerace, and
  D.~Bajoni, ``Quantum implementation of an artificial feed-forward neural
  network,'' {\em Quantum Science and Technology}, vol.~5, no.~4, p.~044010,
  2020.

\bibitem{grant_hierarchical_qclassifiers}
E.~Grant, M.~Benedetti, S.~Cao, A.~Hallam, J.~Lockhart, V.~Stojevic, A.~G.
  Green, and S.~Severini, ``Hierarchical quantum classifiers,'' {\em npj
  Quantum Information}, vol.~4, no.~65, 2018.

\bibitem{cong_convolutional_QNN}
I.~Cong, S.~Choi, and M.~D. Lukin, ``Quantum convolutional neural networks,''
  {\em Nature Physics}, vol.~15, pp.~1273--1278, 2019.

\bibitem{dallaire-demers_gen-adversarial_QNN}
P.-L. Dallaire-Demers and N.~Killoran, ``Quantum generative adversarial
  networks,'' {\em Physical Review A}, vol.~98, no.~1, p.~012324, 2018.

\bibitem{zoufal_QGAN_for_state-preparation}
C.~Zoufal, A.~Lucchi, and S.~Woerner, ``Quantum generative adversarial networks
  for learning and loading random distributions,'' {\em npj Quantum
  Information}, vol.~5, no.~103, 2019.

\bibitem{konar_QFS-Net}
D.~Konar, S.~Bhattacharyya, B.~K. Panigrahi, and E.~C. Behrman,
  ``Qutrit-inspired fully self-supervised shallow quantum learning network for
  brain tumor segmentation,'' {\em IEEE Transactions on Neural Networks and
  Learning Systems}, vol.~33, no.~11, pp.~6331--6345, 2022.

\bibitem{buhrman_swap-test}
H.~Buhrman, R.~Cleve, J.~Watrous, and R.~de~Wolf, ``Quantum fingerprinting,''
  {\em Physical Review Letters}, vol.~87, no.~16, p.~167902, 2001.

\bibitem{havlicek_QKE}
V.~Havl{\'\i}{\v{c}}ek, A.~D. C{\'o}rcoles, K.~Temme, A.~W. Harrow, A.~Kandala,
  J.~M. Chow, and J.~M. Gambetta, ``Supervised learning with quantum-enhanced
  feature spaces,'' {\em Nature}, vol.~567, pp.~209--212, 2019.

\bibitem{schuld_QKE}
M.~Schuld and N.~Killoran, ``Quantum machine learning in feature hilbert
  spaces,'' {\em Physical Review Letters}, vol.~122, no.~4, p.~040504, 2019.

\bibitem{ding_q-inspired_SVM}
C.~Ding, T.-Y. Bao, and H.-L. Huang, ``Quantum-inspired support vector
  machine,'' {\em IEEE Transactions on Neural Networks and Learning Systems},
  vol.~33, no.~12, pp.~7210--7222, 2022.

\bibitem{liu_qspeedup_in_ML}
Y.~Liu, S.~Arunachalam, and K.~Temme, ``A rigorous and robust quantum speed-up
  in supervised machine learning,'' {\em Nature Physics}, vol.~17,
  pp.~1013--1017, 2021.

\bibitem{bravyi_qadvantage_shallow-circ}
S.~Bravyi, D.~Gosset, and R.~K{\"o}nig, ``Quantum advantage with shallow
  circuits,'' {\em Science}, vol.~362, no.~6412, pp.~308--311, 2018.

\bibitem{bravyi_qadvantage_noisy-shallow-circ}
S.~Bravyi, D.~Gosset, R.~K{\"o}nig, and M.~Tomamichel, ``Quantum advantage with
  noisy shallow circuits,'' {\em Nature Physics}, vol.~16, pp.~1040--1045,
  2020.

\bibitem{li_variational_error_minimization}
Y.~Li and S.~C. Benjamin, ``Efficient variational quantum simulator
  incorporating active error minimization,'' {\em Physical Review X}, vol.~7,
  no.~021050, 2017.

\bibitem{temme_error_mitigation}
K.~Temme, S.~Bravyi, and J.~M. Gambetta, ``Error mitigation for short-depth
  quantum circuits,'' {\em Physical Review Letters}, vol.~119, no.~180509,
  2017.

\bibitem{kandala_error_mitigation}
A.~Kandala, K.~Temme, A.~D. C{\'o}rcoles, A.~Mezzacapo, J.~M. Chow, and J.~M.
  Gambetta, ``Error mitigation extends the computational reach of a noisy
  quantum processor,'' {\em Nature}, vol.~567, pp.~491--495, 2019.

\bibitem{stoudenmire_local_feature-map}
E.~M. Stoudenmire and D.~J. Schwab, ``Supervised learning with tensor
  networks,'' in {\em Advances in Neural Information Processing Systems
  (NeurIPS Proceedings)} (D.~Lee, M.~Sugiyama, U.~Luxburg, I.~Guyon, and
  R.~Garnett, eds.), vol.~29, Curran Associates, Inc., 2016.

\bibitem{cover_theorem}
T.~M. Cover, ``Geometrical and statistical properties of systems of linear
  inequalities with applications in pattern recognition,'' {\em IEEE
  Transactions on Electronic Computers}, vol.~EC-14, no.~3, pp.~326--334, 1965.

\bibitem{yanofsky_QC_for_comp-sci}
N.~S. Yanofsky and M.~A. Mannucci, {\em Quantum Computing for Computer
  Scientists}.
\newblock New York, USA: Cambridge University Press, 2008.

\bibitem{haykin_neural-nets}
S.~Haykin, {\em Neural Networks and Learning Machines}.
\newblock New Jersey, USA: Pearson, 2009.

\bibitem{bishop_PRML}
C.~M. Bishop, {\em Pattern Recognition and Machine Learning}.
\newblock New York, USA: Springer, 2006.

\bibitem{rossi_qhypergraph-states}
M.~Rossi, M.~Huber, D.~Bru{\ss}, and C.~Macchiavello, ``Quantum hypergraph
  states,'' {\em New Journal of Physics}, vol.~15, no.~113022, 2013.

\bibitem{qiskit}
M.~S. Anis, Abby-Mitchell, H.~Abraham, AduOffei, R.~Agarwal, G.~Agliardi,
  M.~Aharoni, I.~Y. Akhalwaya, G.~Aleksandrowicz, T.~Alexander, M.~Amy,
  S.~Anagolum, {\em et~al.}, ``Qiskit: An open-source framework for quantum
  computing,'' 2021.

\bibitem{scikit-learn}
F.~Pedregosa, G.~Varoquaux, A.~Gramfort, V.~Michel, B.~Thirion, O.~Grisel,
  M.~Blondel, P.~Prettenhofer, R.~Weiss, V.~Dubourg, J.~Vanderplas, A.~Passos,
  {\em et~al.}, ``Scikit-learn: Machine learning in python,'' {\em Journal of
  Machine Learning Research}, vol.~12, no.~85, pp.~2825--2830, 2011.

\bibitem{numpy}
C.~R. Harris, K.~J. Millman, S.~J. van~der Walt, R.~Gommers, P.~Virtanen,
  D.~Cournapeau, E.~Wieser, J.~Taylor, S.~Berg, N.~J. Smith, R.~Kern, M.~Picus,
  {\em et~al.}, ``Array programming with {NumPy},'' {\em Nature}, vol.~585,
  pp.~357--362, 2020.

\bibitem{yan_any_nonlinear-QN_efficiently}
S.~Yan, H.~Qi, and W.~Cui, ``Nonlinear quantum neuron: A fundamental building
  block for quantum neural networks,'' {\em Physical Review A}, vol.~102,
  no.~052421, 2020.

\bibitem{cerezo_VQA-overview}
M.~Cerezo, A.~Arrasmith, R.~Babbush, S.~C. Benjamin, S.~Endo, K.~Fujii, J.~R.
  McClean, K.~Mitarai, X.~Yuan, L.~Cincio, and P.~J. Coles, ``Variational
  quantum algorithms,'' {\em Nature Reviews Physics}, vol.~3, pp.~625--644,
  2021.

\bibitem{huang_meta-VQA}
R.~Huang, X.~Tan, and Q.~Xu, ``Learning to learn variational quantum
  algorithm,'' {\em IEEE Transactions on Neural Networks and Learning Systems},
  2022.
\newblock early access.

\end{thebibliography}

\end{document}